\documentclass[smallextended]{svjour3}                     
\smartqed  
\usepackage{graphicx}
 \journalname{J Low Temp Phys}

\begin{document}

\title{Quantum correlations of few dipolar bosons in a double-well trap}
\author{Michele Pizzardo$^1$, Giovanni Mazzarella$^{1,2}$, and Luca Salasnich$^{1,2,3}$}
\institute{$^1$ Dipartimento di Fisica e Astronomia "Galileo Galilei", Universit\`a di Padova, Via F. Marzolo 8, 35131 Padova, Italy\\
$^2$ CNISM, Consorzio Interuniversitario per le Scienze Fisiche della Materia, Via F. Marzolo 8, 35131 Padova, Italy\\
$^3$Istituto Nazionale di Ottica (INO) del Consiglio Nazionale delle Ricerche (CNR), Sezione di Sesto Fiorentino, Via Nello Carrara 1, 50019 Sesto Fiorentino, Italy\\
G. Mazzarella [corresponding author]. \email{mazzarella@pd.infn.it}, Tel.:+39 049 8277137 }
\date{\today}

\maketitle
\begin{abstract}
We consider $N$ interacting dipolar bosonic atoms at zero temperature in a double-well potential. This system is described by the two-space-mode extended Bose-Hubbard (EBH) Hamiltonian which includes (in addition to the familiar BH terms) the nearest-neighbor interaction, correlated hopping and bosonic-pair hopping. For systems with $N=2$ and $N=3$ particles we calculate analytically both the ground state and the Fisher information, the coherence visibility, and the entanglement entropy that characterize the correlations of the lowest energy state. The structure of the ground state crucially depends on the correlated hopping $K_c$. On one hand we find that this process makes possible the occurrence of Schr\"odinger-cat states even if the onsite interatomic attraction is not strong enough to guarantee the formation of such states. On the other hand, in the presence of a strong onsite attraction, sufficiently large values of $|K_c|$ destroys the cat-like state in favor of a delocalized atomic coherent state.
\end{abstract}

\maketitle

\section{Introduction}

Ultracold and interacting dilute alkali-metal vapors trapped by one-dimensional double-well potentials \cite{oliver} offer the opportunity to study the formation of macroscopic coherent states \cite{smerzi,stringari,anglin,mahmud,anna} and macroscopic Schr\"odinger-cat states \cite{cirac,dalvit,huang,carr,brand,main,miguelfr}.
The two-site Bose-Hubbard (BH) Hamiltonian \cite{twomode:milburn} efficiently describes the microscopic dynamics of such systems. When the boson-boson interaction is repulsive and the number of bosons is even, the crossover from a delocalized atomic coherent state to a (fully incoherent) localized Fock state (the so called twin Fock state with the particles equally shared between the two wells) takes place by increasing the interatomic coupling strength \cite{stringari,anglin,mahmud,anna,main,galante}. For attractively interacting bosons, the two-spatial mode BH model predicts the formation of a macroscopic Schr\"odinger-cat state \cite{cirac,dalvit,huang,carr,brand,main} when the interatomic attraction becomes sufficiently large. Finally, when the attraction between the bosons is sufficiently strong the collapse should take place \cite{sb,io-e-boris}. In the Schr\"odinger-cat states context, very recently, we have considered the possibility to induce such states in assisted way, thanks to the interaction of the bosonic matter with the radiation field inside an optical resonator, with onsite attractions weaker than those required without cavity radiation \cite{rosson2015}. On the repulsive side, the above described crossover is reminiscent of the quantum phase transition with optical-lattice-confined bosons theoretically predicted in \cite{jaksch} and experimentally observed by Greiner and co-workers \cite{greiner}. This quantum phase transition - induced by varying the depth of the optical potential - is a transition from the superfluid phase (where the hopping dominates the Hamiltonian: in this case each atom is spread out over the entire lattice) to the Mott insulator one (where the onsite interaction dominates the Hamiltonian: in this case, exact numbers of atoms are localized at individual lattice sites). Note that in the presence of strong onsite repulsions, an even-odd (boson number) difference exists. With such interactions, as commented above, when the number of bosons is even the lowest energy state is a separable twin Fock state, while when the number of particles is odd the ground state is a symmetric linear superposition of two Fock states with non-fully populated wells \cite{galante}. This difference, which tends to become less relevant for larger particle numbers, is a well known Mott insulators feature, as discussed, for example, in \cite{oguri}.

The above considerations hold for bosons with negligible dipole or electric moments so that the dipole-dipole interatomic interaction can be be safely neglected.
This is not the case of dipolar bosonic atoms, as for example chromium $^{52}$Cr, characterized by very large atomic magnetic dipoles. The study of dipolar quantum gases began just with $^{52}$Cr \cite{cr}, which has a magnetic momentum of $6\mu_B$ ($\mu_B$ is the Bohr magneton), and recently important results have been achieved with lanthanide atoms like erbium  $^{168}$Er \cite{er} and dysprosium $^{164}$Dy \cite{dy} that have dipole moments of $7\mu_B$ and $10 \mu_B$, respectively. To date, as for what concerns dipolar bosons confined in few-site potentials important theoretical efforts have been devoted to investigate such systems in the presence of double \cite{xiong,blume,abad,mazzdell}, triple \cite{peter,wuner,lahaye2,dmps,muntsa}, and four-well potentials \cite{mazzpennaq}. Remarkably, Abad and co-workers \cite{abad} have shown that the dipolar interaction makes possible self-inducing a double-well potential structure. Moreover, in the context of dipolar bosons confined in triple-well shaped potentials, we have analyzed \cite{dmps} the problem also from the quantum correlations point of view as well by calculating the entanglement entropy \cite{bwae}, and in Ref. \cite{muntsa} the authors have addressed the physics of the problem by stressing the role played by the anisotropic nature of the dipole-dipole interaction potential.

Motivated by the concrete possibility to isolate single atomic ions \cite{heidelberg,wineland1,bergquist,cheinet} and manipulate quantum gases at single-atom level \cite{cheinet,diedrich,monroe,haroche-raimond,nobel},
we focus on the behavior of few trapped bosonic dipolar atoms at zero temperature.
In the present work, then, we aim to study the ground state of a system consisting of a low number $N$ of dipolar bosonic atoms confined in a symmetric double-well trap and to characterize such a state from the quantum correlations point of view. To do this we use the two-site extended Bose-Hubbard model (EBH). The Hamiltonian of this model includes, in addition to the standard nearest-neighbor hopping and onsite interaction, also the terms describing interaction, correlated, and bosonic-pair hopping with bosons in nearest-neighbor wells. In fact, these (beyond-onsite) terms --widely studied for bosons loaded in optical lattices \cite{mgi,eckolt,dutta1,dutta2}--  that originate from the contribution (to the second quantized Hamiltonian) describing two-body interactions (involving the product of four field operators, see Sec. 2) cannot be neglected since the long-range nature of the dipole-dipole interaction. One thus refers to the correlated hopping as density-induced \cite{observation} or collisionally-induced tunneling. Note that in 2012 Maluckov {\it et al.} \cite{boris} have studied, in the mean-field limit, the ground state of dipolar Bose-Einstein condensates in optical lattices.

We diagonalize the two-site extended Bose-Hubbard Hamiltonian by analytically calculating the eigenvector and the eigenvalue of the lowest energy state state with $N=2$ and $N=3$ bosons. Hence, we provide analytical formulas for the parameters that describe the correlation properties of the ground state of the dipolar bosons. Note that in Ref. \cite{galante}, we have analytically calculated the ground state and correlation-characterizing parameters for a two-site Bose-Hubbard model with $N=2,3,4$ (nondipolar) bosons. In the present paper, we calculate the Fisher information $F$ \cite{braunstein,pezze}. This is related to the fluctuation of the number of bosons in a single well and achieves its maximum in correspondence to the Schr\"odinger-cat state. We consider the coherence visibility $\alpha$ \cite{stringari,anglin,anna,baym} which measures the coherence induced by the single-particle tunneling --across the central barrier-- that attains its maximum value in correspondence to the atomic coherent state. Finally, we drive our attention on the entanglement entropy $S$ \cite{bwae} which characterizes the genuine quantum correlations of the ground state from the bipartition perspective. To make simpler, but not trivial, our analysis and motivated by the recent review of Dutta and co-workers \cite{dutta2} and careful analysis by Xiong and Fischer \cite{uwe} in the context of the interaction-induced coherence in few-site-trapped bosons systems, we assume that the only active beyond-onsite process is the correlated hopping (with the nearest-neighbor interaction reabsorbed in an effective onsite process). Thus we study the ground state and the parameters $F$, $\alpha$, $S$ by widely exploring the density-induced tunneling range - once fixed the onsite interaction - which, as discussed in \cite{dutta2}, may change sign depending on the trap characteristics. In this way, we analyze the role of the correlated hopping in determining the form of the ground state sustained by the two-mode EBH. We point out that even if the onsite interaction is not strong enough for the formation of the Schr\"odinger-cat state (attractive onsite interactions) or Fock states (repulsive onsite interactions), the presence of the collisionally-induced tunneling makes possible such circumstances. This is corroborated by a complete analysis of the Fisher information, coherence visibility, and entanglement entropy performed by varying the onsite interaction and keeping fixed the correlated hopping. We investigate, moreover, the role of the density-induced tunneling in the occurrence of the atomic coherent state as ground state. To do this we fix the onsite interaction in such a way to have a cat-like-shaped (strong attractions) or Fock states-structured (strong repulsions) ground state. Then, by amply investigating the density-induced tunneling range, we observe that the correlated hopping tends to destroy the two ground states above in favor of a delocalized atomic coherent state. Also in this case, a detailed analysis of the ground-state quantum correlations indicators gives us the possibility to complete and support our results.

\section{The system}

We consider $N$ identical interacting dipolar bosonic atoms at zero temperature. We suppose that these atoms are confined by an external potential $V_{trap}(\bf{r})$ achieved by superimposing an isotropic harmonic confinement -- generated in the radial (transverse) plane ($y-z$) -- to a symmetric a double-well $V_{DW}$ -- created in the axial ($x$) direction -- namely
$V_{trap}({\bf r}) =V_{DW}(x)+ m\omega_{\bot}^2(y^2+z^2)/2$,
where $m$ is the mass of the bosons and $\omega_{\bot}$ the trapping frequency in the transverse directions. The system can be treated as quasi-one-dimensional since the above radial harmonic confinement is assumed to be very strong. Under the condition that the energy per particle in the axial direction is much smaller than the transverse level spacing $\hbar \omega_{\bot}$, one can assume that the bosons stay in the ground state of the transverse harmonic oscillator, i.e. $w(y,z)=\exp[-(y^2+z^2)/2l_{\bot}^2]/(l_\bot \sqrt{\pi})$ with $l_{\bot}=\sqrt{\hbar/m\omega_{\bot}}$ being the transverse harmonic oscillator length.
%
The model describing our system can be derived from the bosonic-field Hamiltonian
\begin{eqnarray}
\label{system:ham0}
\hat{H} &=& \int d^{3}{\bf r}\,\hat{\Psi}^\dagger({\bf r})\,(-\frac{\hbar^2}{2m}\nabla^2+V_{trap}({\bf r}))\,\hat{\Psi}({\bf r}) \nonumber\\
&+&\frac{1}{2}\int d^{3} {\bf r}d^{3} {\bf r'}\hat{\Psi}^\dagger({\bf r})\hat{\Psi}^\dagger({\bf r'}) V({\bf r}-{\bf r'})
\hat{\Psi}({\bf r'})\hat{\Psi}({\bf r})
\;.\end{eqnarray}
The field operators $\hat{\Psi}({\bf r})$ and $\hat{\Psi}^\dagger({\bf r})$ that annihilates and creates, respectively, a boson at the position ${\bf r}$ satisfy the bosonic commutation rules: $[\hat{\Psi}({\bf r}),\hat{\Psi}^\dagger({\bf r'})]=\delta^{(3)}({\bf r}-{\bf r'})$, and $[\hat{\Psi}({\bf r}),\hat{\Psi}({\bf r'})]=0=[\hat{\Psi}({\bf r})^\dagger,\hat{\Psi}^\dagger({\bf r'})]$.
Potential $V({\bf r}-{\bf r'}) =$ $ g\, \delta^3 ({\bf r}-{\bf r'})+V_{dd}({\bf r}-{\bf r'})$,
describing boson-boson interactions, is the sum of a short-range (sr) $g$-dependent contact
potential ($g=4\pi\hbar^2 a_s/m$ with $a_s$ the interatomic s-wave scattering length)
and of a long-range dipole-dipole (dd) potential
\begin{equation}
\label{interaction}
V_{dd}({\bf r}-{\bf r'})=\gamma \frac{1-3 \cos^2 \theta}{|{\bf r}-{\bf r'}|^3}\;.
\end{equation}
The coupling of dipoles through the relevant magnetic moment $\mu$ (electric moment $d$)
is embodied in $\gamma=\mu_0\mu^2/4\pi$ ($\gamma=d^2/4\pi \varepsilon_0$) where
$\mu_0$ ($\varepsilon_0$) is the vacuum magnetic susceptibility (vacuum dielectric constant).
The relative position of the particles is given by the vector ${\bf r}-{\bf r'}$.
For external (electric or magnetic) fields large enough the boson dipoles are aligned along
the same direction, so that $\theta$ is the angle between the vector ${\bf r}-{\bf r'}$ and
the dipole orientation.
%
Bosonic field operator $\hat{\Psi}({\bf r})$ can be expanded in terms of annihilation operators $\hat{a}_{k}$
\begin{eqnarray}
\label{expansion}
&&\hat{\Psi}({\bf r})=\sum_{k=L,R}
\Phi_{k}({\bf r})\; \hat{a}_{k}\;
\;.\end{eqnarray}
%
Operators $\hat{a}_{k}$ and $\hat{a}^{\dagger}_{k}$  obey the usual bosonic algebra, that is $[\hat{a}_{k},\hat{a}^{\dagger}_{l}]=\delta_{k,l}$. Functions $\Phi_{k}$ are orthonormal between each other and, owing to the form of the external potential $V_{trap}$, they may be written as the product of two functions: a function depending on the radial coordinates ($y,z$) (i.e., the ground state of the transverse harmonic oscillator $w(y,z)$) and the other one on the axial coordinate ($x$), say $\phi_k(x)$ ($k=L,R$), so that $\Phi_{k}({\bf r})=\phi_{k}(x)w(y,z)$. Here $\phi_L(x)$ and $\phi_R(x)$ are orthonormal real single-particle wavefunctions localized about the minima of the left ($L$) and right ($R$) wells. In this scheme we interpret $\hat{a}_{k}$ ($\hat{a}^{\dagger}_{k}$) as operator annihilating (creating) a boson in the well $k$.


%
%
By using the expansion (\ref{expansion}) -- and its Hermitian conjugate -- at the right-hand side of Eq. (\ref{system:ham0}) and the spatial symmetry of $V_{DW}$, one achieves the following two-space-mode (note that we often use the term two-site in place of two-space-mode) extended Bose-Hubbard (EBH) Hamiltonian:
\begin{eqnarray}
\label{ham:ebh}
&&\hat{H} = -J(\hat{a}_L^\dagger\hat{a}_R + \hat{a}_R^\dagger\hat{a}_L) +\frac{U_0}{2}\big(\hat{n}_L(\hat{n}_L-1)+\hat{n}_R(\hat{n}_R-1)\big)\nonumber\\
&+&U_1 \hat{n}_L\,\hat{n}_R+K_c\,(\hat{a}_L^\dagger\hat{n}_L\hat{a}_R+\hat{a}_R^\dagger\hat{n}_L\hat{a}_L+\hat{a}_R^\dagger\hat{n}_R\hat{a}_L+\hat{a}_L^\dagger\hat{n}_R\hat{a}_R)\nonumber\\
&+&K_p\,(\hat{a}_L^\dagger\hat{a}_L^\dagger \hat{a}_R \hat{a}_R+\hat{a}_R^\dagger\hat{a}_R^\dagger \hat{a}_L \hat{a}_L)
\;.\end{eqnarray}
Operator $\hat{n}_k = \hat{a}^\dagger_k\hat{a}_k$ counts bosons in the $k$th well. The Hamiltonian (\ref{ham:ebh}) commutes with the total number operator $\hat{N}=\hat{n}_L+\hat{n}_R$.
Quantity $J$ is the interwell tunnel matrix element. By using the fact that the $\Phi$'s are orthonormal, the explicit form of $w(y,z)$ (see above), and by integrating over the transverse directions, it is possible to provide a formula which states the dependence of the hopping amplitude on the microscopic parameters of the system:
\begin{equation}
\label{j}
J =-\int_{-\infty}^{+\infty} dx\,\phi_{L}(x)\,\big(-\frac{\hbar^2}{2m}\frac{d^2}{dx^2}+V_{DW}(x)\big)\,\phi_R(x)
\;.\end{equation}
Amplitudes $U_0$, $U_1$, $K_c$ and $K_p$ derive from the second row of Eq. (\ref{system:ham0}), so that the processes that they describe are interaction-induced phenomena. These amplitudes are given by the sum of two contribution. The first contribution is due to the short-range part \cite{mazzmoratti} of the interaction potential $V({\bf r}-{\bf r'})$, while the second one to the dipole-dipole interaction. As for what concerns the short-range contributions we are able (by following the same path followed for the hopping amplitude) to write down formulas similar to that for $J$ [Eq. (\ref{j})]. $U_0$ measures the strength of the boson-boson interaction in the same well (onsite or intrawell interaction)
\begin{equation}
\label{u0}
U_0 =\tilde g\int_{-\infty}^{+\infty}dx (\phi_k(x))^4+U_{kkkk}
\;.\end{equation}
The interaction between bosons in adjacent wells (nearest-neighbor interaction) is characterized by strength $U_1$ given by
\begin{equation}
\label{u1}
U_1 =2\tilde g\int_{-\infty}^{+\infty}dx (\phi_k(x))^2(\phi_l(x))^2+(U_{klkl}+U_{kllk})
\;.\end{equation}
$K_c$ and $K_p$ are the amplitudes of correlated (or density- or collisionally-induced) and pair hopping processes and read, respectively
\begin{equation}
\label{kc}
K_c =\tilde g\int_{-\infty}^{+\infty}dx (\phi_k(x))^3(\phi_l(x))+U_{kkkl}
\;,\end{equation}
\begin{equation}
\label{kp}
K_p =\frac{\tilde g}{2}\int_{-\infty}^{+\infty}dx (\phi_k(x))^2(\phi_l(x))^2+\frac{1}{2}\,U_{kkll}
\;.\end{equation}
In these equations $k,l=L,R$ ($k \ne l$), and $\tilde g=g/2\pi l_\bot^2$ [with $l_{\bot}$ and $g$ defined before and after Eq. (\ref{system:ham0}), respectively]
and
\begin{eqnarray}
\label{ddamplitudes}
&&U_{ijkl}=\int\, d^{3} {\bf r}\,d^{3} {\bf r'}\Phi_{i}({\bf r})\Phi_{j}({\bf r'}) V_{dd}({\bf r}-{\bf r'})
\Phi_{k}({\bf r'})\Phi_{l}({\bf r})\;,
\end{eqnarray}
where $i,j,k,l=L,R$, and $V_{dd}({\bf r}-{\bf r'})$ is given by Eq. (\ref{interaction}).

Due to the conservation of the total boson-number, it can be easily proved that the nearest-neighbor interaction can be reabsorbed in the onsite term which will then describe an effective onsite interaction having amplitude $U=U_0-U_1$. We have thus finally the following Hamiltonian:
\begin{eqnarray}
\label{ham:effebh}
&&\hat{H} = -J(\hat{a}_L^\dagger\hat{a}_R + \hat{a}_R^\dagger\hat{a}_L) +\frac{U}{2}\big(\hat{n}_L(\hat{n}_L-1)+\hat{n}_R(\hat{n}_R-1)\big)\nonumber\\
&+&K_c\,(\hat{a}_L^\dagger\hat{n}_L\hat{a}_R+\hat{a}_R^\dagger\hat{n}_L\hat{a}_L+\hat{a}_R^\dagger\hat{n}_R\hat{a}_L+\hat{a}_L^\dagger\hat{n}_R\hat{a}_R)\nonumber\\
&+&K_p\,(\hat{a}_L^\dagger\hat{a}_L^\dagger \hat{a}_R \hat{a}_R+\hat{a}_R^\dagger\hat{a}_R^\dagger \hat{a}_L \hat{a}_L)
\;.\end{eqnarray}
At this point some observations are in order. As commented above, the amplitudes $U_0$, $U_1$, $K_c$ and $K_p$ [see Eqs. (\ref{u0})-(\ref{kp})] are given by the sum of two contributions, with the short-range contribution which affects mainly $U_0$ (which involves bosons on the same site), whereas for $U_1$, $K_c$ and $K_p$ (that involve bosons in adjacent wells) basically only the dipole-dipole interaction contributes \cite{lahaye2,dmps}. In this scenario, the effective on-site interaction of amplitude $U$ [see the first row of Eq. (\ref{ham:effebh})] can be controlled by separately tuning the s-wave scattering length (short-range potential) and the features of the dipole-dipole interaction potential. In this respect, for example, we have demonstrated \cite{dmps} that the condition $U<0$ (necessary to have the cat-like state as ground state of dipolar bosons in ring-geometry-triple-well potentials) can be met with $U_0$ and $U_1$ positive (by-passing thus the collapse of the bosonic cloud \cite{sb,io-e-boris}). Moreover, from the two last rows of the Hamiltonian (\ref{ham:effebh}), one can see that the dipolar effects are explicitly embodied in the collisionally-induced tunneling (of amplitude $K_c$) and in the pair-bosonic hopping (of amplitude $K_p$) terms.

{\it Quantum analysis}. We are assuming that the system is at zero temperature so that only the ground state of the two-site extended Bose-Hubbard Hamiltonian $\hat{H}$ [Eq. (\ref{ham:effebh})] is populated. To find the lowest energy state of this Hamiltonian we have to solve the eigenproblem
\begin{equation}
\label{eigenproblem}
\hat {H} |E_j\rangle = E_j |E_j \rangle
\end{equation}
for a fixed number $N$ of particles. The Hamiltonian $\hat {H}$ can be represented by a $(N+1)\times(N+1)$ matrix in the Fock basis $|i,N-i\rangle=|i\rangle_L \otimes |N-i\rangle_R$ (with $\otimes$ denoting the tensor product) with $i=0,...,N$ [$i$ ($N-i$) is the left(right)-boson population]. For each eigenvalue $E_j$, where $j=0,1,...,N$, the corresponding eigenstate $|E_j\rangle$ has the form
\begin{equation}
\label{eigenstate}
|E_j\rangle=\sum_{i=0}^{N}\,c_{i}^{(j)} \, |i,N-i\rangle \;
\end{equation}
with $|c_i^{(j)}|^2$ the probability to find $i$ ($N-i$) bosons in the left (right) well when the system is in the $j$th eigenstate of the two-site EBH Hamiltonian.
Note that since the left-right symmetry of our Hamiltonian, one has that
\begin{equation}
\label{lrs}
\langle E_j| \hat{n}_L| E_j \rangle=\langle E_j| \hat{n}_R| E_j \rangle
\;.\end{equation}
We are interested only in the ground state: $j=0$.
In the absence of the beyond-onsite terms (i.e., those having amplitude $K_c$, $K_p$), Eq. (\ref{ham:effebh}) gives back the familiar two-site Bose-Hubbard Hamiltonian. The ground state has been analyzed in details in \cite{main,galante} where we have studied the changes of structure of the ground state determined by the interplay between the onsite interaction $U$ and the hopping amplitude $J$ controlled by the ratio $\zeta =U/J$. At this point let us briefly recall some (remarkably) limit cases.
\begin{itemize}
\item $\zeta=0$. The ground state is the atomic coherent state \cite{arecchi}
\begin{equation}
\label{ACS}
|ACS\rangle = \frac{1}{\sqrt{N!}}(\frac{1}{\sqrt{2}}\big(\hat{a}_L^\dagger+\hat{a}_R^\dagger)\big)^N|0,0\rangle
\;,\end{equation}
where $|0,0\rangle=|0\rangle_L\otimes|0\rangle_R$ is the tensor product between the vacuum of the operator $\hat{a}_L$ and the vacuum of $\hat{a}_R$, that is there are no particles in the left well and no particles in the right well.
\item $U>0$ : $\zeta\rightarrow+\infty$. In the case of a strong repulsive onsite interaction and with an even number $N$ of bosons, as well known, the ground state tends to the twin Fock state
\begin{equation}
\label{fock}
|FOCK\rangle = \bigg|\frac{N}{2},\frac{N}{2}\bigg\rangle
\;.\end{equation}
If $N$, instead, is odd, when  $\zeta\rightarrow+\infty$ the ground state tends to \cite{galante}
\begin{equation}
\label{pseudo:fock}
|pseudoFOCK\rangle = \frac{1}{\sqrt{2}}\Big(\Big|\frac{N-1}{2},\frac{N+1}{2}\Big\rangle + \Big|\frac{N+1}{2},\frac{N-1}{2}\Big\rangle\Big)
\;.\end{equation}
\item $U<0$ : $\zeta\rightarrow-\infty$. In the case of a strong attractive onsite interaction, the ground state tends to the superposition state
\begin{equation}
\label{cat}
|CAT\rangle=\frac{1}{\sqrt{2}}(|N,0\rangle+|0,N\rangle)
\;.\end{equation}
\end{itemize}
This state, frequently called NOON state, is the boson-version of the Schr\"odinger cat state \cite{cirac,dalvit,huang,carr,brand,main}.

In the forthcoming part of the paper, we analyze the role of the beyond-onsite contributions of the Hamiltonian (\ref{ham:effebh}) in achieving the states listed above as ground state of this Hamiltonian. In particular, to make simpler but not trivial this analysis and motivated by the recent review by Dutta {\it et al.} \cite{dutta2} we assume that, among the crosswell terms, the correlated hopping is the only active process (i.e., $K_p=0$).

\section{Analysis parameters}
In this section we introduce the parameters that we use to study the correlations of the ground state of the two-site extended BH Hamiltonian (\ref{ham:effebh}). These parameters are the Fisher information, the coherence visibility, and the entanglement entropy.

\begin{itemize}
\item {\it Fisher Information.}\\
The quantum Fisher information $F_{QFI}$ is the quantity \cite{braunstein,pezze}
\begin{equation}
\label{qfi}
F_{QFI} =(\Delta\hat{n}_{L,R})^2 = \langle(\hat{n}_L-\hat{n}_R)^2\rangle - (\langle\hat{n}_L-\hat{n}_R\rangle)^2
\;,\end{equation}
where the expectation values are taken with respect to the ground state $|E_0\rangle$.
The right-hand side of Eq. (\ref{qfi}) measures the fluctuations of the interwell population imbalance due
to the particles tunneling between the two wells. Quantity $F_{QFI}$ can be used to characterize the collective transfer of bosons across the central barrier.
In terms of the coefficients $c_{i}^{(0)}$, $F_{QFI}$ reads:
\begin{equation}
F_{QFI} = \sum_{i=0}^N (2i-N)^2|c_{i}^{(0)}|^2
\;.\end{equation}
It is convenient to normalize $F_{QFI}$ at its maximum value $N^2$ by defining the Fisher information $F$ as
\begin{equation}
\label{fi}
F=\frac{F_{QFI}}{N^2}
\;,\end{equation}
so that we have a quantity ranging in $[0,1]$:
\begin{equation}
\label{fi2}
F_{QFI} = \frac{1}{N^2}\sum_{i=0}^N (2i-N)^2|c_{i}^{(0)}|^2
\end{equation}
which is equal to $1$ for the NOON state (\ref{cat}).

\item {\it Coherence visibility.}\\
In ultracold atom physics, the investigation of the coherence properties in terms of the momentum distribution $n(p)$ is a customary task. This quantity is the Fourier transform of the one-body density matrix $\rho_1(x,x')$ \cite{stringari,anglin,anna}:
\begin{equation}
\label{np}
n(p)=\int dx dx'\exp\big(-ip(x-x')\big)\,\rho_1(x,x') \;,
\end{equation}
where
\begin{equation}
\rho_1(x,x')=\langle \hat{\Psi}(x)^{\dagger}\hat{\Psi}(x')\rangle \;.
\end{equation}
Operators $\hat{\Psi}(x)$ and $\hat{\Psi}^{\dagger}(x)$ (obeying the standard bosonic commutation rules) annihilates and creates,
respectively, a boson at the point $x$, and the average $\langle...\rangle$ is intended calculated in the ground state.
By following Refs. \cite{stringari,anglin,anna}, it is possible to show that $n(p)$ is
\begin{equation}
\label{npexp}
n(p)= n_0(p) \bigg(1 +\alpha \cos\big(pd \big) \bigg) \;.
\end{equation}
Function $n_0(p)$ [depending on the shape of the double-well potential $V_{DW}(x)$] is the momentum distribution in the fully incoherent regime, and
$d$ is the distance between the two minima of $V_{DW}(x)$. Quantity $\alpha$ measures the visibility of the interference fringes and is given by \cite{baym}
\begin{equation}
\label{visibility}
\alpha=\frac{2\,|\langle \hat{a}^{\dagger}_L\hat{a}_R\rangle|}{N} \;
\end{equation}
with the expectation value taken with respect to the ground state. $\alpha$ characterizes the degree of coherence, between the two wells, associated to the left-right (and back) tunneling across the barrier.

We can write the coherence visibility (\ref{visibility}) in terms of the coefficients $c_{i}^{(0)}$:
\begin{equation}
\label{visibility2}
\alpha = \frac{2}{N}|\sum_{i=0}^N c_{i}^{(0)} \bar c_{i+1}^{(0)} \sqrt{(i+1)(N-i)}|
\;,\end{equation}
where $\bar c$ is the complex conjugate of $c$. $\alpha$ is maximum, that is $1$, for the atomic coherent state (\ref{ACS}).

\item {\it Entanglement entropy.}\\
Finally, it is interesting to analyze the genuine quantum correlations of the ground state $|E_0\rangle$. In particular, we study the quantum entanglement
of $|E_0\rangle$ from the bipartition perspective with the two wells playing the role of the two partitions.  When the system is in $|E_0\rangle$, the
density matrix $\hat{\rho}$ is
\begin{equation}
\label{dm}
\hat{\rho} =|E_0\rangle\langle E_0|
\;.\end{equation}
The system is in a pure state so that an excellent measure of the entanglement between the two wells is provided by the entanglement entropy $S$ \cite{bwae} which is the von Neumann entropy of the reduced density matrix $\hat{\rho}_{L(R)}$ defined by
\begin{equation}
\hat{\rho}_{L(R)} =Tr_{R(L)} \hat{\rho}
\;.\end{equation}
The latter is the matrix obtained by tracing out, from the total density matrix (\ref{dm}), the right (left) well [note that $\hat{\rho}_{L}=\hat{\rho}_{R}$]. By using the definition of trace of a matrix, we have that the entanglement entropy
\begin{equation}
\label{ee0}
S=-Tr \hat{\rho}_{L(R)} \log_{2}\hat{\rho}_{L(R)}
\;,\end{equation}
in terms of the coefficient $c_{i}^{(0)}$ reads
\begin{equation}
\label{ee}
S=-\sum_{i=0}^{N}|c_{i}^{(0)}|^2\log_{2}|c_{i}^{(0)}|^2
\;.\end{equation}
For a given number of bosons $N$, the theoretical maximum value of $S$ is $\log_2(N+1)$ that would correspond to the situation in which the quantities $|c_{i}^{(0)}|^2$ are all equal to $1/(N+1)$.
\end{itemize}

\section{The ground state: analytical results for a few-dipolar atom system}
In this section, we calculate the ground state of the two-site extended Bose-Hubbard Hamiltonian when $N=2$ and $N=3$. For these two cases, we present analytical results for the eigenvalues and eigenvectors of the lowest energy state and calculate analytically the Fisher information (\ref{fi}), the coherence visibility (\ref{visibility}), and the entanglement entropy (\ref{ee}). As we have said at the end of Sec. 2, we study the particular case with $K_p=0$. We then analyze the structure of the ground state and characterize this state by studying $F$, $\alpha$, $S$ in terms of the scaled onsite interaction $\zeta=U/J$ and scaled density-induced tunneling $\kappa=K_c/J$.

{\it Ground state with $N=2$ and $N=3$: analytical results for eigenvectors and eigenvalues}. Let us start with $N=2$. The coefficients $c_{i}^{(0)}$ of the ground state
$$|E_0\rangle=c_{0}^{(0)} |0,2\rangle+c_{1}^{(0)} |1,1\rangle+c_{2}^{(0)} |2,0\rangle$$
are
\begin{eqnarray}
\label{gs2}
&&c_{0}^{(0)}=\frac{\sqrt{A_2-\zeta}}{2\sqrt{A_2}}\nonumber\\
&&c_{1}^{(0)}=-\frac{\sqrt{A_2}\big(A_2+\zeta\big)}{4\sqrt{2}(\kappa-1)\sqrt{A_2-\zeta}}\nonumber\\
&& A_2=\sqrt{16(\kappa-1)^2+\zeta^2}
\;\end{eqnarray}
with $c_{2}^{(0)}=c_{0}^{(0)}$. The associated energy $E_0$ reads
\begin{equation}
\label{energy2}
E_0=\frac{1}{2}(\zeta-\sqrt{\zeta^2+16\kappa^2-32\kappa+16})
\;.\end{equation}
When $N=3$, the ground state has the form
$$|E_0\rangle=c_{0}^{(0)} |0,3\rangle+c_{1}^{(0)} |1,2\rangle+ c_{2}^{(0)} |2,1\rangle+c_{3}^{(0)}|3,0\rangle.$$
In this situation we have shown that two cases are possible. The first case corresponds to $\kappa<\frac{1}{2}$. We have that
\begin{eqnarray}
\label{gs31}
&&c_{0}^{(0)}=\frac{\sqrt{3}}{2\,A_{3,<}}\nonumber\\
&&c_{1}^{(0)}=\frac{(1-2\kappa)\,A_{3,<}}{2\,B_{3,<}}\nonumber\\
&&A_{3,<}=\sqrt{\frac{\zeta  \left(B_{3,<}-4 \kappa +2\right)-(2 \kappa -1) \left(B_{3,<}-8 \kappa +4\right)+\zeta ^2}{(1-2 \kappa )^2}}\nonumber\\
&&B_{3,<}=\sqrt{\zeta ^2-4 (\zeta +4) \kappa +2 \zeta +16 \kappa ^2+4}\;\nonumber\\
\end{eqnarray}
with $c_{3}^{(0)}=c_{0}^{(0)}$ and $c_{2}^{(0)}=c_{1}^{(0)}$. The eigenvalue $E_0$ associated to this state is
\begin{equation}
\label{energy31}
E_0=-1+2(\kappa+\zeta)-\sqrt{4(1-2\kappa)^2+2\zeta(1-2\kappa+\zeta)}
\;.\end{equation}
The second case is that corresponding to $\kappa>\frac{1}{2}$. One has
\begin{eqnarray}
\label{gs32}
&&c_{0}^{(0)}=\frac{1}{A_{3,>}}\nonumber\\
&&c_{1}^{(0)}=\frac{B_{3,>}+\zeta +2 \kappa -1}{\sqrt{3}\,(2\kappa-1)\,A_{3,>}}\nonumber\\
&&A_{3,>}=\sqrt{\frac{2 \left(B_{3,>}+\zeta +2 \kappa -1\right)^2}{3 (1-2 \kappa )^2}+2}\nonumber\\
&&B_{3,>}=\sqrt{\zeta ^2+4 (\zeta -4) \kappa -2 \zeta +16 \kappa ^2+4}\;\nonumber\\
\end{eqnarray}
with $c_{3}^{(0)}=c_{0}^{(0)}$ and $c_{2}^{(0)}=-c_{1}^{(0)}$. The associated $E_0$ reads
\begin{equation}
\label{energy32}
E_0=1-2(\kappa-\zeta)-\sqrt{4(1-2\kappa)^2-2\zeta(1-2\kappa-\zeta)}
\;.\end{equation}

{\it Ground state with $N=2$ and $N=3$: analytical formulas for $F$, $\alpha$, $S$}. Here we provide analytical formulas for the Fisher information, the coherence visibility, and the entanglement entropy when the number of bosons in the system is equal to $2$ and $3$.

Let us start with the Fisher information. We use at the right-hand side of Eq. (\ref {fi2}) the expressions for the coefficients $c_{i}^{(0)}$ given by Eq. (\ref{gs2}) when $N=2$, and when $N=3$ by Eq. (\ref{gs31}), $\kappa<1/2$, and by Eq. (\ref{gs32}), $\kappa>1/2$ (note that the $c_i$'s are real for any $N$, so that $\bar c_{i}=c_{i}$). For $N=2$ we obtain
\begin{equation}
\label{f2}
F=\frac{1}{2}-\frac{\zeta}{2\,A_2}
\;,\end{equation}
where $A_2$ is provided by the third row of Eq. (\ref{gs2}).

When $N=3$ and $\kappa<\frac{1}{2}$ one gets
\begin{equation}
\label{f31}
F=\frac{5\,B_{3,<}-4\,\zeta +8\,\kappa -4}{9\,B_{3,<}}
\;,\end{equation}
[with $B_{3,<}$ provided by the last row of Eq. (\ref{gs31})] while when $\kappa>\frac{1}{2}$, we have
\begin{equation}
\label{f32}
F=\frac{5\,B_{3,>}-4\,\zeta -8\,\kappa +4}{9\,B_{3,>}}
\;,\end{equation}
where $B_{3,>}$ is provided by the fourth row of Eq. (\ref{gs32}).

To obtain the coherence visibility $\alpha$, we use at the right-hand side of Eq. (\ref{visibility2}) the form of the $c_{i}^{(0)}$'s provided by Eq. (\ref{gs2}) when $N=2$, and when $N=3$ by Eq. (\ref{gs31}), $\kappa<1/2$, and by Eq. (\ref{gs32}), $\kappa>1/2$. When $N=2$, we get
\begin{equation}
\label{alpha2}
\alpha= \frac{4\left|\kappa-1\right|}{A_2}
\;,\end{equation}
where $A_2$ is provided by the third row of Eq. (\ref{gs2}).
When $N=3$ and $\kappa<\frac{1}{2}$, we have
\begin{equation}
\label{alpha31}
\alpha=\frac{\left|B_{3,<}+\zeta -8\,\kappa +4\right|}{3\,B_{3,<}}
\;,\end{equation}
[with $B_{3,<}$ being provided by the last row Eq. (\ref{gs31})] while for $\kappa>\frac{1}{2}$, one obtains
\begin{equation}
\label{alpha32}
\alpha=\frac{\left|-B_{3,>}-\zeta -8\,\kappa +4\right|}{3\,B_{3,>}}
\;\end{equation}
with $B_{3,>}$ provided by the last row of Eq. (\ref{gs32}).

Finally, by using the expressions of the coefficients $c_{i}^{(0)}$ (\ref{gs2}) in Eq. (\ref{ee}), we calculate the entanglement entropy $S$ for $N=2$, and obtain
\begin{eqnarray}
\label{s2}
&&S=S_2\Bigg[\frac{\left(A_2+\zeta\right)^2\log_2 \left(\frac{1}{2} \left(\frac{\zeta}{A_2}+1\right)\right)}{(\kappa-1)^2}+16 \log_2 B_2\Bigg]\nonumber\\
&&S_2=-\frac{B_2}{{8\log_2 2}}\nonumber\\
&& B_2=\frac{1}{4}\left(1-\frac{\zeta}{A_2}\right)
\;\end{eqnarray}
with $A_2$ given by the third row of Eq. (\ref{gs2}).
By employing the expressions of $c_{i}^{(0)}$ provided by Eqs. (\ref{gs31}), $\kappa<1/2$, and (\ref{gs32}), $\kappa>1/2$, in Eq. (\ref{ee}), we calculate the entanglement entropy $S$ for $N=3$. When $\kappa<\frac{1}{2}$, one has the following expression:
\begin{eqnarray}
\label{s31}
&&S=\frac{\Bigg[ \Bigg(2 \left(B_{3,<}+\zeta -2 \kappa +1\right)^2
\log_2 \left[\frac{B_{3,<}+\zeta -2 \kappa +1}{4\,B_{3,<}}\right]\Bigg)
+6 \log_2 \left(\frac{B_{3,<}-\zeta +2 \kappa -1}{4\,B_{3,<}}\right)\Bigg]}{S_{3,<}}\nonumber\\
&&S_{3,<}=-4 \log_2 2 \Bigg(\zeta (B_{3,<}-4 \kappa +2)
-(2 \kappa -1)(B_{3,<}-8 \kappa +4)+\zeta ^2\Bigg)\;\nonumber\\
\end{eqnarray}
with $B_{3,<}$ provided by the last row of Eq. (\ref{gs31}). When $\kappa>\frac{1}{2}$, one gets
\begin{eqnarray}
\label{s32}
&&S=\frac{\Bigg[\frac{2}{(1-2\,\kappa )^2}\Bigg(\left(B_{3,>}+\zeta +2 \kappa -1\right)^2
\log_2 \left[\frac{B_{3,>}+\zeta +2\,\kappa -1}{4\,B_{3,>}}\right]\Bigg)
+6\,\log_2 \left(\frac{1}{\frac{2 \left(B_{3,>}+\zeta +2\,\kappa -1\right)^2}{3 (1-2 \kappa )^2}+2}\right)\Bigg]}{S_{3,>}}\nonumber\\
&&S_{3,>}=\log_2 2 \left(\frac{2 \left(B_{3,>}+\zeta +2\,\kappa -1\right)^2}{3 (1-2 \kappa )^2}+2\right)\;,\nonumber\\
\end{eqnarray}
where $B_{3,>}$ is given by the fourth row of Eq. (\ref{gs32}).

{\it Discussion}. In this paragraph we discuss to what extent the correlated hopping determines the structure of the ground state. Let us start with Fig. 1 and Fig. 2, where we show distribution $|c_i^{(0)}|^2$ of the lowest energy state with $N=2$ and $N=3$, respectively.
The left-top (bottom) plot of Figs. 1-2 describes the situation when the correlated hopping is absent and the onsite interatomic attraction (repulsion) does not guarantee the formation of the NOON state (\ref{cat}) (twin Fock state (\ref{fock}) when $N=2$; pseudo-Fock state (\ref{pseudo:fock}) when $N=3$). The right-top (bottom) plot --of Figs. 1-2--  corresponds to the same value of the onsite attraction (repulsion) of the left plot --of the same figures-- but with a nonzero correlated hopping. The nonvanishing correlated hopping makes possible the occurrence of the NOON state (twin Fock state when $N=2$; pseudo-Fock state when $N=3$).
\begin{figure}[htbp]
\centerline{\includegraphics[width=4.3cm,clip]{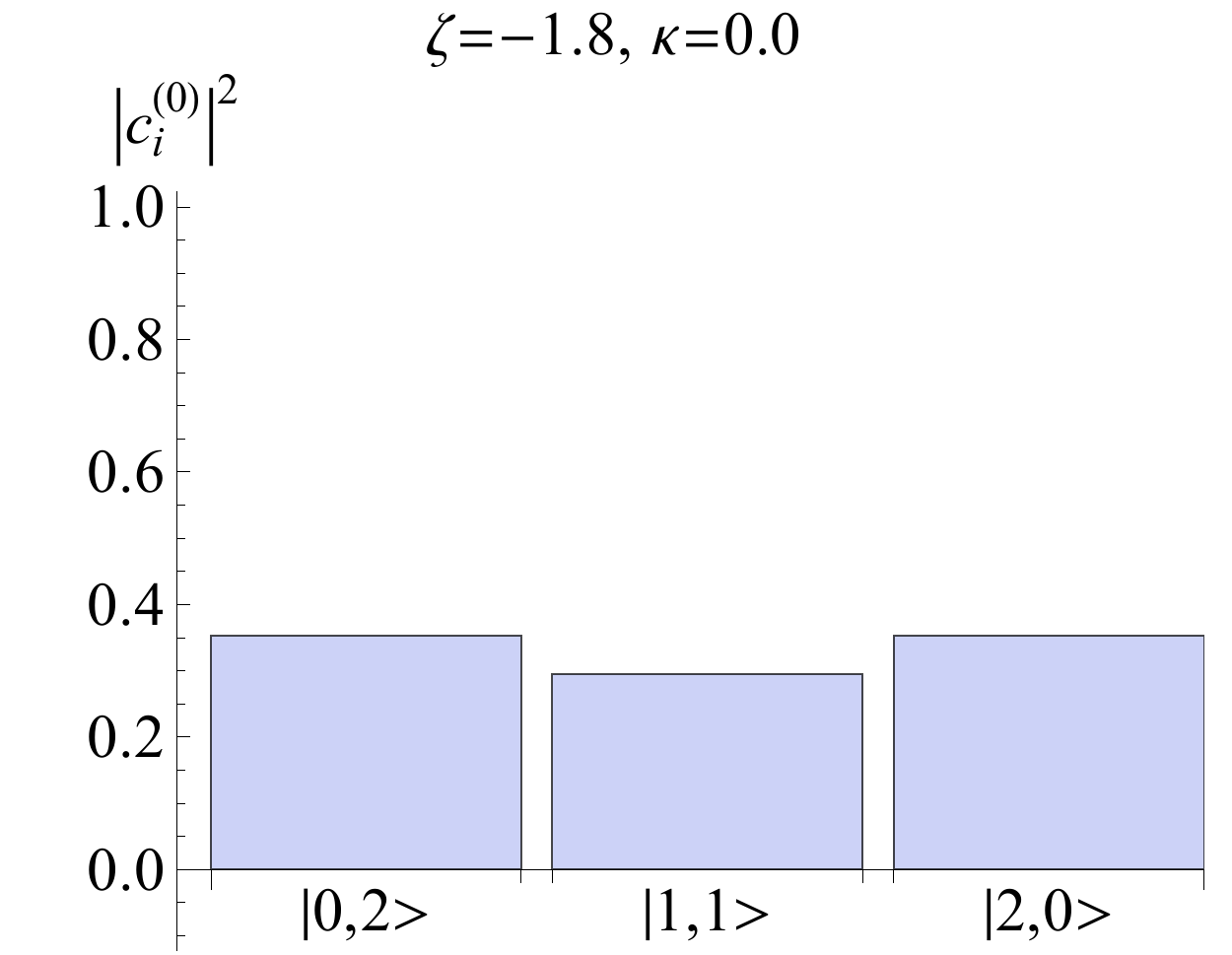}
\includegraphics[width=4.3cm,clip]{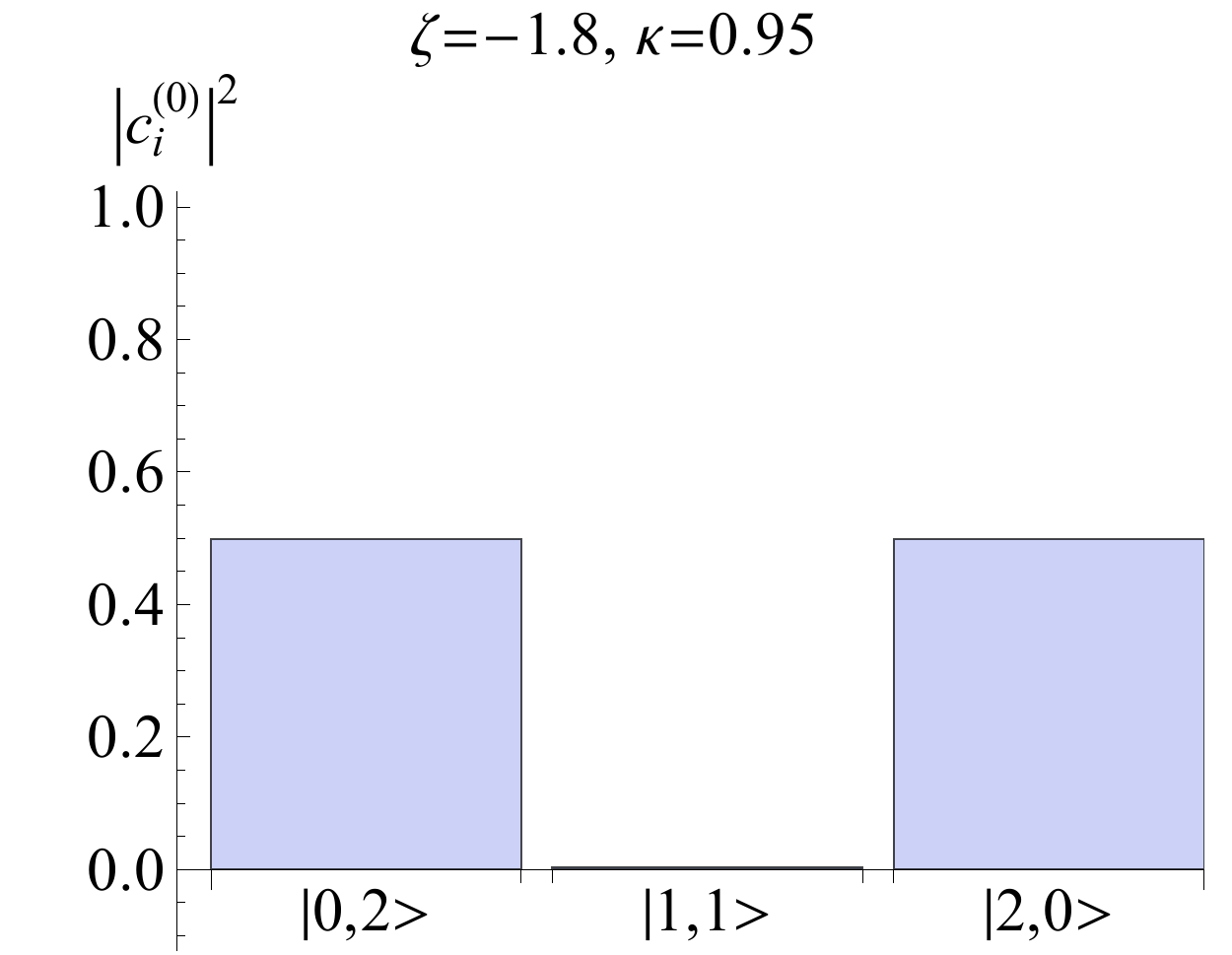}
}
\centerline{\includegraphics[width=4.3cm,clip]{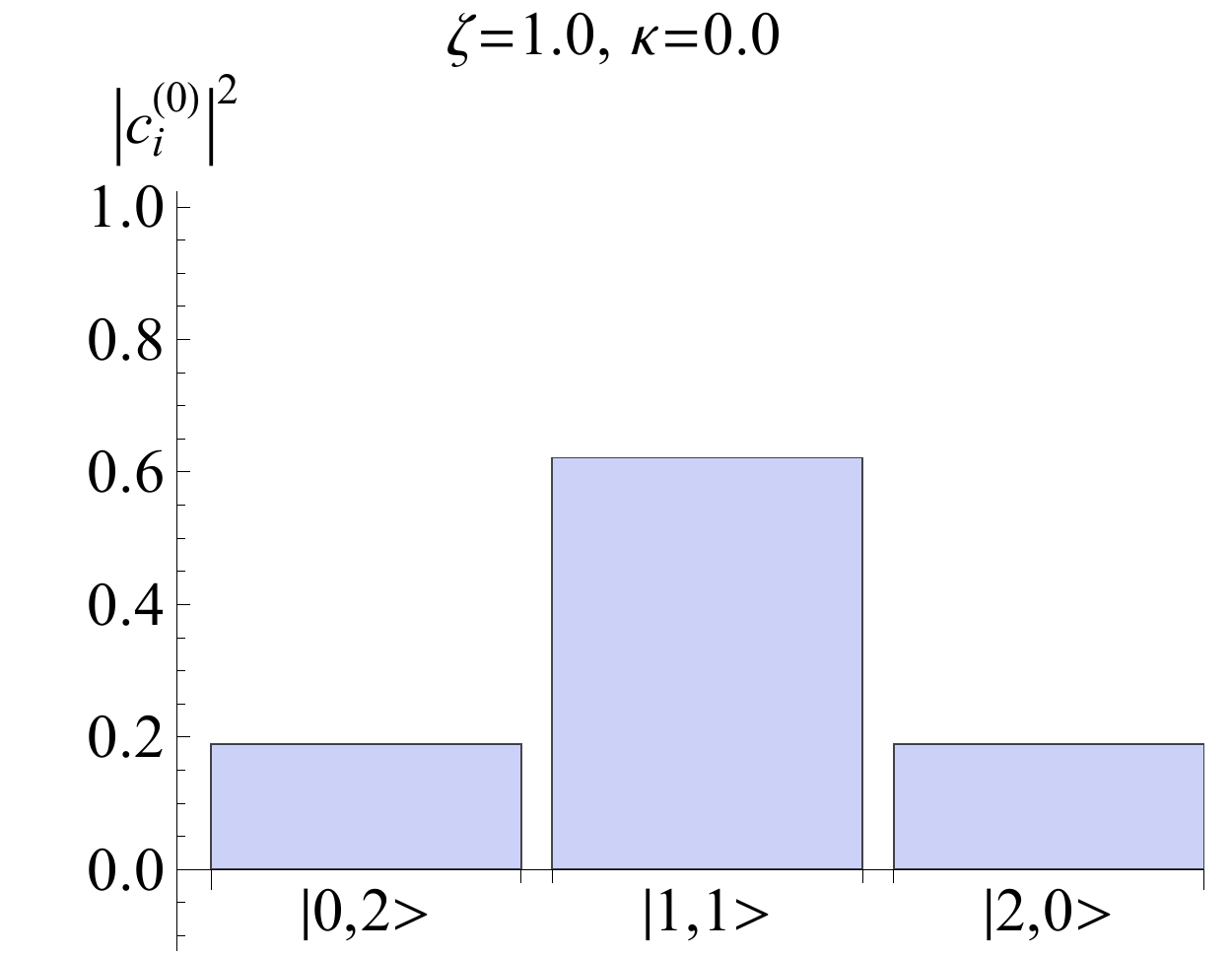}
\includegraphics[width=4.3cm,clip]{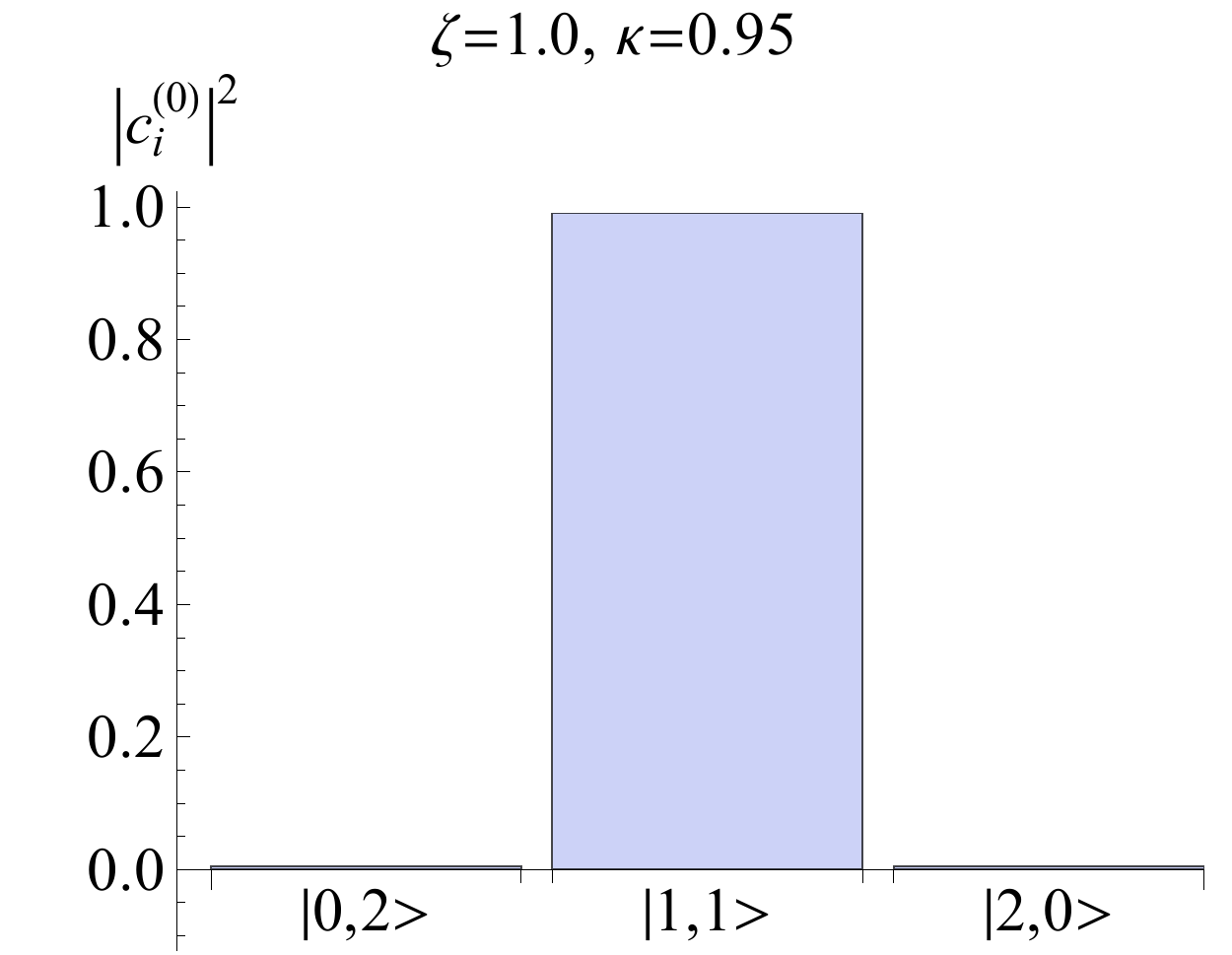}
}
\caption{$N=2$. Horizontal axis: ket $|i,N-i\rangle$. Vertical axis: $|c_{i}^{(0)}|^2$. Left column: $\zeta=U/J \neq 0$, $\kappa=K_c/J=0$.
Right column: $\zeta=U/J \neq 0$, $\kappa=K_c/J \neq 0$. All the quantities are dimensionless.}
\label{fig1}
\end{figure}
\begin{figure}[htbp]
\centerline{\includegraphics[width=4.3cm,clip]{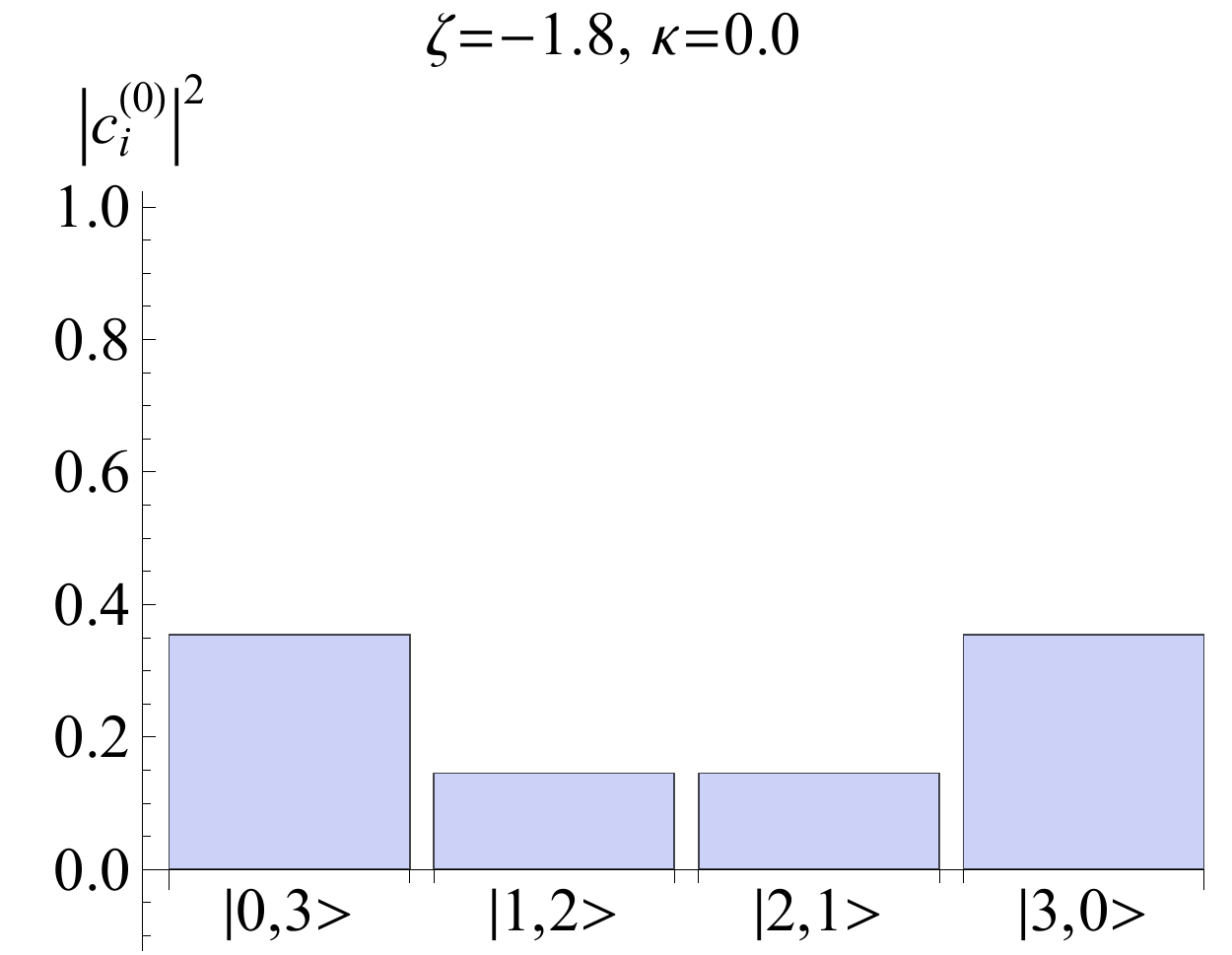}
\includegraphics[width=4.3cm,clip]{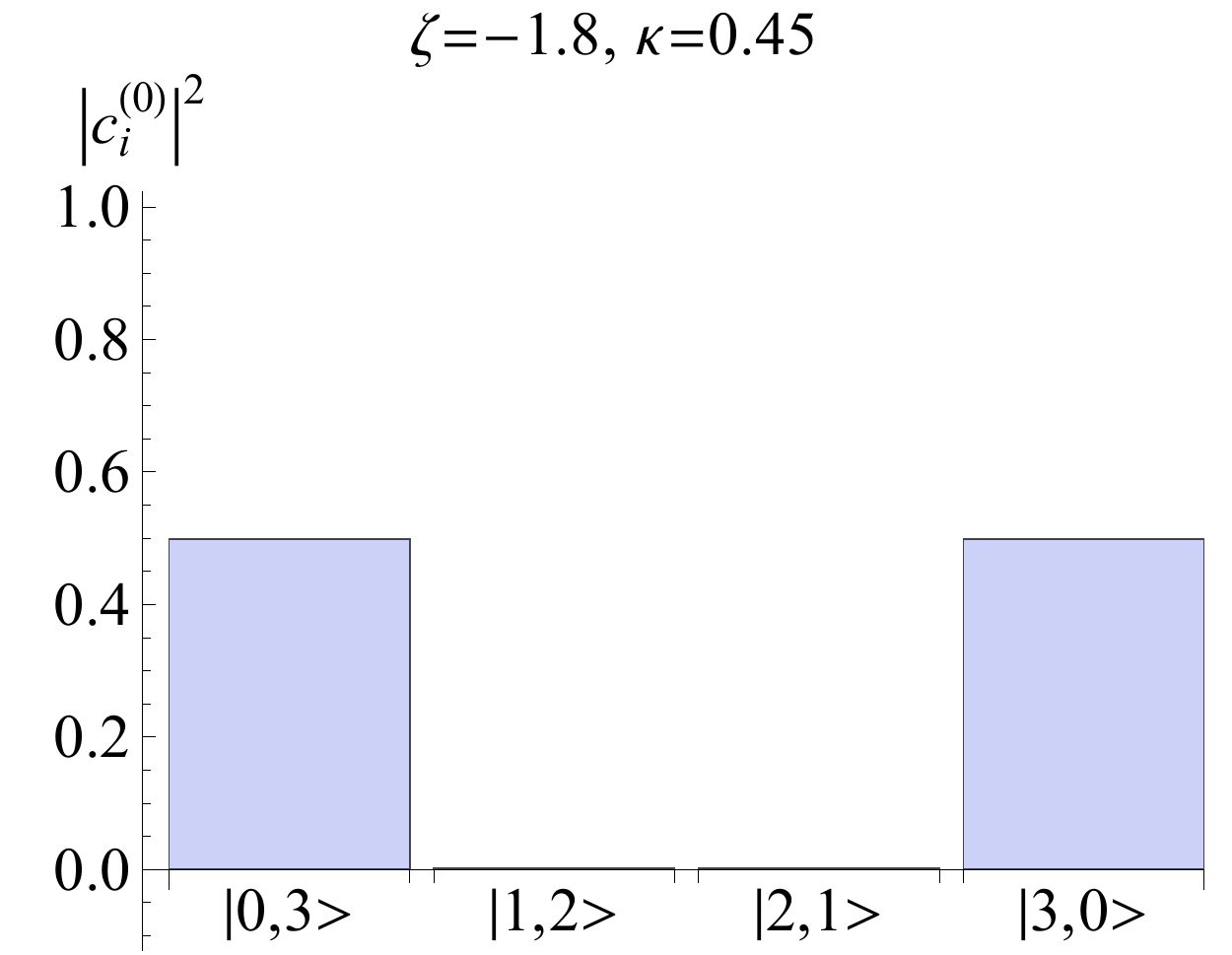}
}
\centerline{\includegraphics[width=4.3cm,clip]{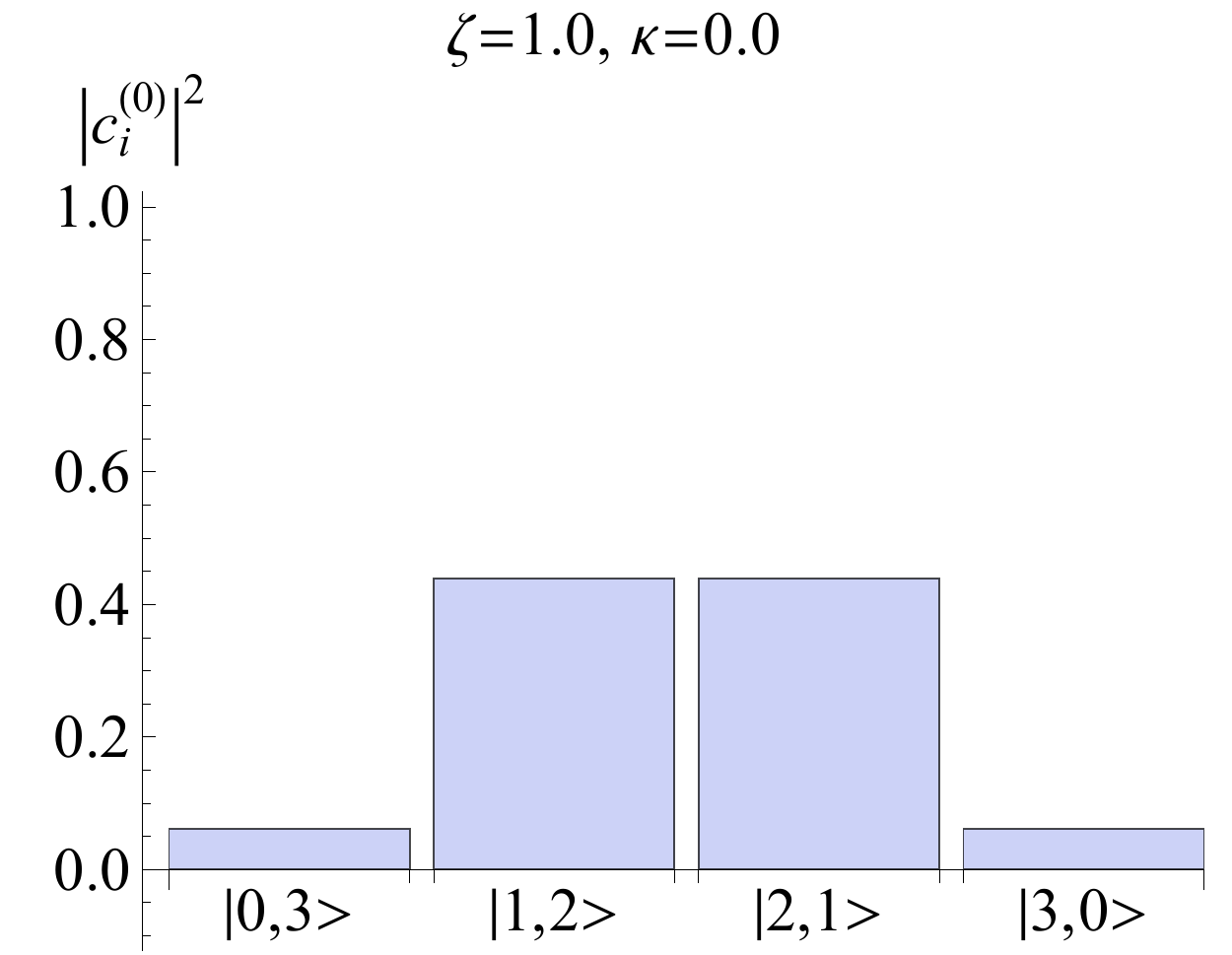}
\includegraphics[width=4.3cm,clip]{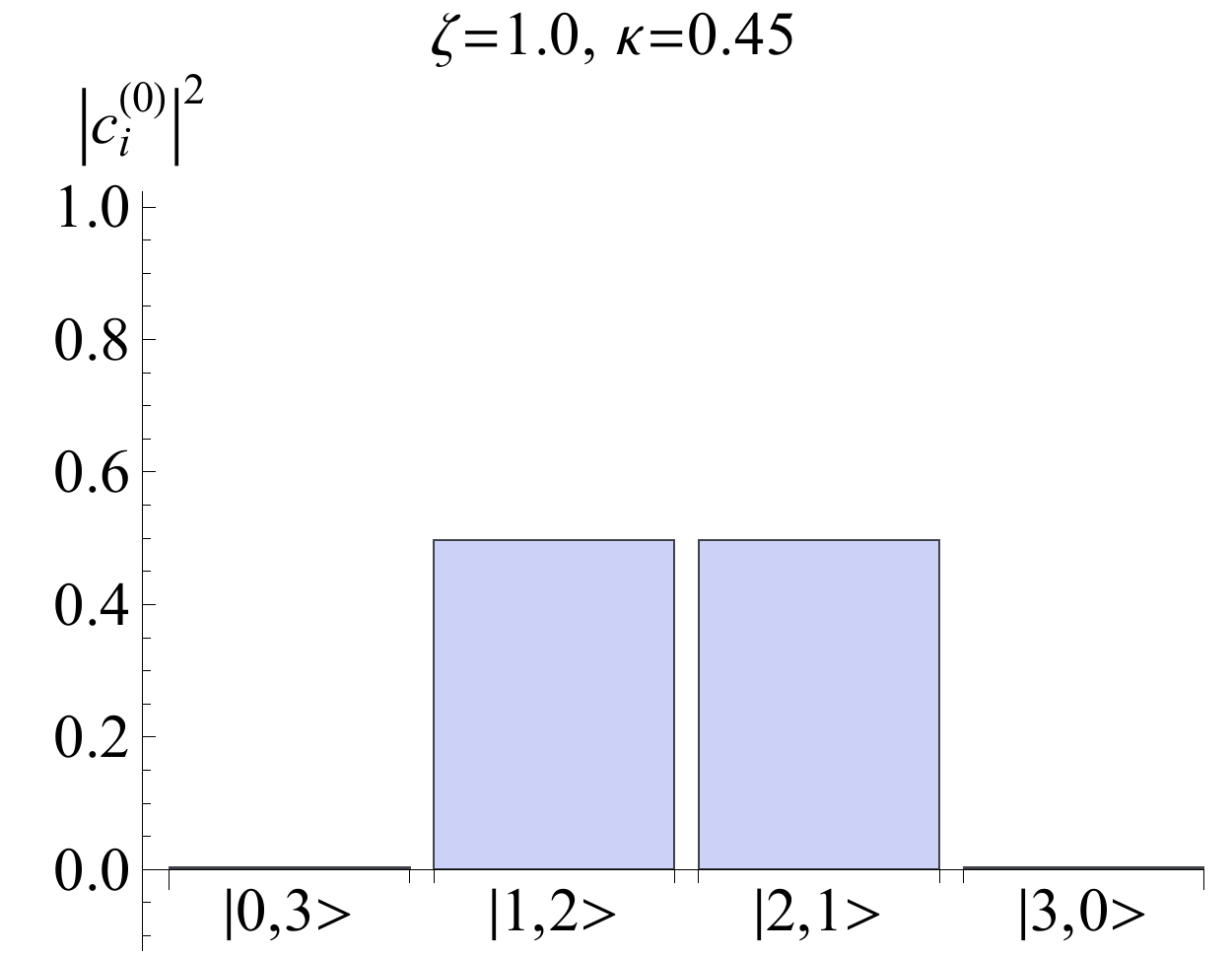}
}
\caption{$N=3$. Horizontal axis: ket $|i,N-i\rangle$. Vertical axis: $|c_{i}^{(0)}|^2$. Left column: $\zeta=U/J \neq 0$, $\kappa=K_c/J=0$.
Right column: $\zeta=U/J \neq 0$, $\kappa=K_c/J \neq 0$. All the quantities are dimensionless.}
\label{fig2}
\end{figure}
This scenario is supported by the analysis of the correlation properties of the ground state. We have studied the Fisher information $F$, the coherence visibility $\alpha$, and the entanglement entropy $S$ as functions of the scaled onsite interaction $\zeta=U/J$ both when $\kappa=0$ ($\kappa=K_c/J$), Fig. 3 (corresponding to the left column of Figs. 1-2), and when $\kappa \neq 0$, Fig. 4 (corresponding to the right column of Figs. 1-2).
\begin{figure}[htbp]
\centerline{\includegraphics[width=4.3cm,clip]{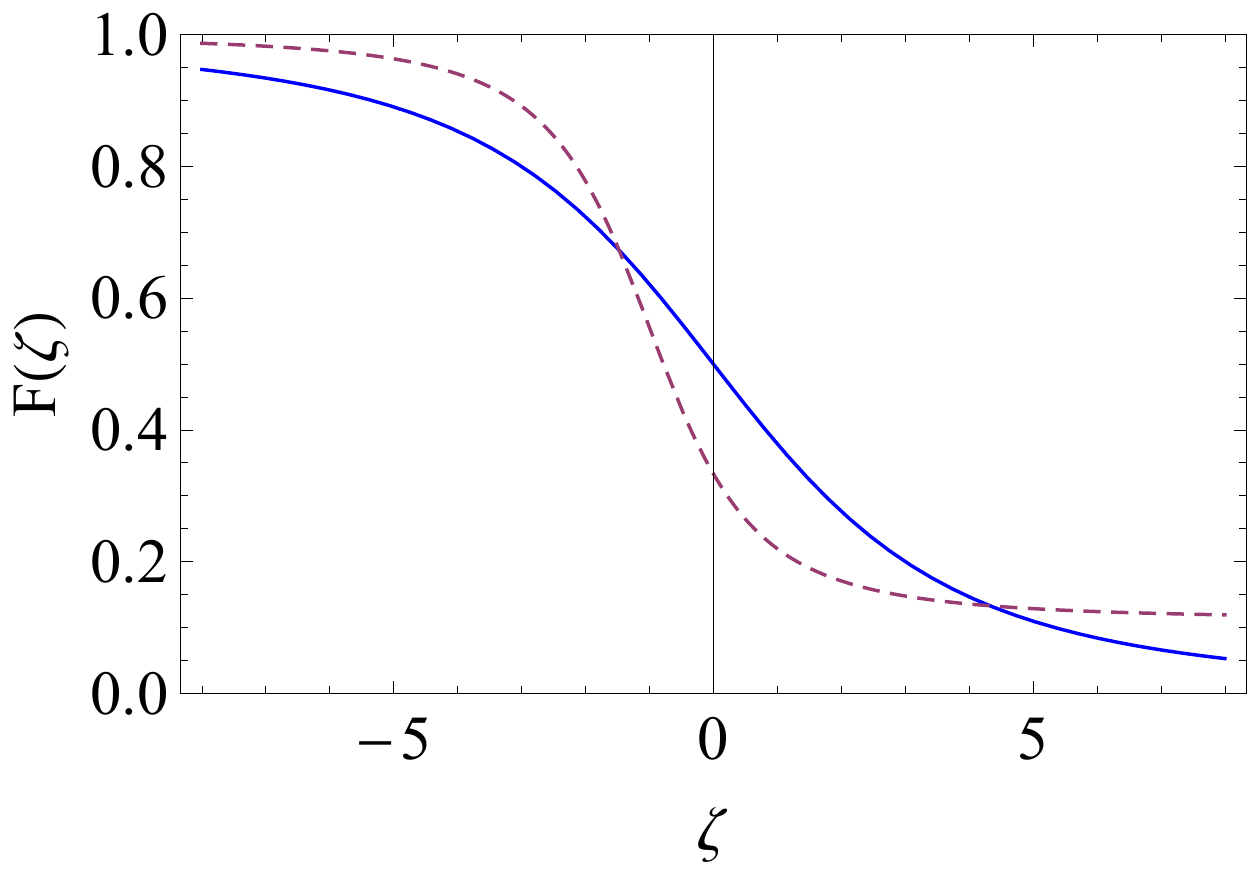}
\includegraphics[width=4.3cm,clip]{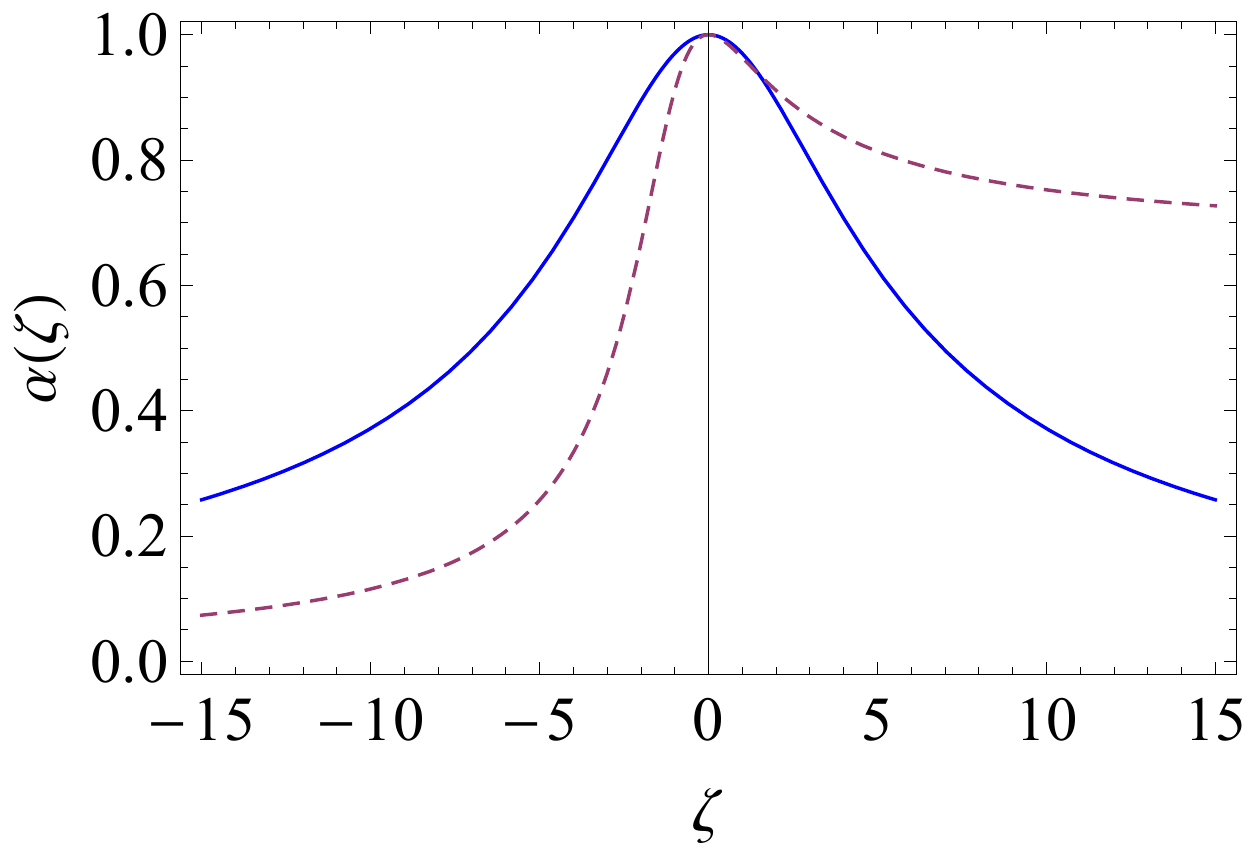}
\includegraphics[width=4.3cm,clip]{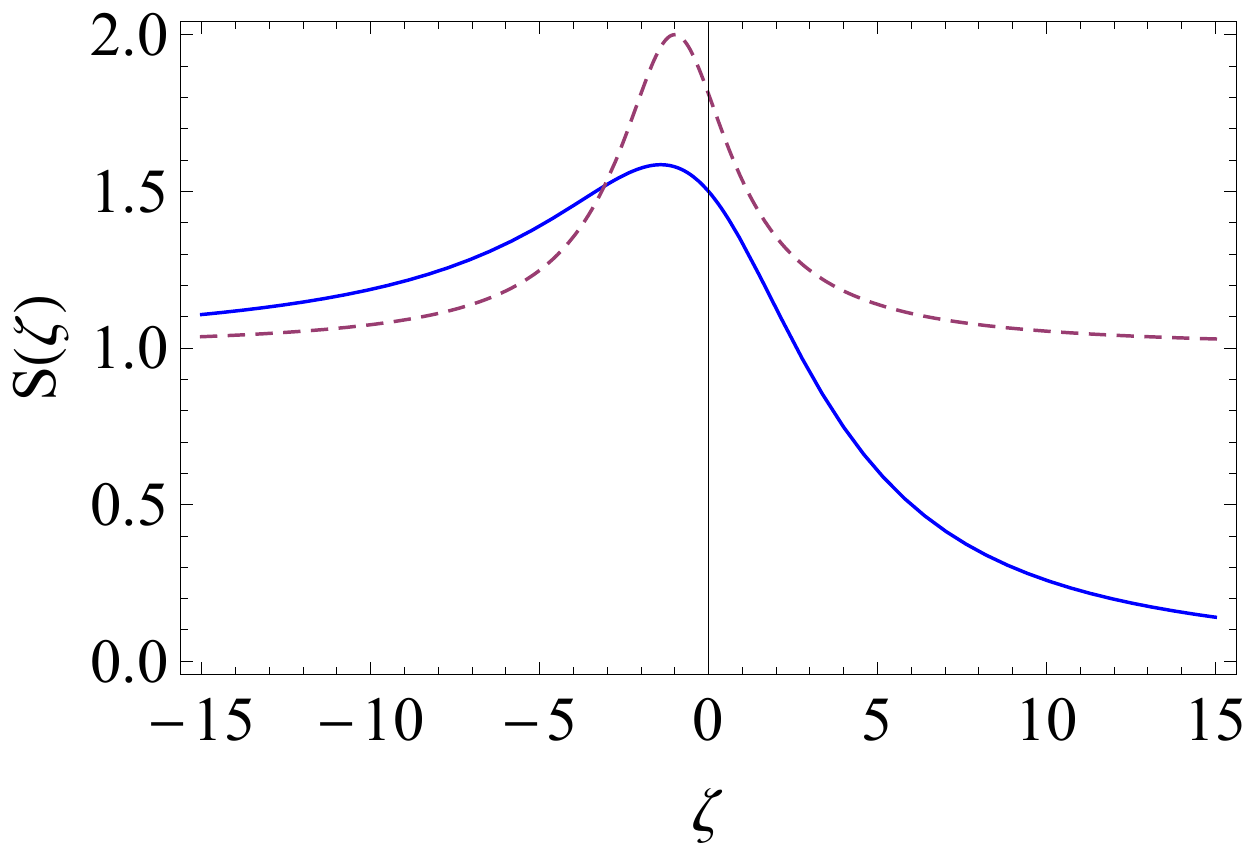}}
\caption{Left panel: Fisher information $F$ vs scaled onsite interaction $\zeta=U/J$. Middle panel: coherence visibility $\alpha$ vs $\zeta$. Right panel: entanglement entropy $S$ vs $\zeta$. In all the panels: $\kappa=K_c/J=0$.
In each panel the solid line corresponds to $N=2$, the dashed line to $N=3$. All the quantities are dimensionless.}
\label{fig3}
\end{figure}
\begin{figure}[htbp]
\centerline{\includegraphics[width=4.3cm,clip]{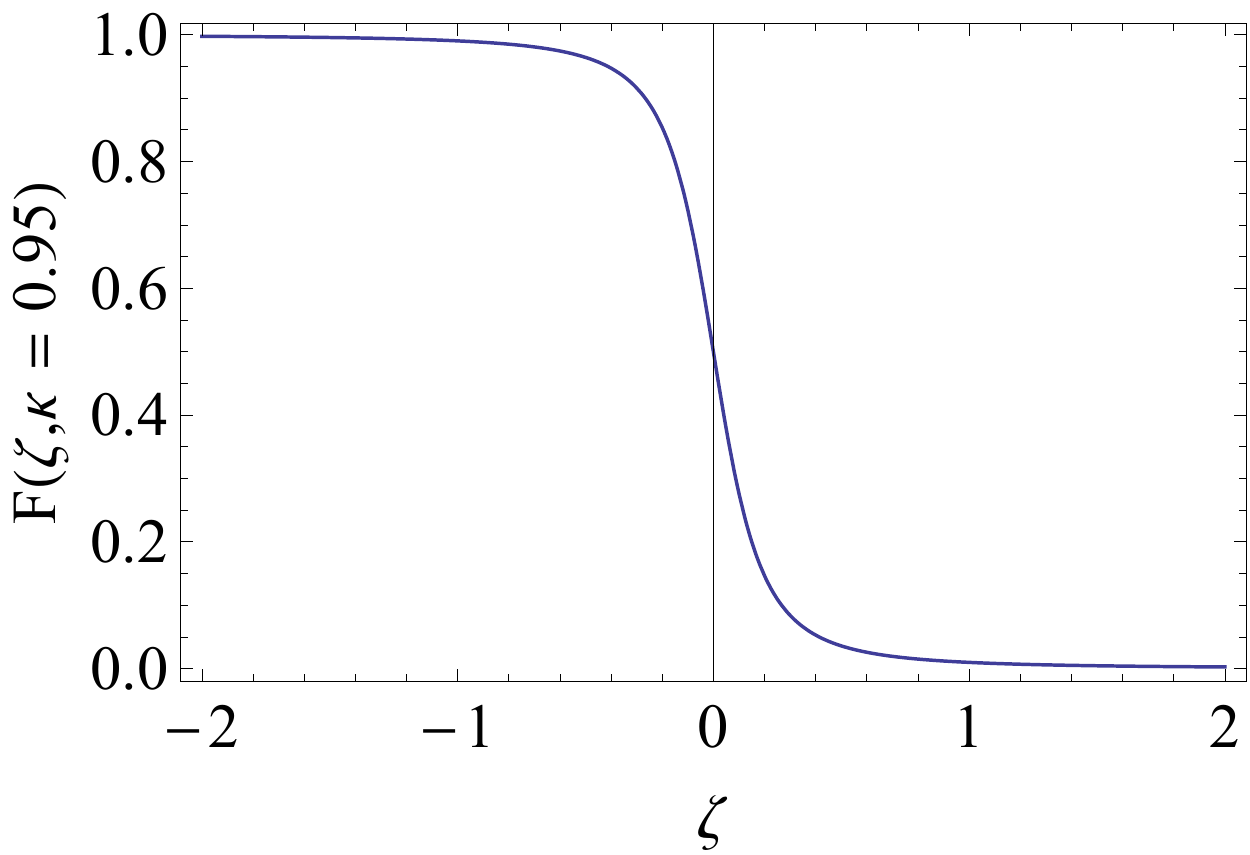}
\includegraphics[width=4.3cm,clip]{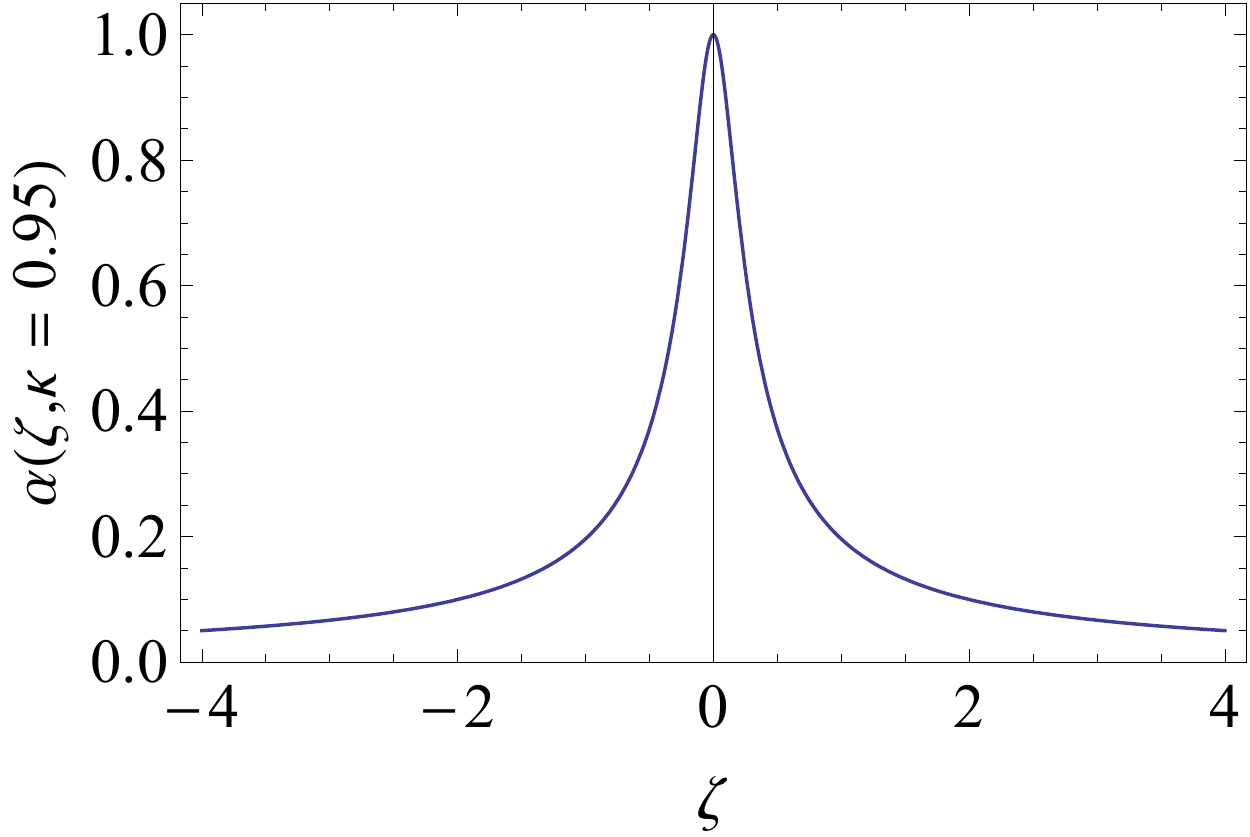}
\includegraphics[width=4.3cm,clip]{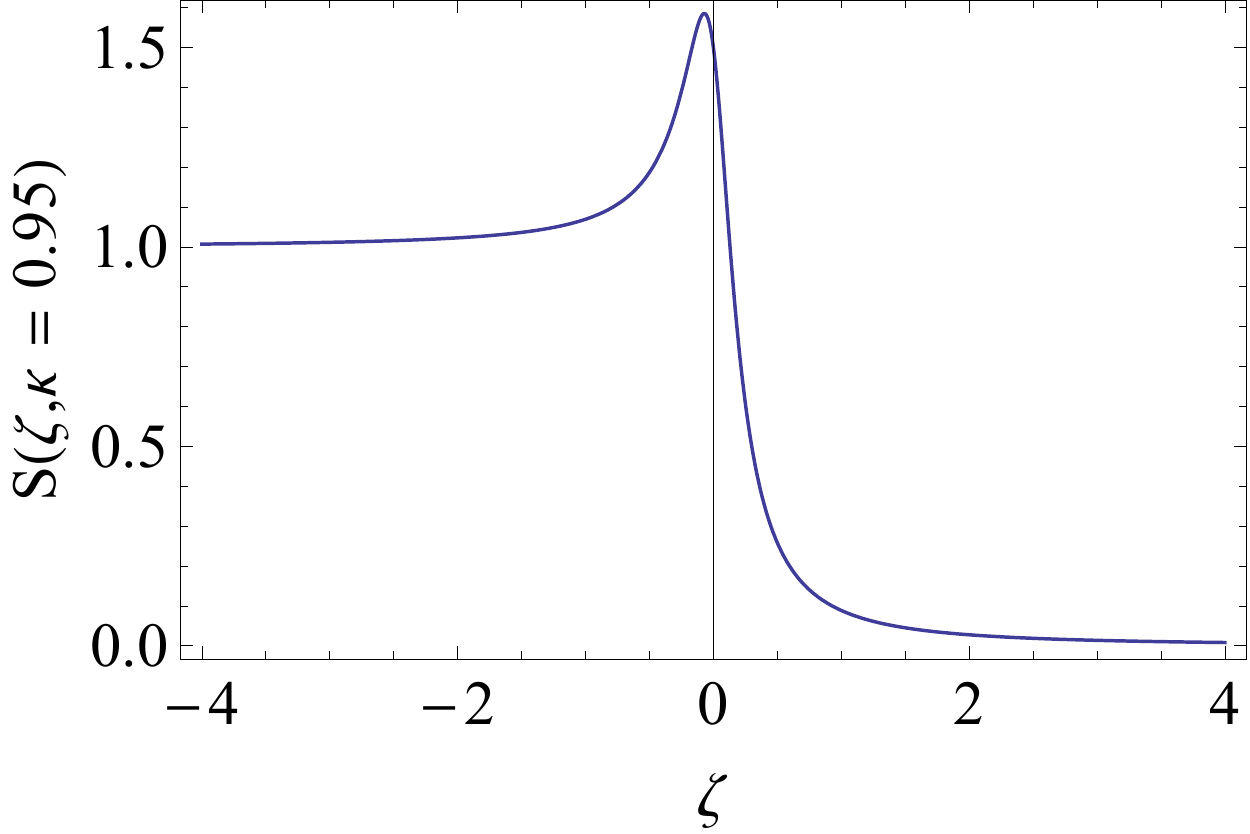}}
\centerline{
\includegraphics[width=4.3cm,clip]{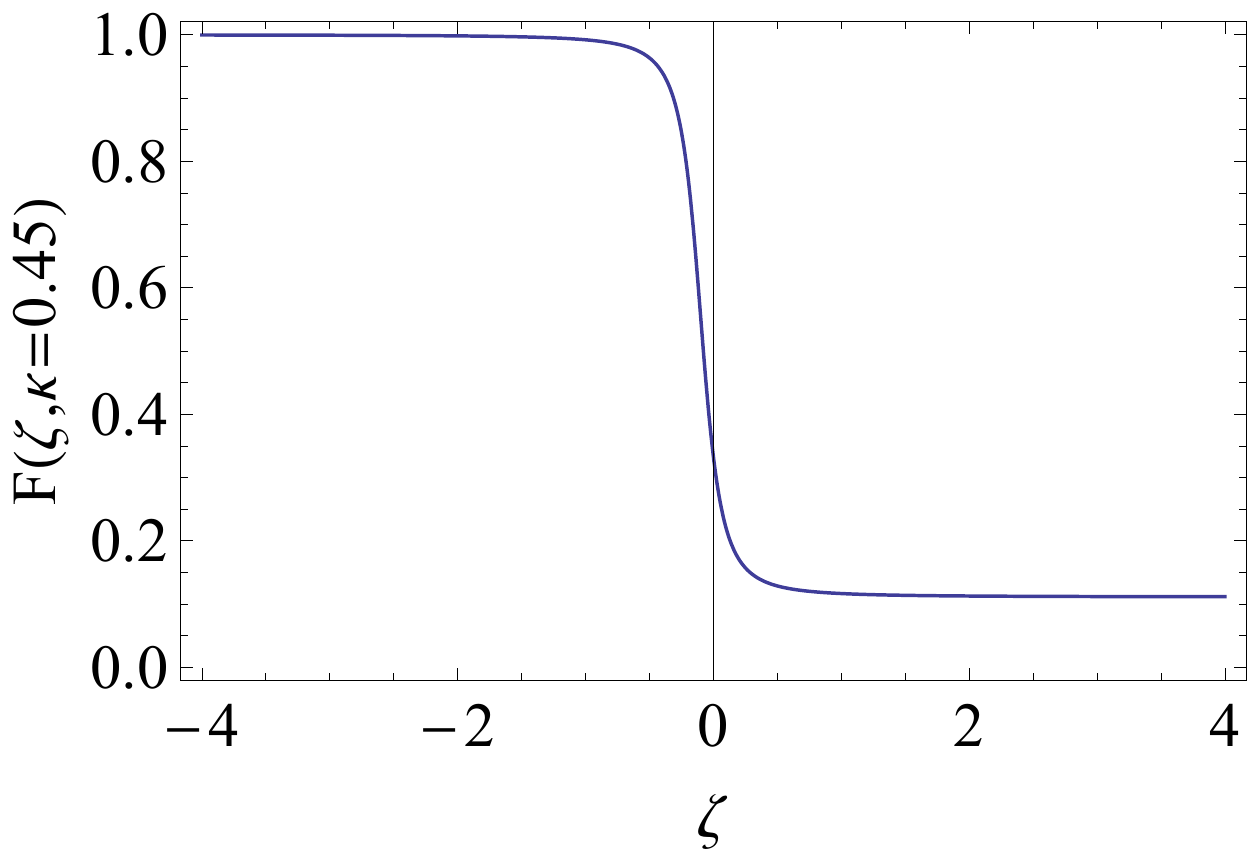}
\includegraphics[width=4.3cm,clip]{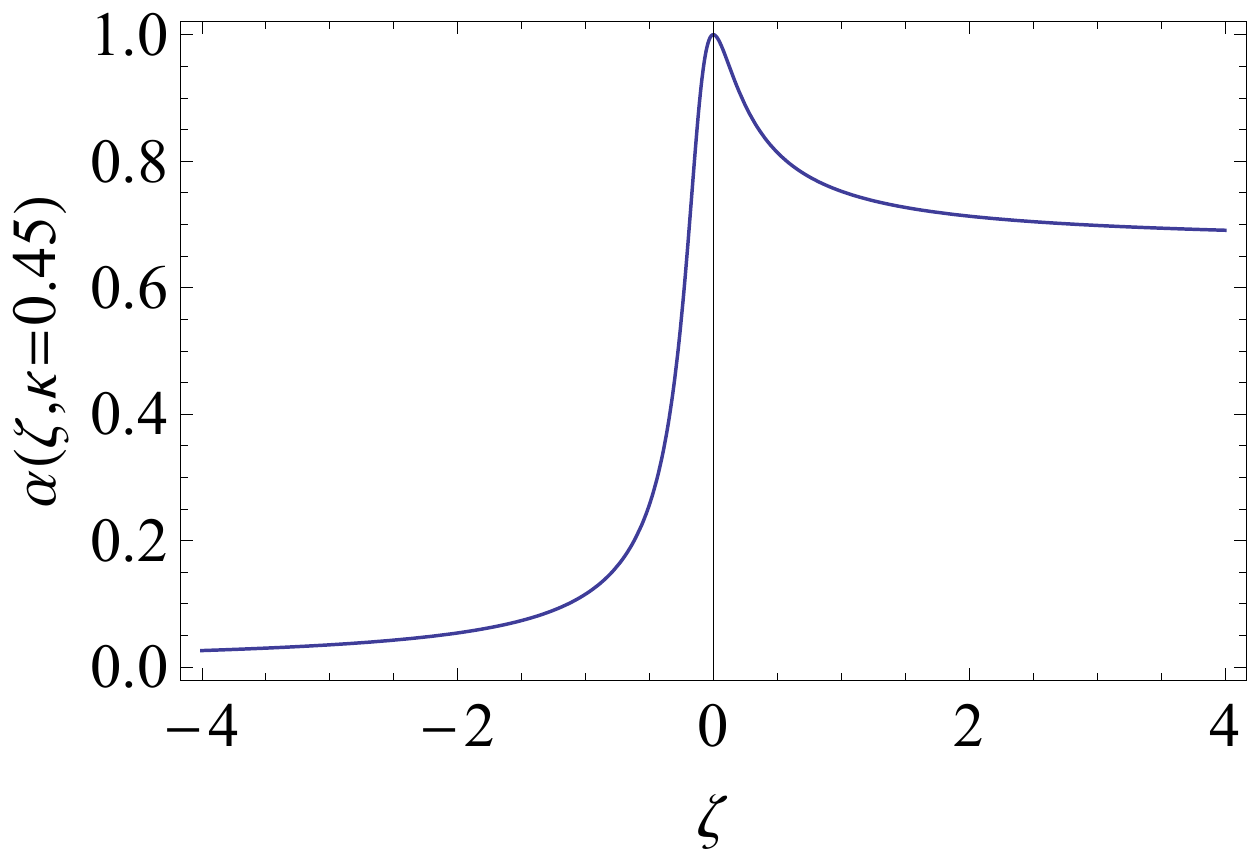}
\includegraphics[width=4.3cm,clip]{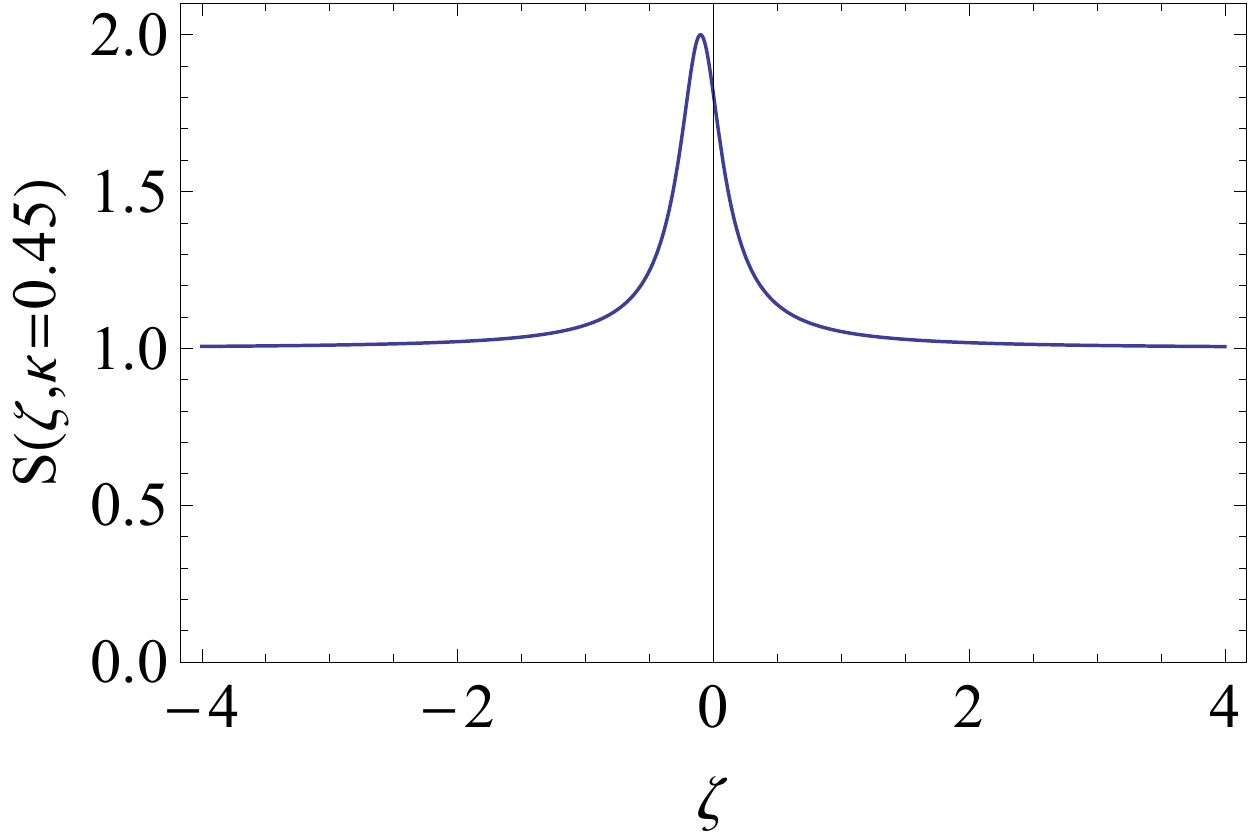}
}
\caption{Top ($N=2$, $\kappa=K_c/J=0.95$). Bottom ($N=3$, $\kappa=K_c/J=0.45$). Left panels: Fisher information $F$ vs scaled onsite interaction $\zeta=U/J$. Middle panels: coherence visibility $\alpha$ vs $\zeta$. Right panels: entanglement entropy $S$ vs $\zeta$.
All the quantities are dimensionless.}
\label{fig4}
\end{figure}
To fix the ideas, we focus on $N=2$ (similar arguments being valid when $N=3$). Let us compare the solid lines in Fig. 3 with the plots in the top panel in Fig. 4 and start with $F$ (we recall that $F$ attains its maximum value, $1$, in correspondence to the NOON state). In Fig. 3 we see that when $\zeta=-1.8$ and $\kappa=0$ (uniform distribution of $|c_{i}^{(0)}|^2$ of the left-top plot of Fig. 1) $F$ is about $0.7$. Instead, the left-top panel in Fig. 4 shows that when $\zeta=-1.8$ and $\kappa=0.95$ (double-peak structure --which is the structure of a cat-like state--  of the right top-plot of Fig. 1) $F$ is almost one. We continue with $S$ (we recall that $S$ is zero for the twin Fock state). We observe that in correspondence to the single-peak state ($\zeta=1$ and $\kappa=0.95$, right-bottom plot of Fig. 1, practically a twin Fock state), $S$ (right-top plot in Fig. 4) is much smaller than $S$ (solid line of the right panel in Fig. 3) pertaining to the ground state that one has when only $\zeta$ is finite (left-bottom plot of Fig. 1).
At this point, a conclusive comment is in order about the difference between the case with $\kappa=0$ and that with nonzero $\kappa$'s. We proceed from Figs. 3-4. If the behaviors of $F$, $\alpha$, and $S$ displayed in Fig. 4 are qualitatively the same of those reported in Fig. 3 (described also in \cite{galante}, where the interatomic onsite interaction is the only active process), from a quantitative point view there is a (crucial) difference. This consists in the fact that when $\kappa \neq 0$, the remarkable behaviors [$\zeta<0$: $F=1$, $S=1$ associated with the formation of the cat-like state (\ref{cat}); $\zeta>0$: $S=0$, $F=0$ associated with the formation of the twin Fock state (\ref{fock}) when $N=2$, and $S=1$, $F=0$ associated with the formation of the pseudo-Fock state (\ref{pseudo:fock}) when $N=3$] appear for $|\zeta|$'s smaller than those required when $\kappa=0$.
\begin{figure}[h]
\centerline{\includegraphics[width=4.cm,clip]{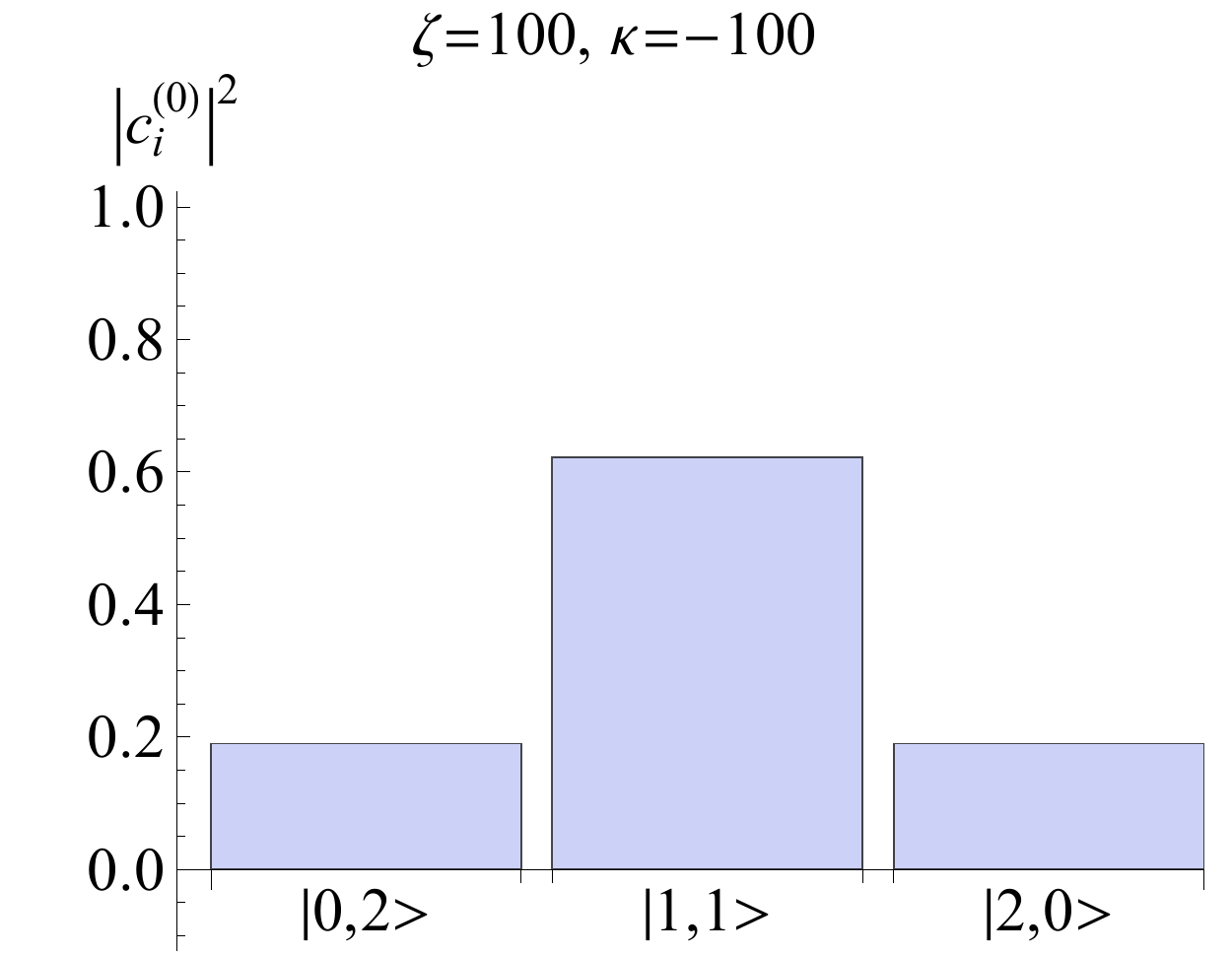}
\includegraphics[width=4.cm,clip]{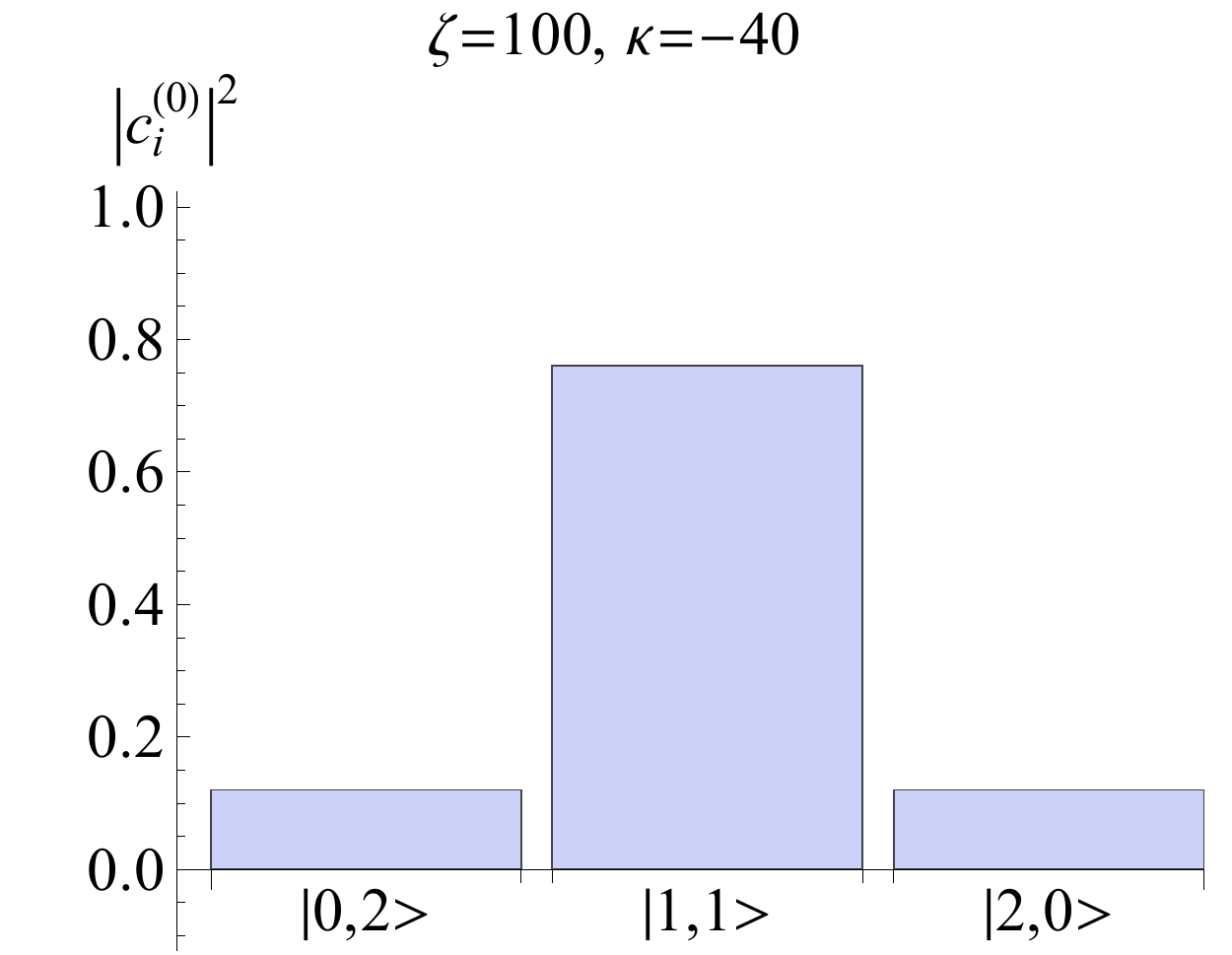}
\includegraphics[width=4.cm,clip]{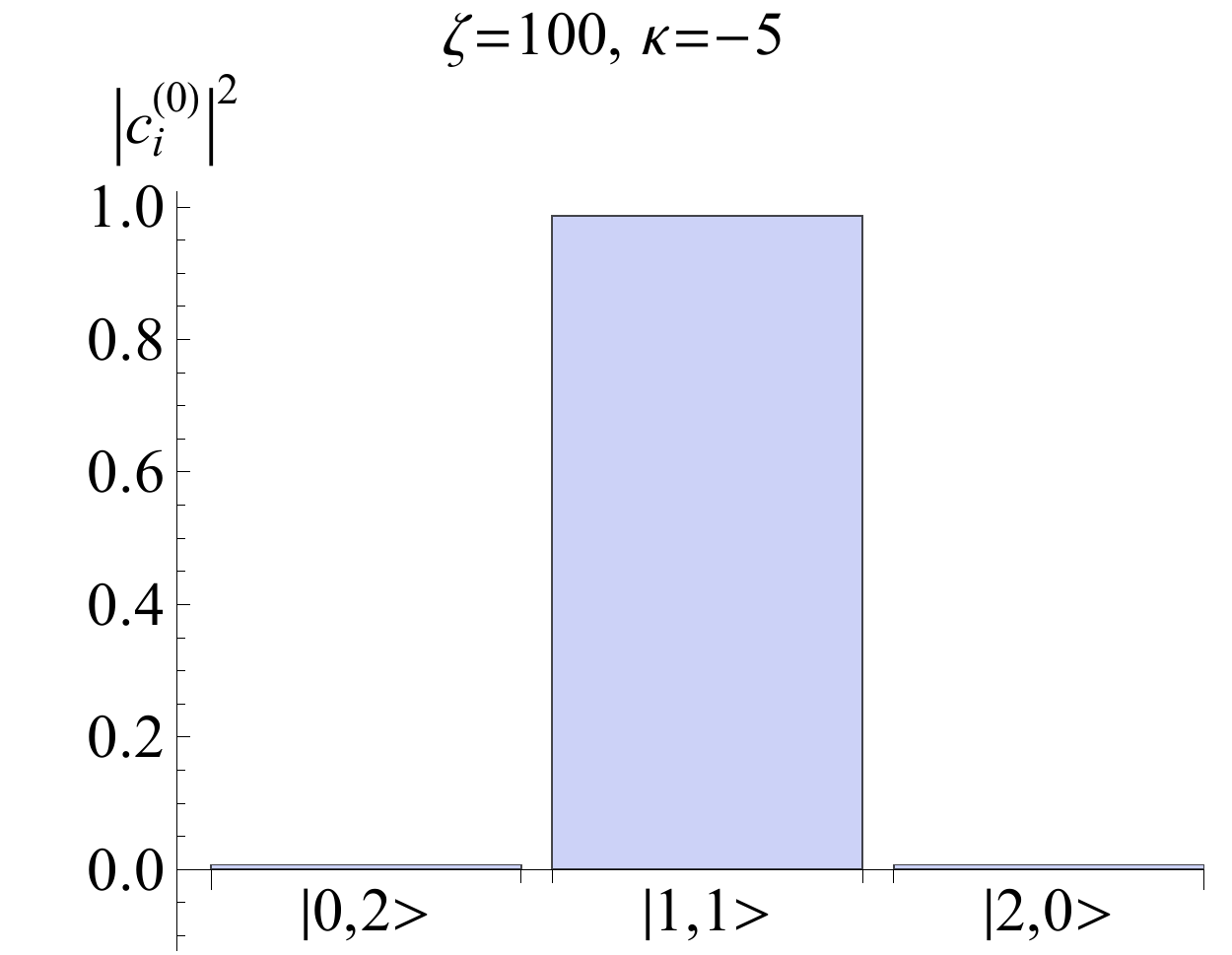}
\includegraphics[width=4.cm,clip]{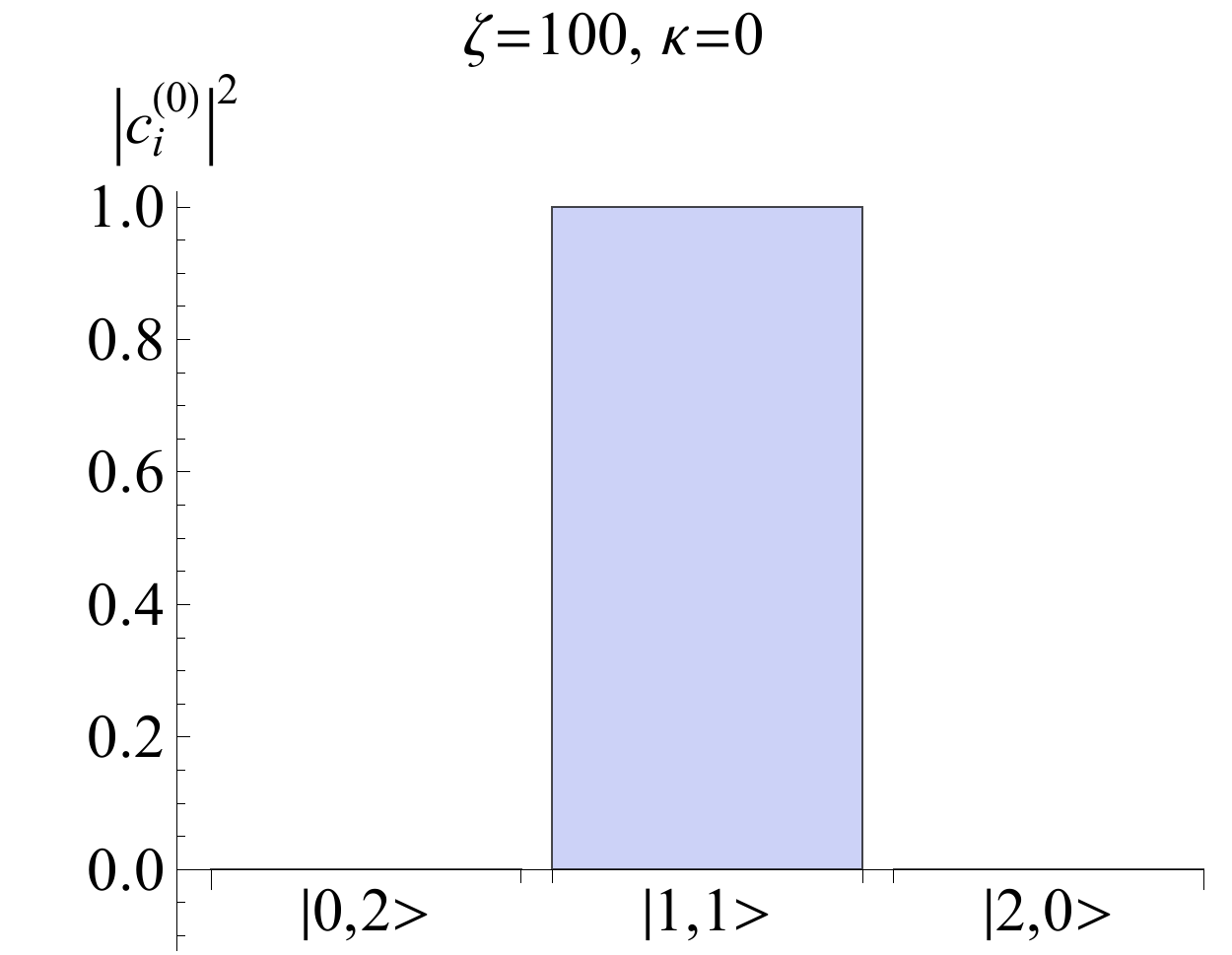}}
\centerline{
\includegraphics[width=4.cm,clip]{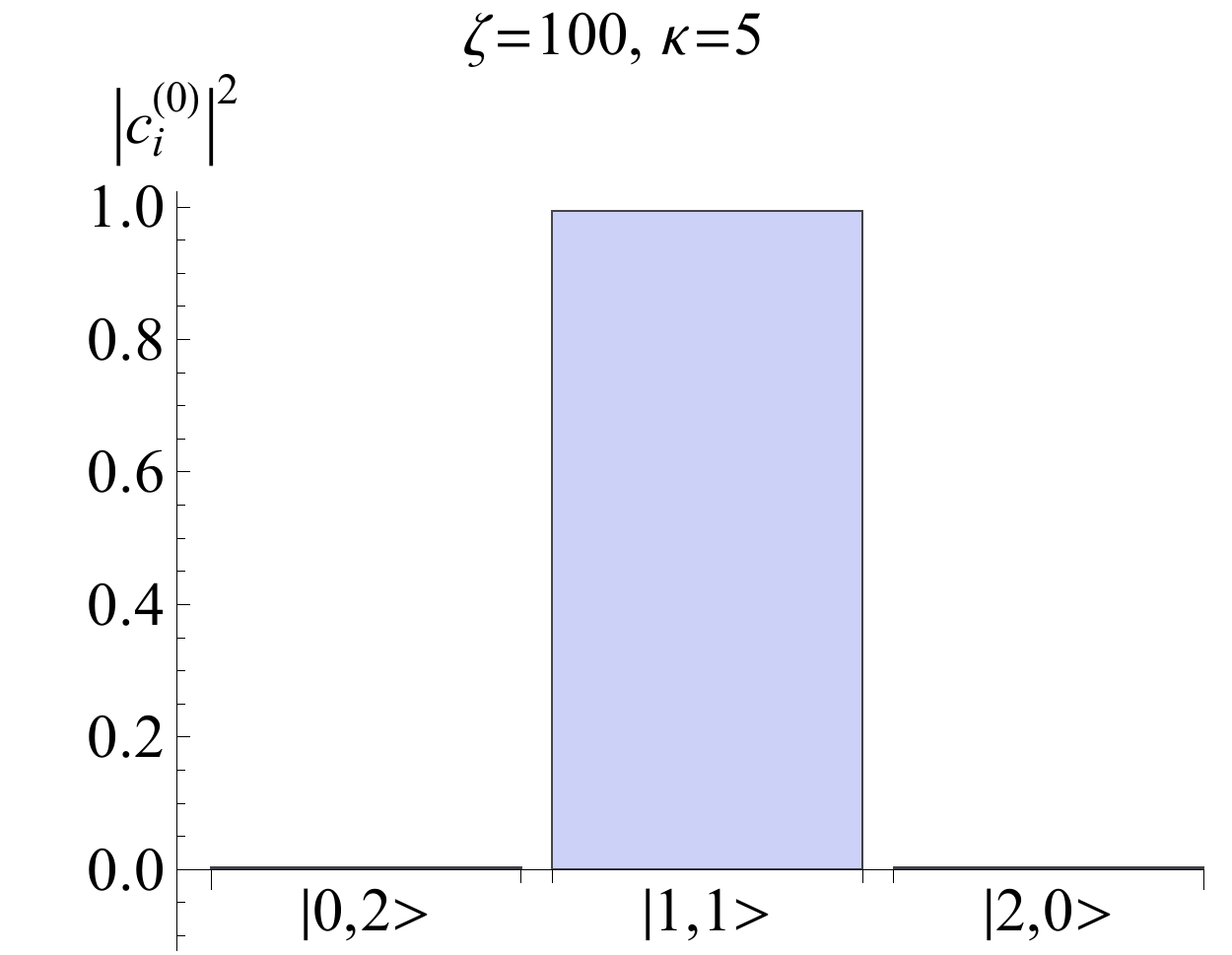}
\includegraphics[width=4.cm,clip]{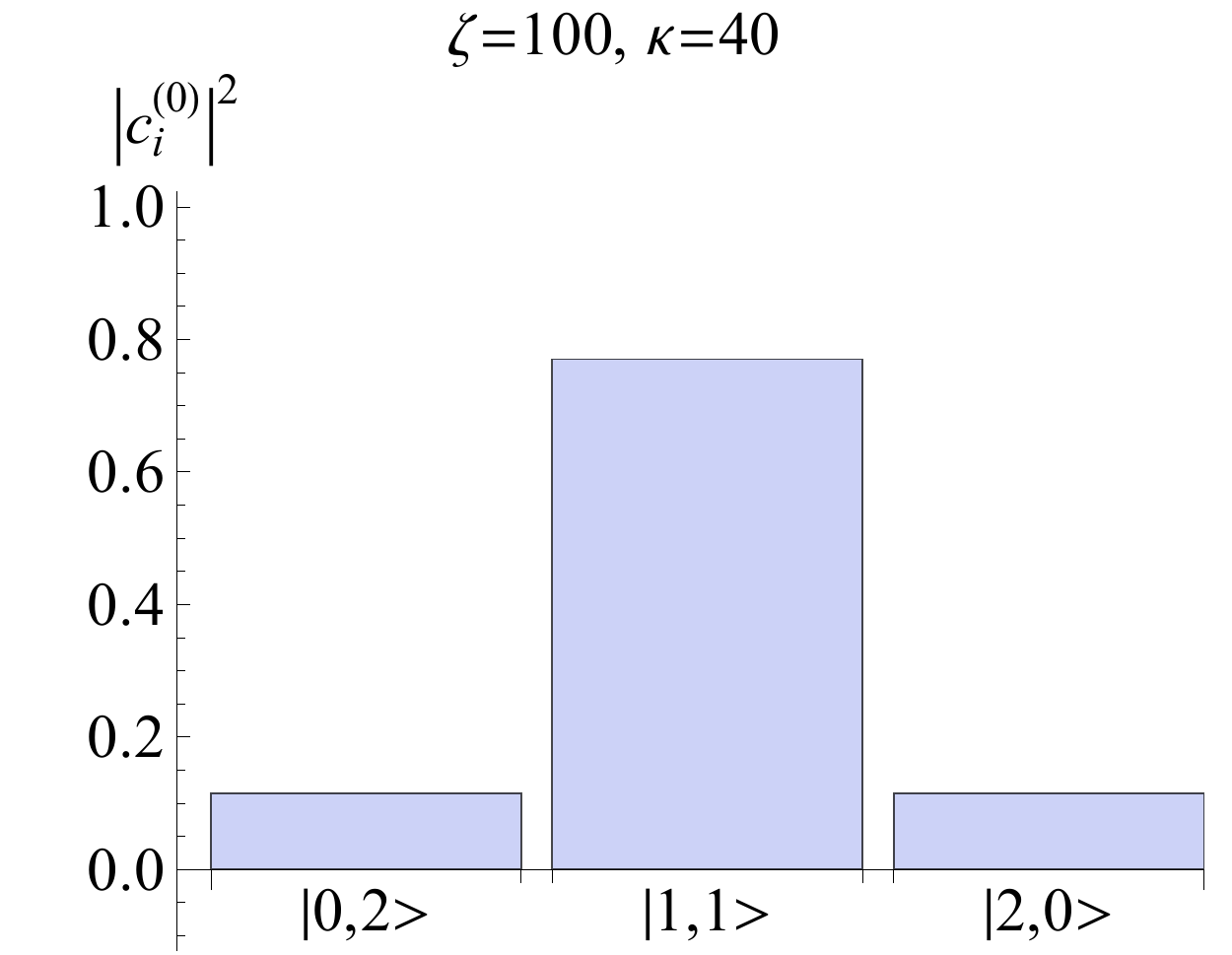}
\includegraphics[width=4.cm,clip]{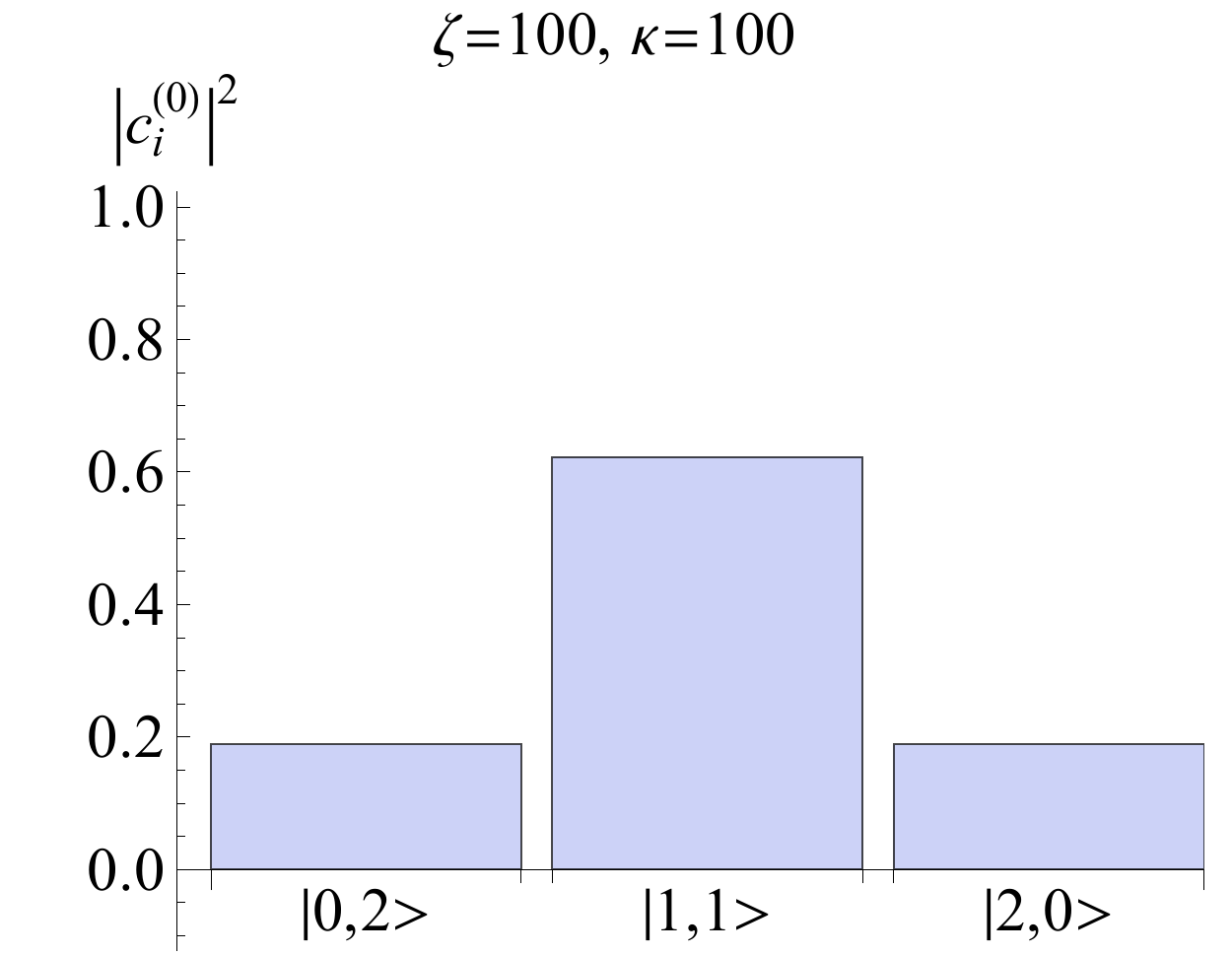}
\includegraphics[width=4.cm,clip]{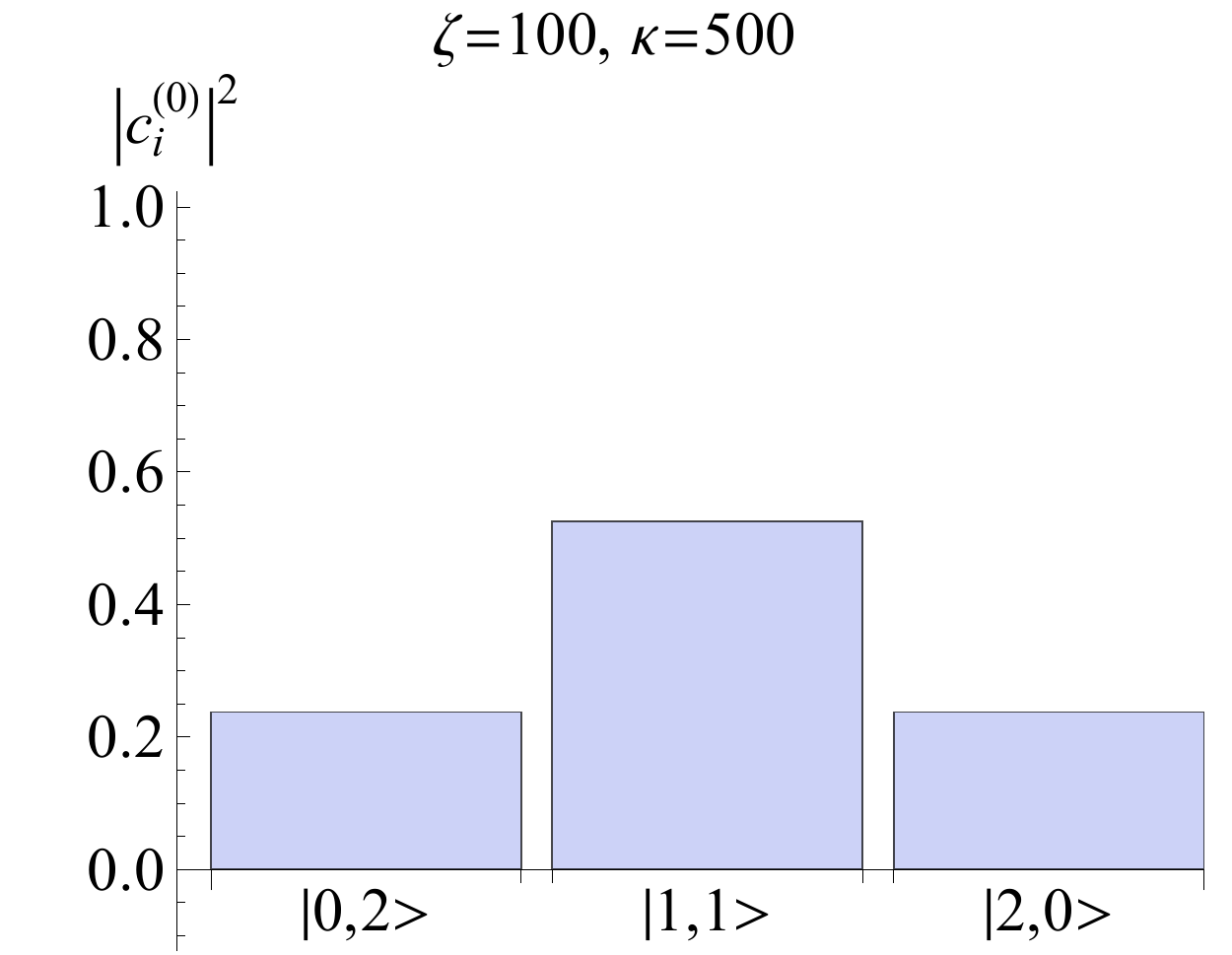}
}
\caption{$N=2$. Horizontal axis: ket $|i,N-i\rangle$. Vertical axis: $|c_{i}^{(0)}|^2$. The scaled onsite interaction $\zeta=U/J$ is fixed (strong repulsion). The scaled correlated hopping $\kappa=K_c/J$ changes. Top-bottom: in the first row ($\kappa \le 0$), $|\kappa|$ decreases from left to right; in the second row ($\kappa >0$), $\kappa$ increases from left to right. All the quantities are dimensionless.}
\label{fig5}
\end{figure}
\begin{figure}[h]
\centerline{\includegraphics[width=4.cm,clip]{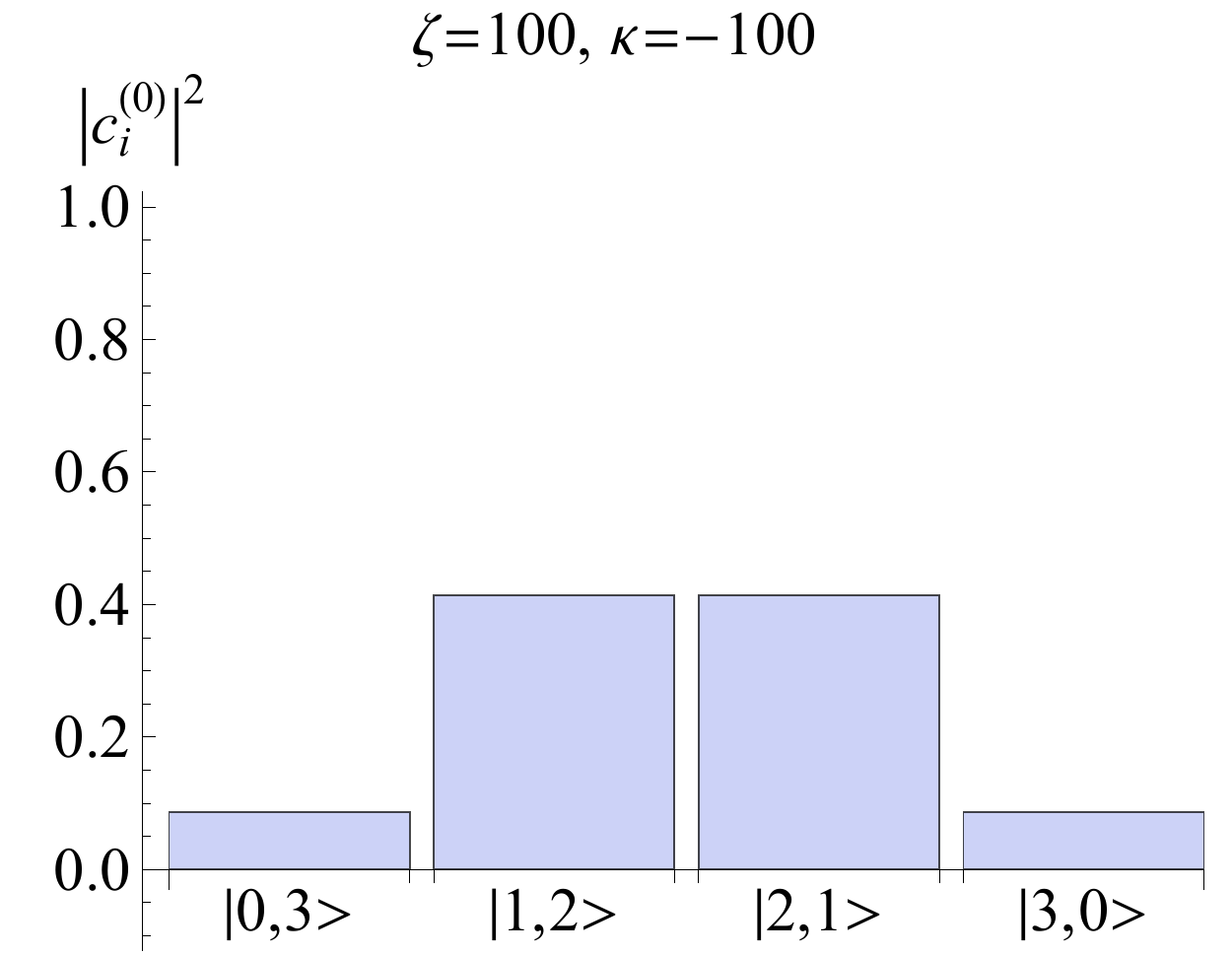}
\includegraphics[width=4.cm,clip]{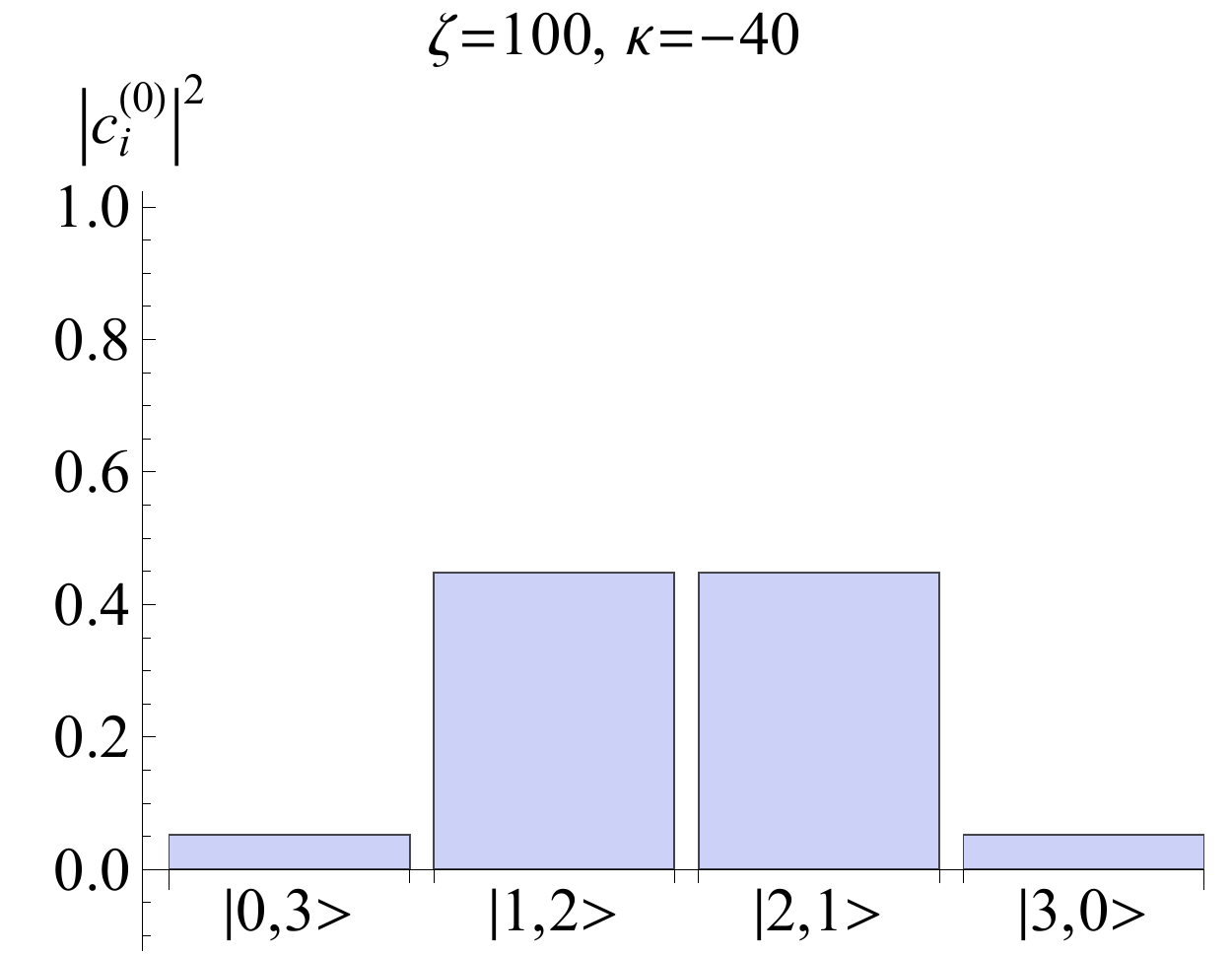}
\includegraphics[width=4.cm,clip]{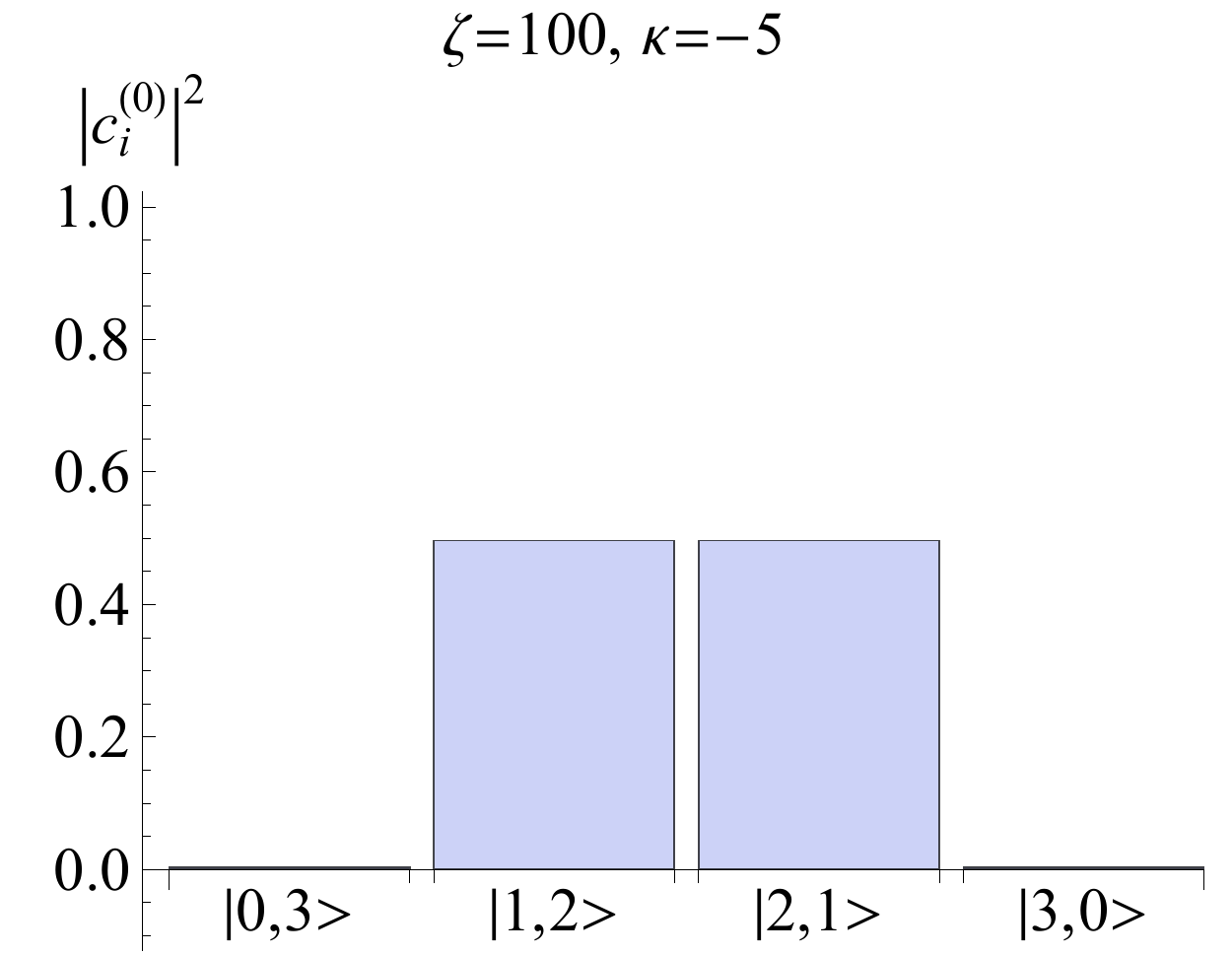}
\includegraphics[width=4.cm,clip]{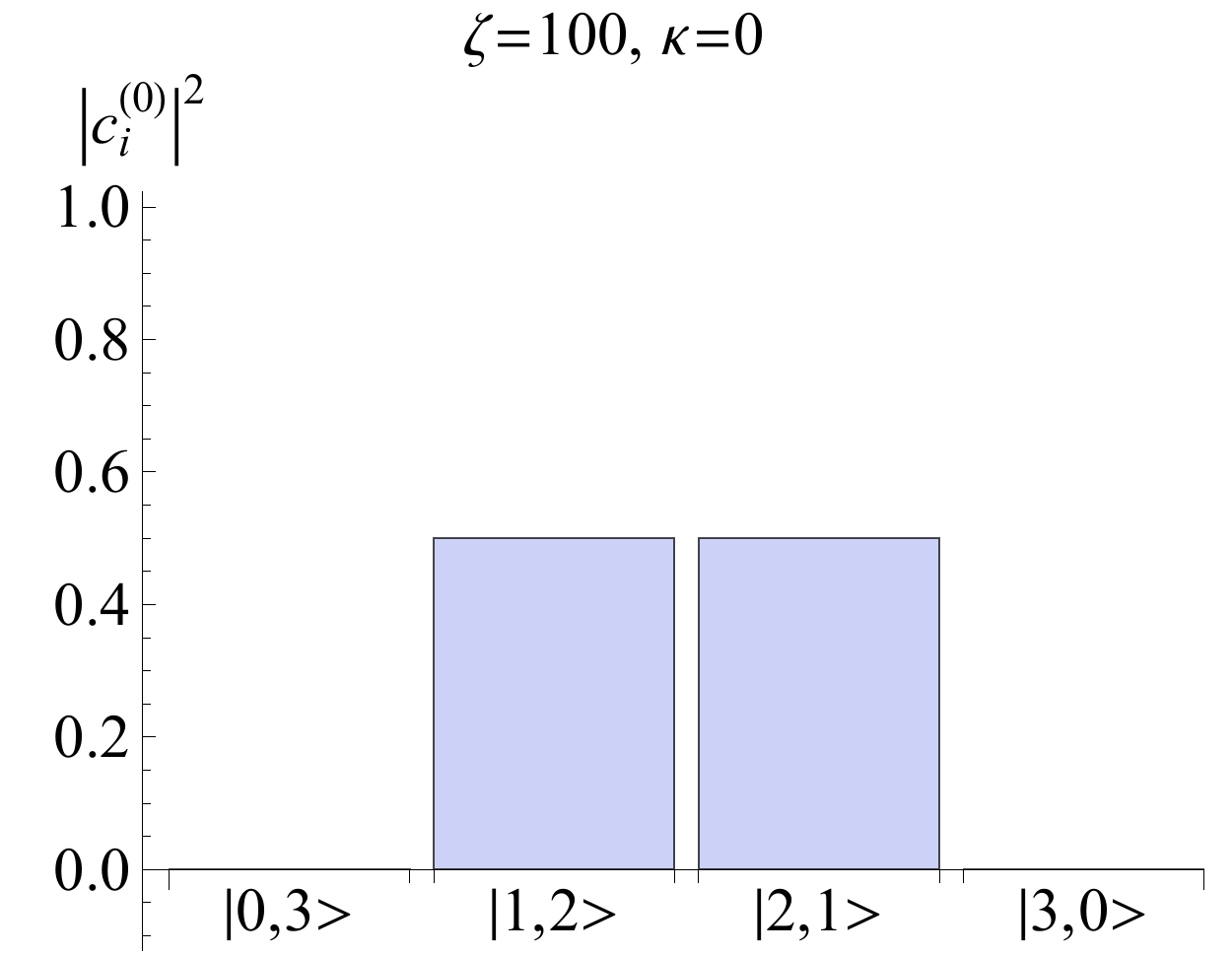}}
\centerline{
\includegraphics[width=4.cm,clip]{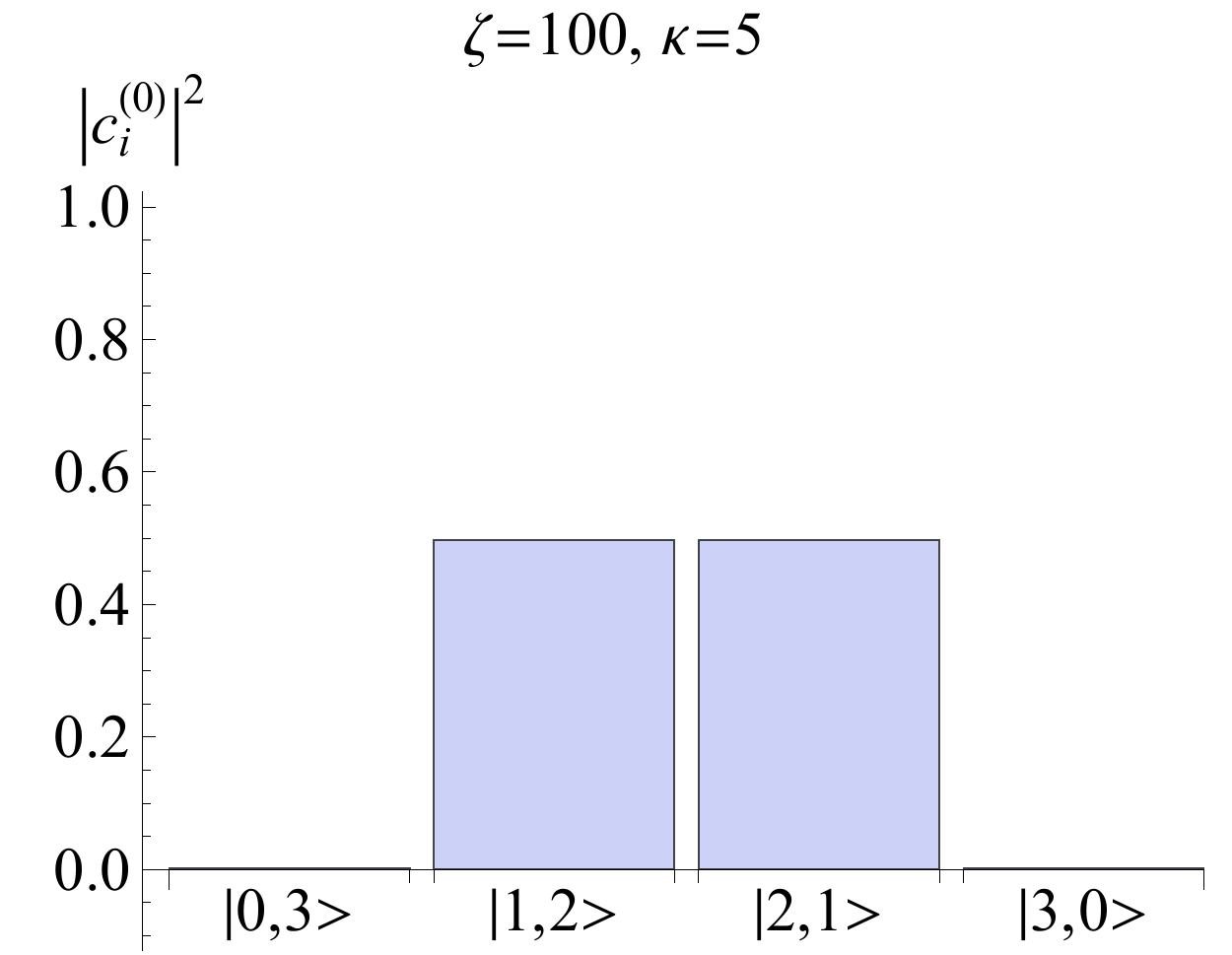}
\includegraphics[width=4.cm,clip]{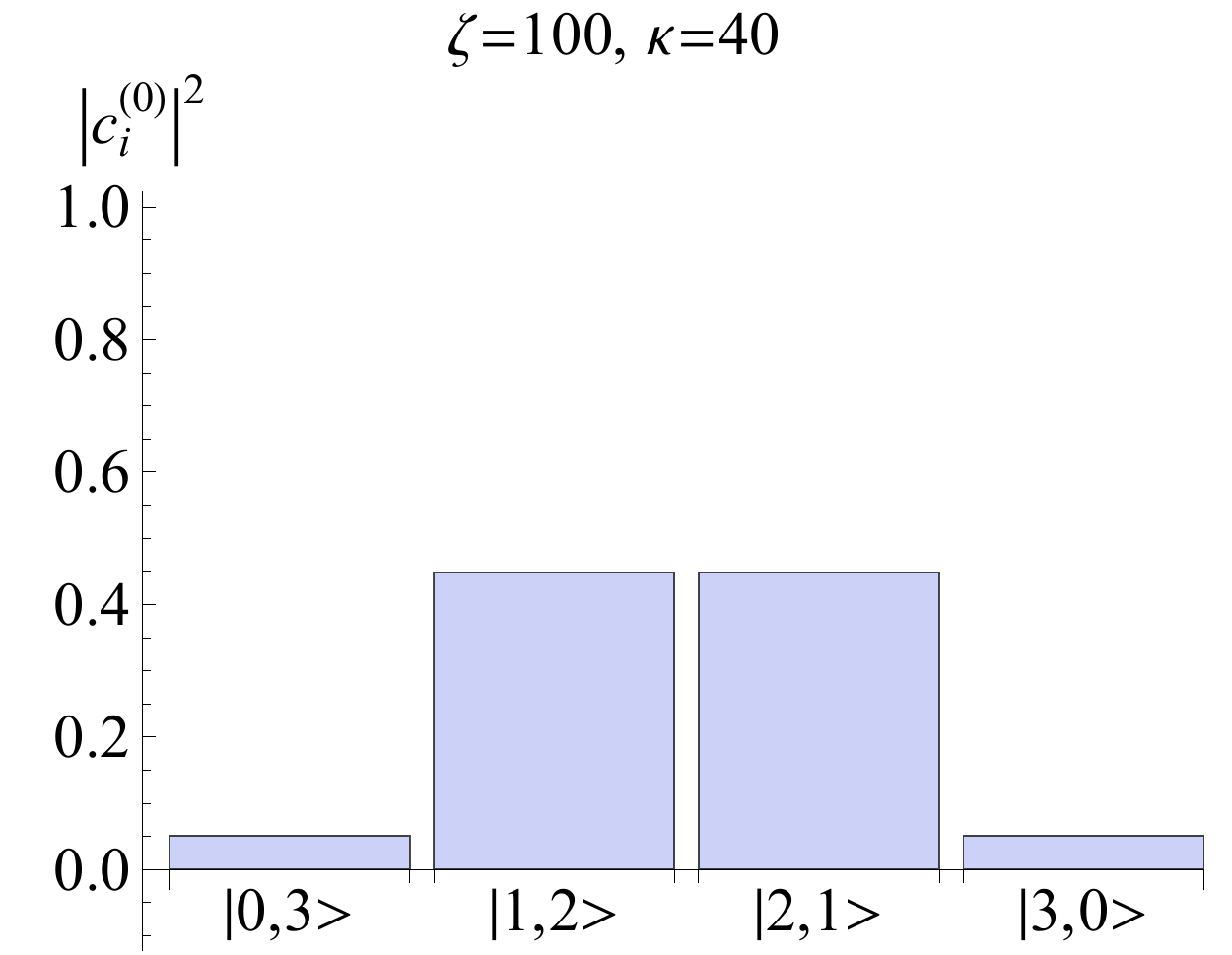}
\includegraphics[width=4.cm,clip]{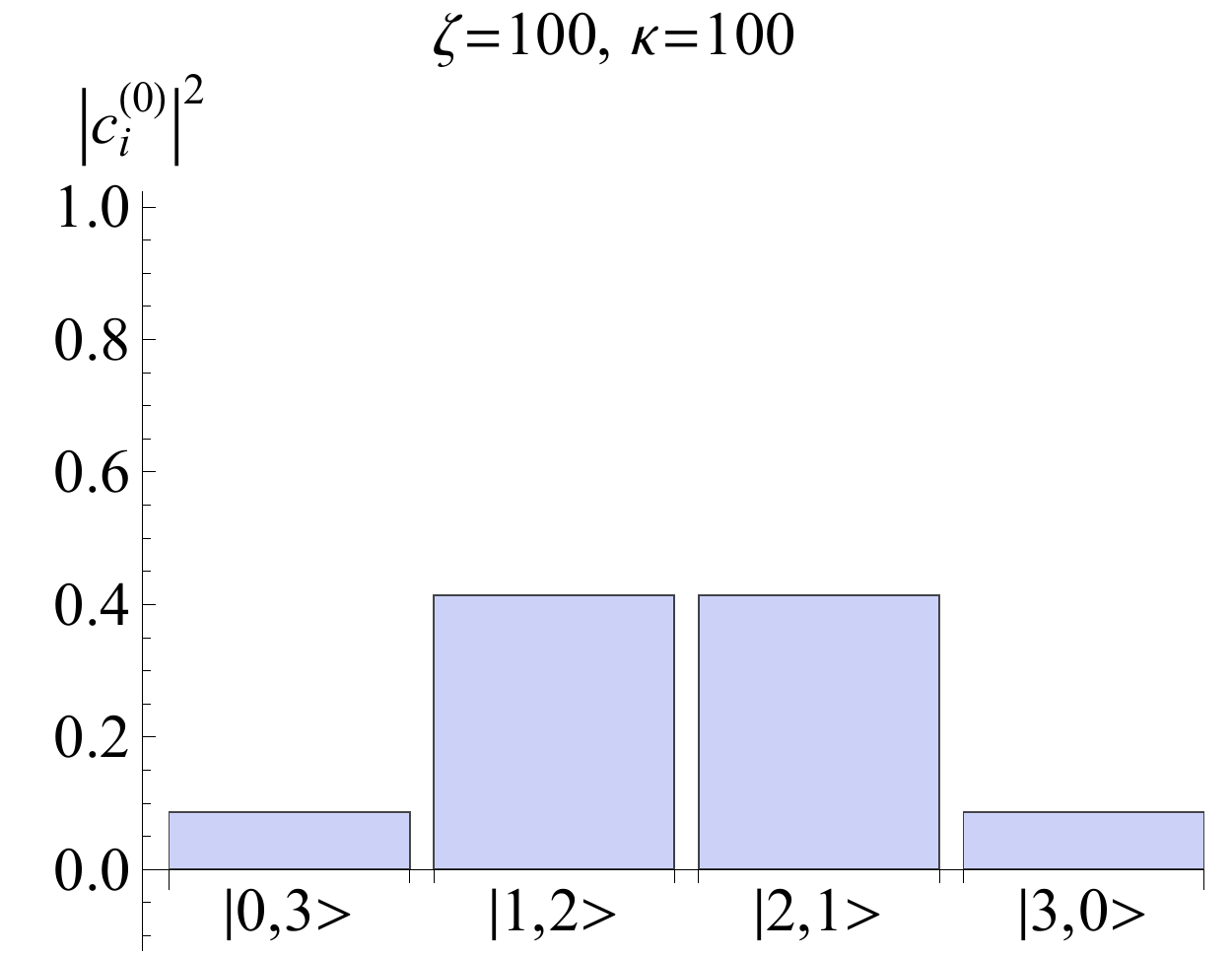}
\includegraphics[width=4.cm,clip]{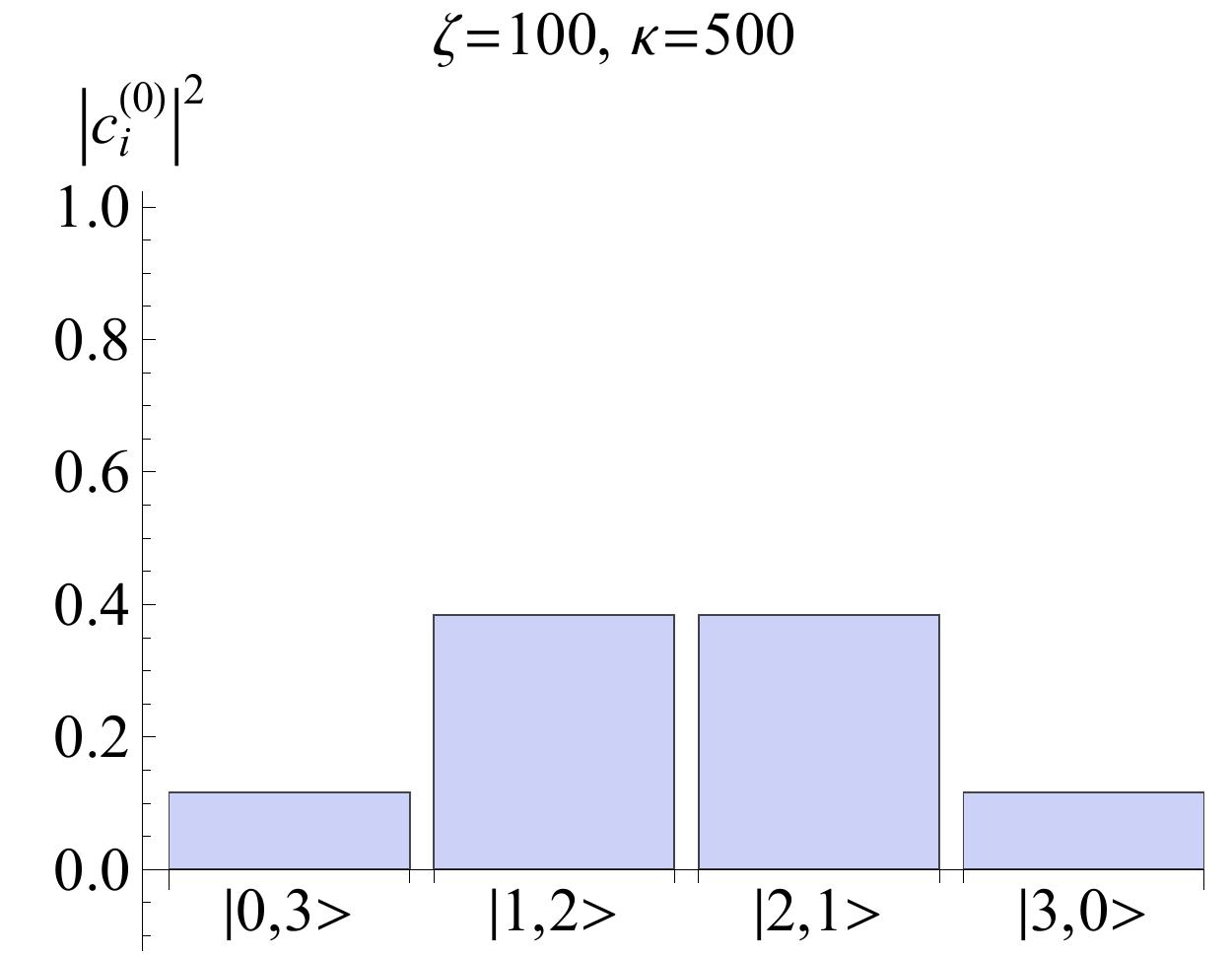}
}
\caption{$N=3$. Horizontal axis: ket $|i,N-i\rangle$. Vertical axis: $|c_{i}^{(0)}|^2$. The scaled onsite interaction $\zeta=U/J$ is fixed (strong repulsion). The scaled correlated hopping $\kappa=K_c/J$ changes. Top-bottom: in the first row ($\kappa \le 0$), $|\kappa|$ decreases from left to right; in the second row ($\kappa >0$), $\kappa$ increases from left to right. All the quantities are dimensionless.}
\label{fig6}
\end{figure}
\begin{figure}[h]
\centerline{\includegraphics[width=4.cm,clip]{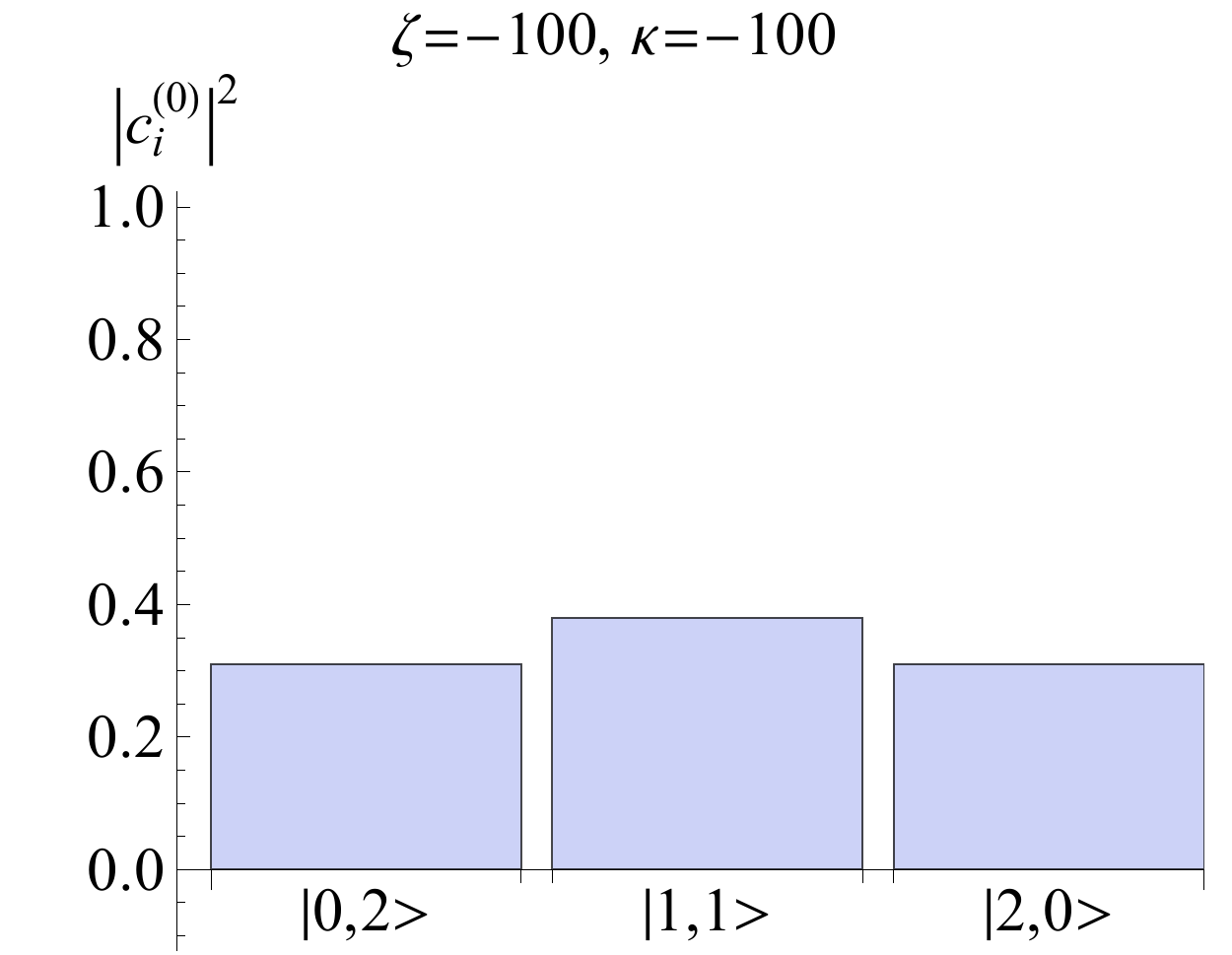}
\includegraphics[width=4.cm,clip]{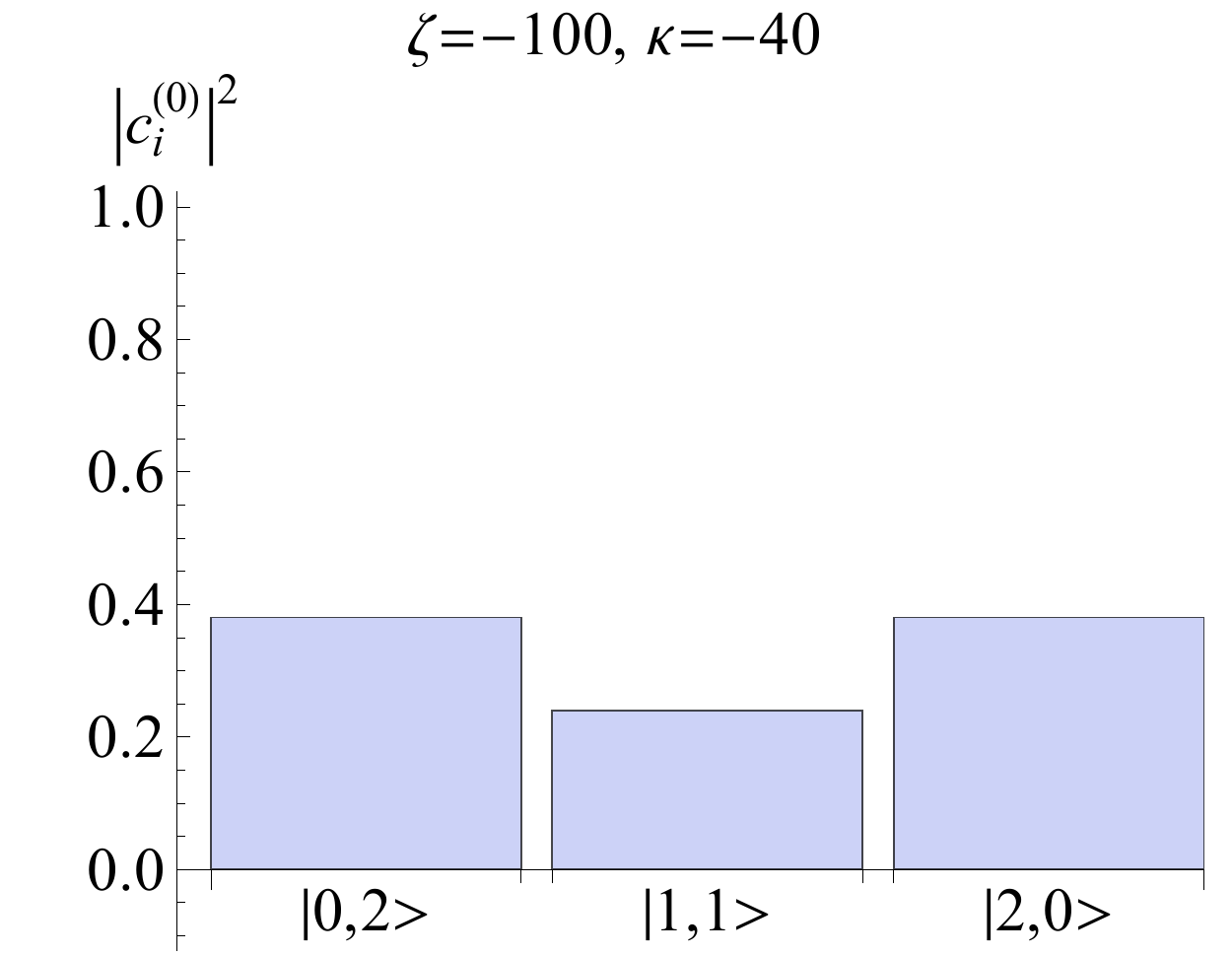}
\includegraphics[width=4.cm,clip]{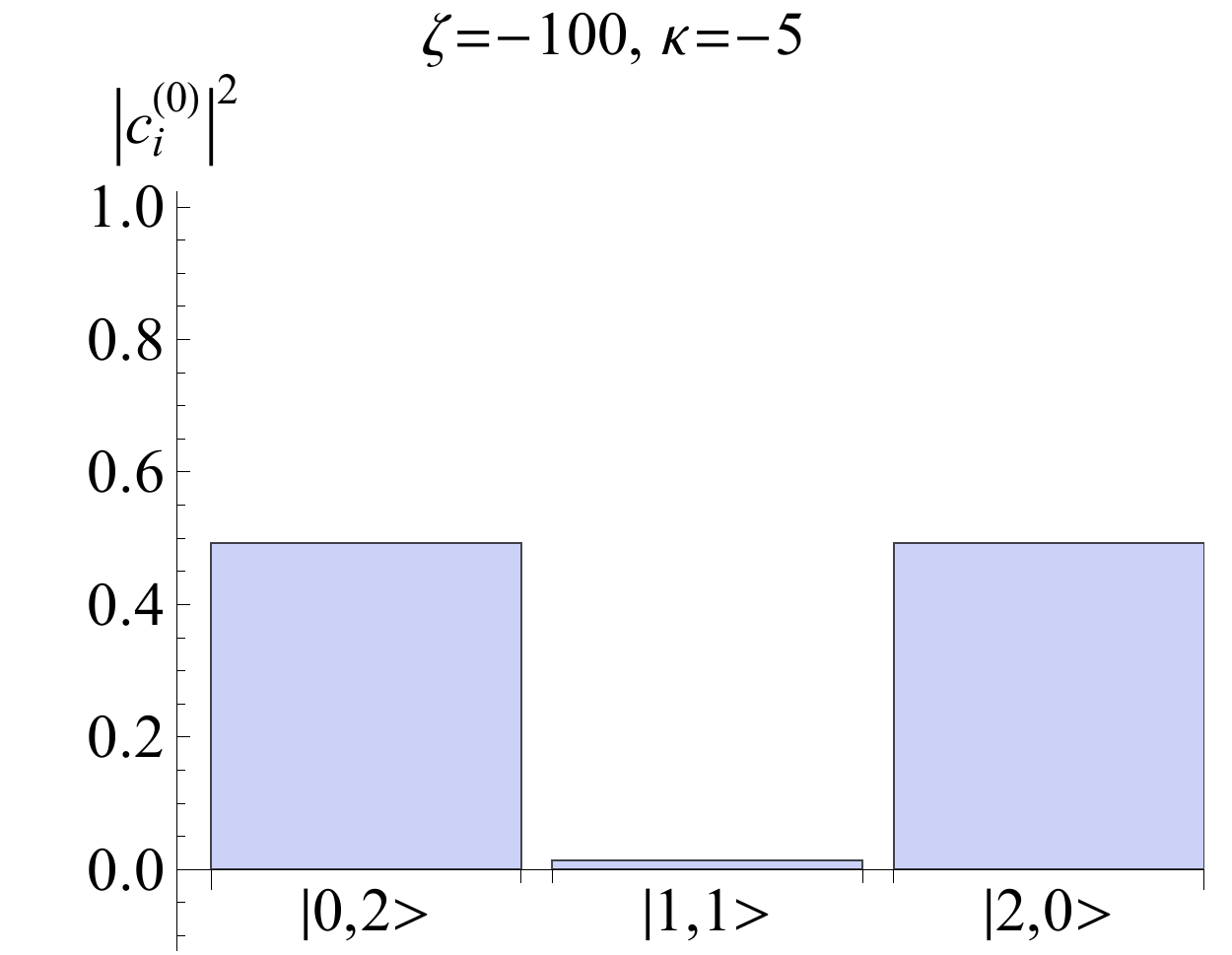}
\includegraphics[width=4.cm,clip]{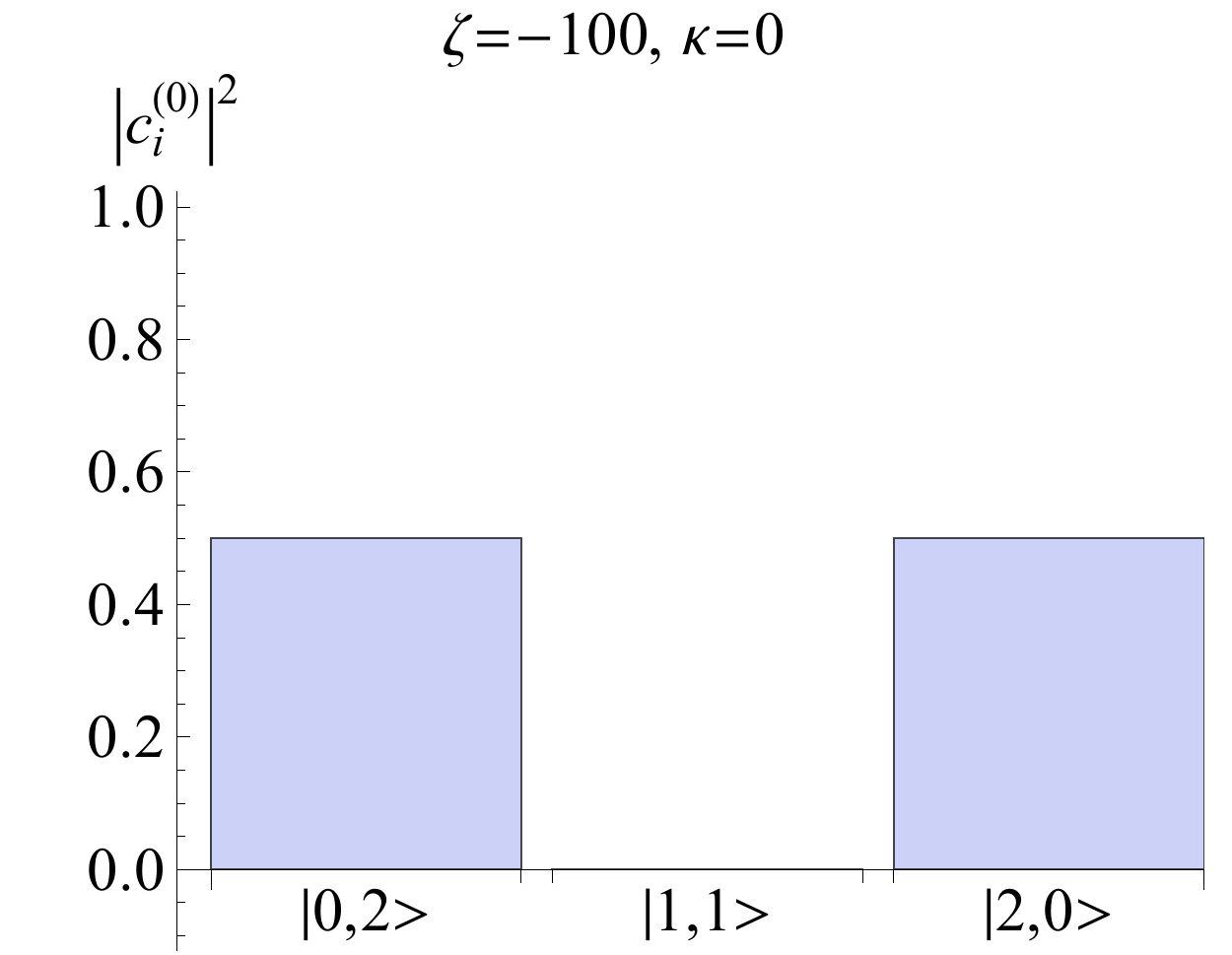}}
\centerline{
\includegraphics[width=4.cm,clip]{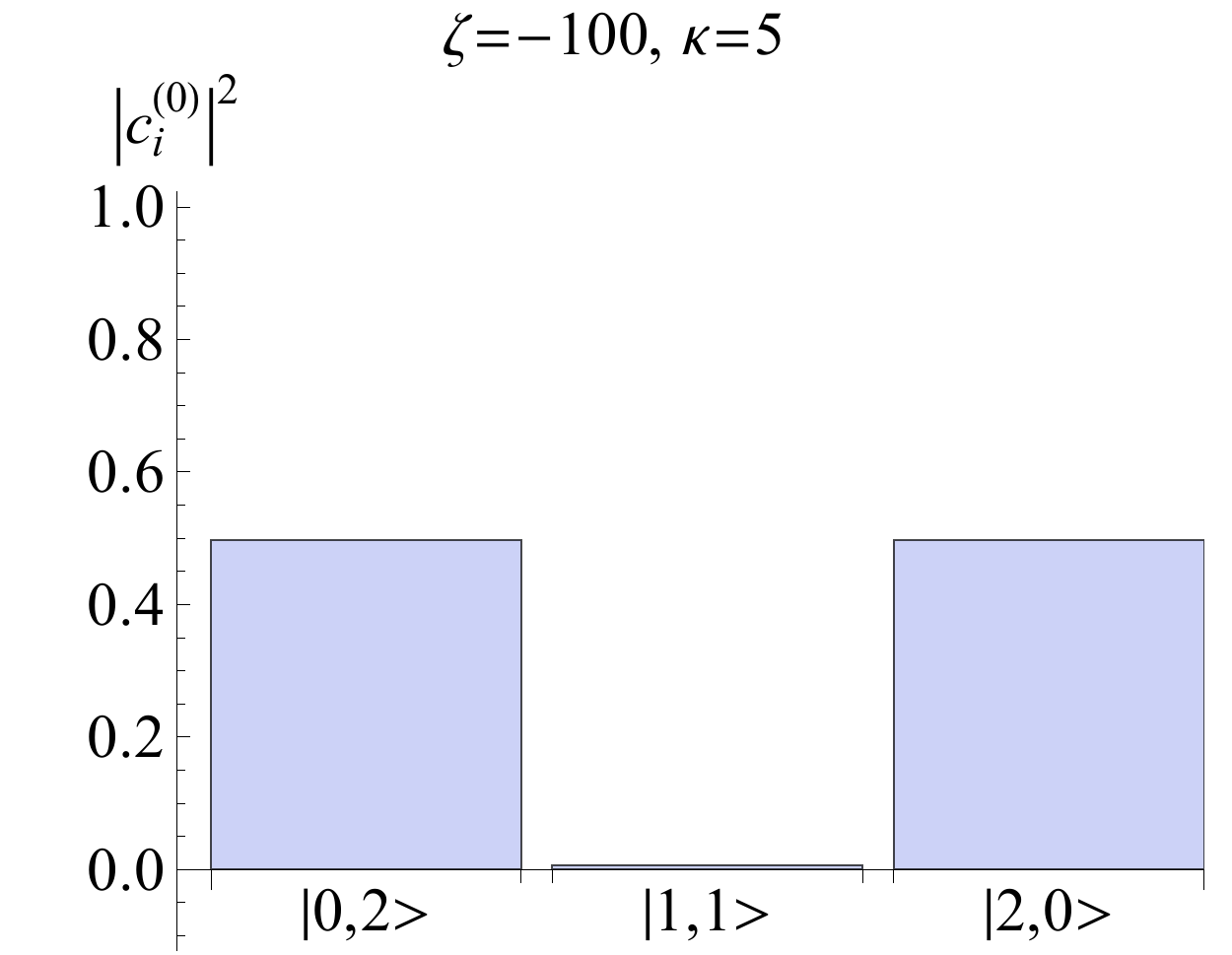}
\includegraphics[width=4.cm,clip]{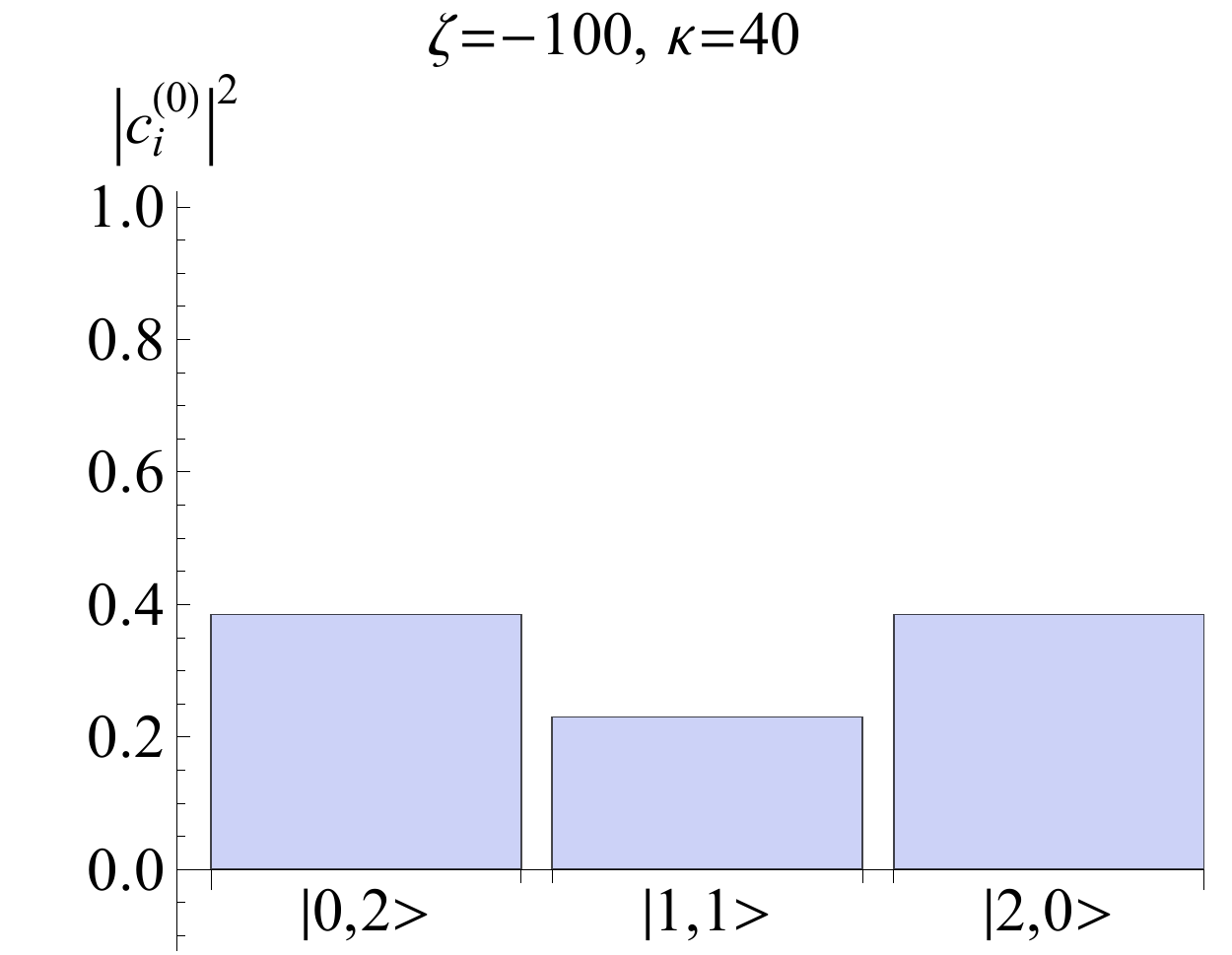}
\includegraphics[width=4.cm,clip]{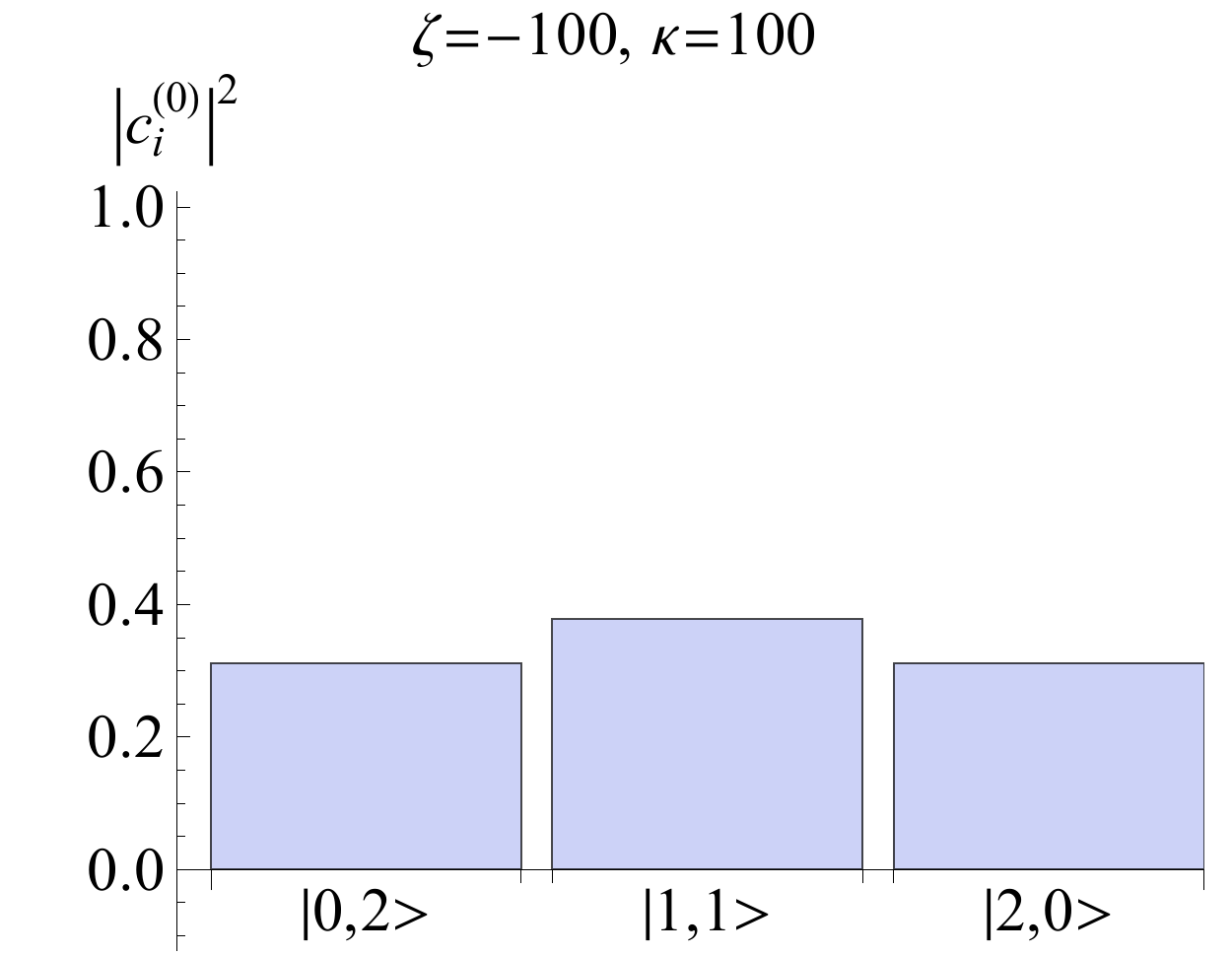}
\includegraphics[width=4.cm,clip]{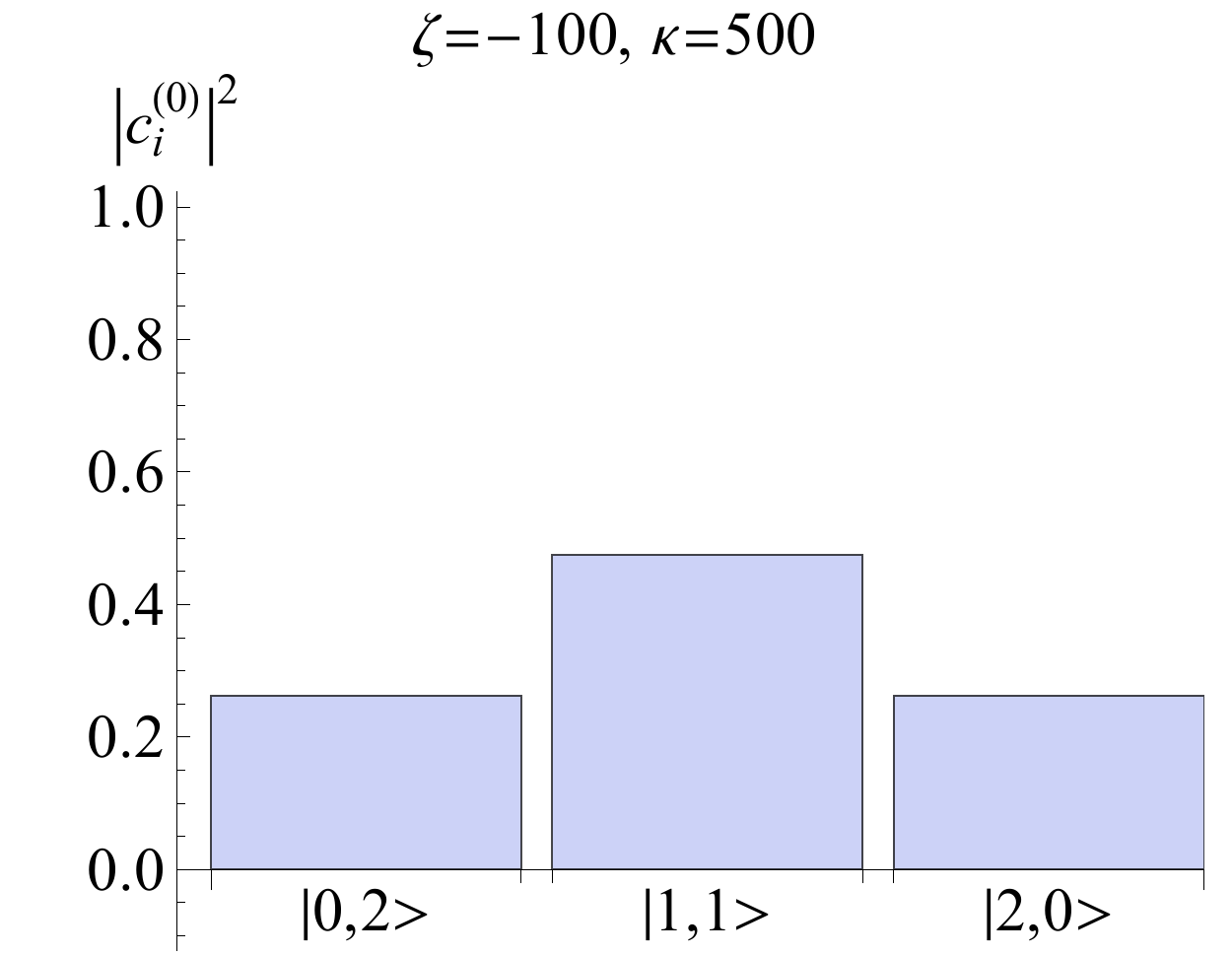}}
\caption{$N=2$. Horizontal axis: ket $|i,N-i\rangle$. Vertical axis:$|c_{i}^{(0)}|^2$. The scaled onsite interaction $\zeta=U/J$ is fixed (strong attraction). The scaled correlated hopping $\kappa=K_c/J$ changes. Top-bottom: in the first row ($\kappa \le 0$), $|\kappa|$ decreases from left to right; in the second row ($\kappa >0$), $\kappa$ increases from left to right. All the quantities are dimensionless.}
\label{fig7}
\end{figure}
\begin{figure}[h]
\centerline{\includegraphics[width=4.cm,clip]{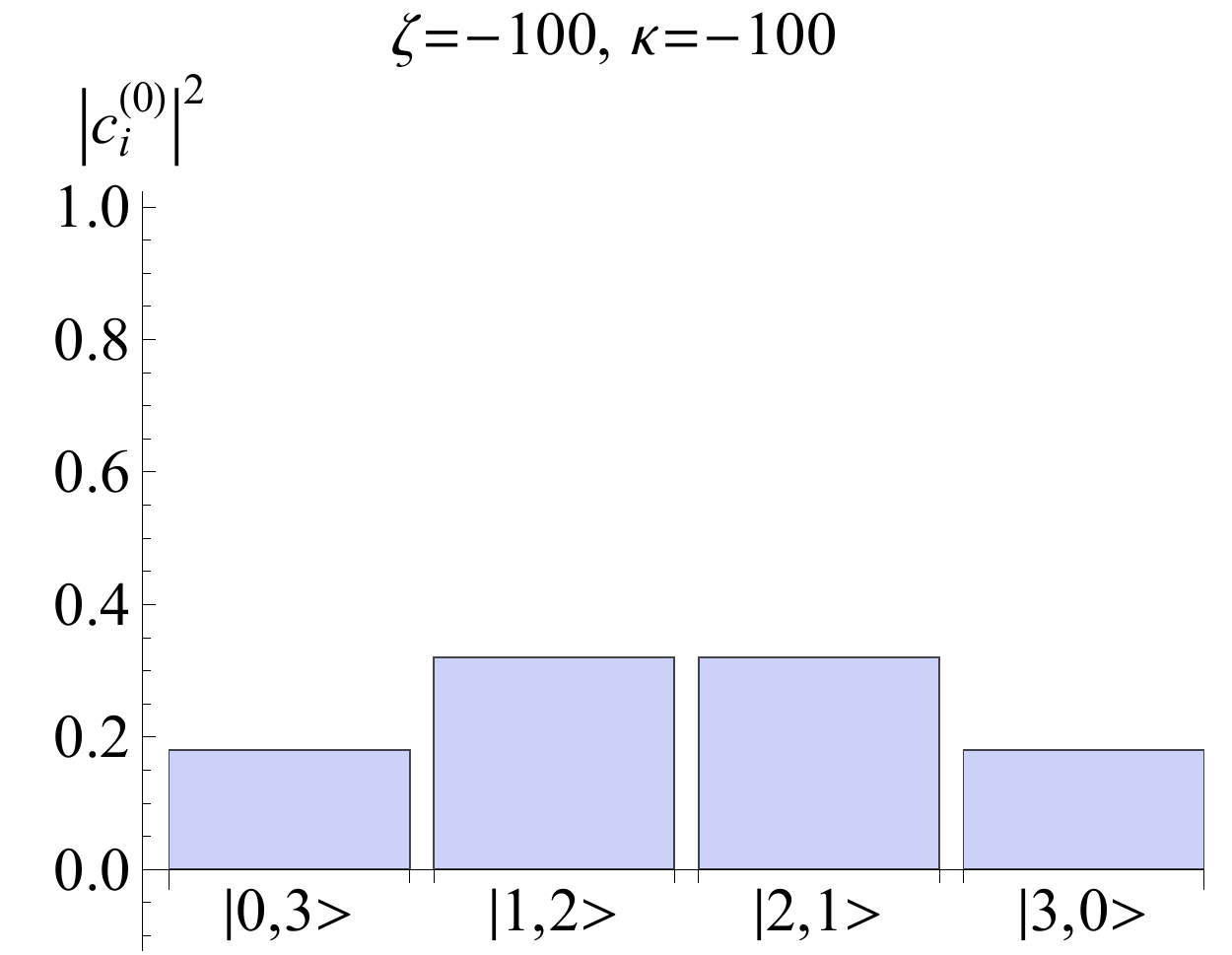}
\includegraphics[width=4.cm,clip]{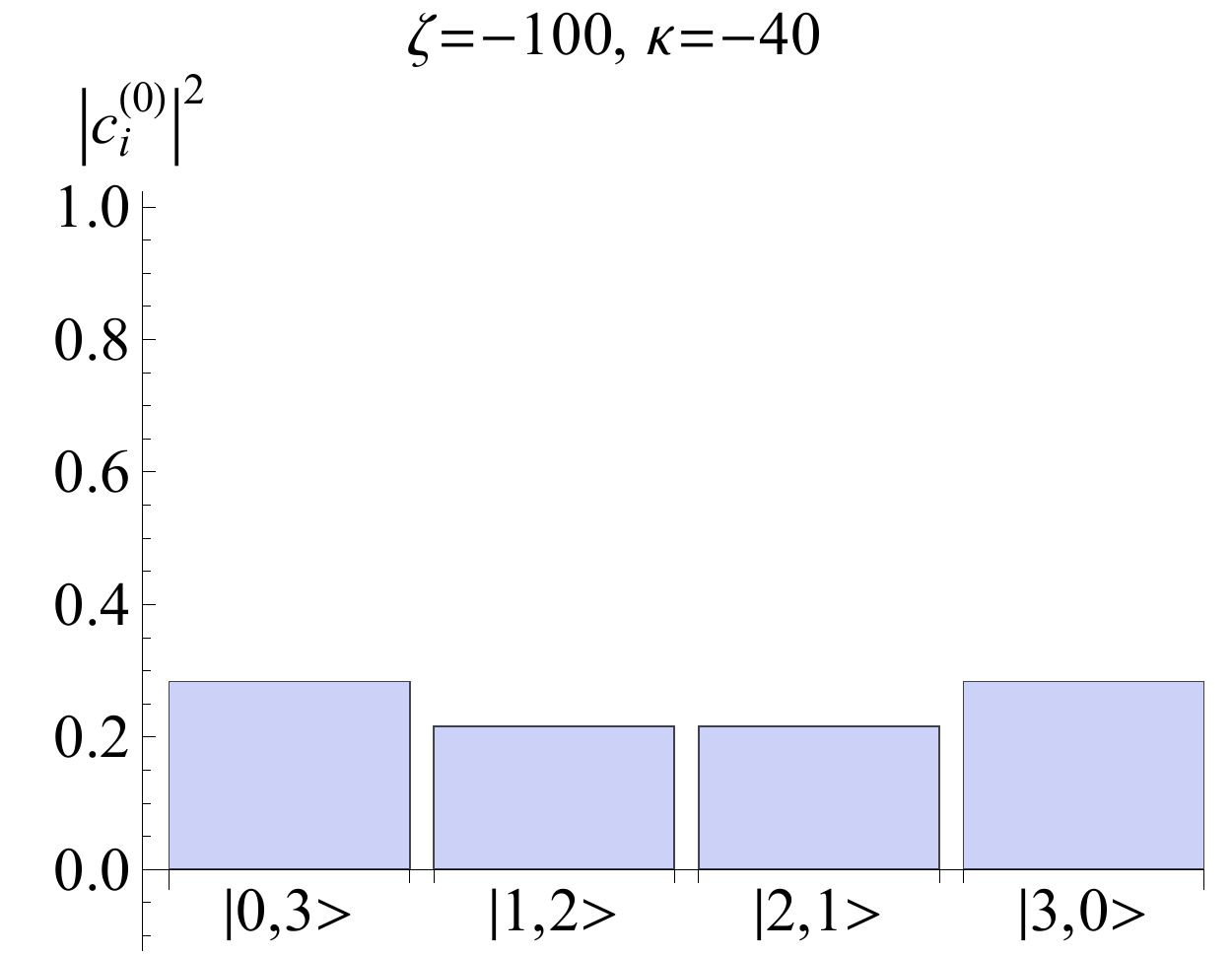}
\includegraphics[width=4.cm,clip]{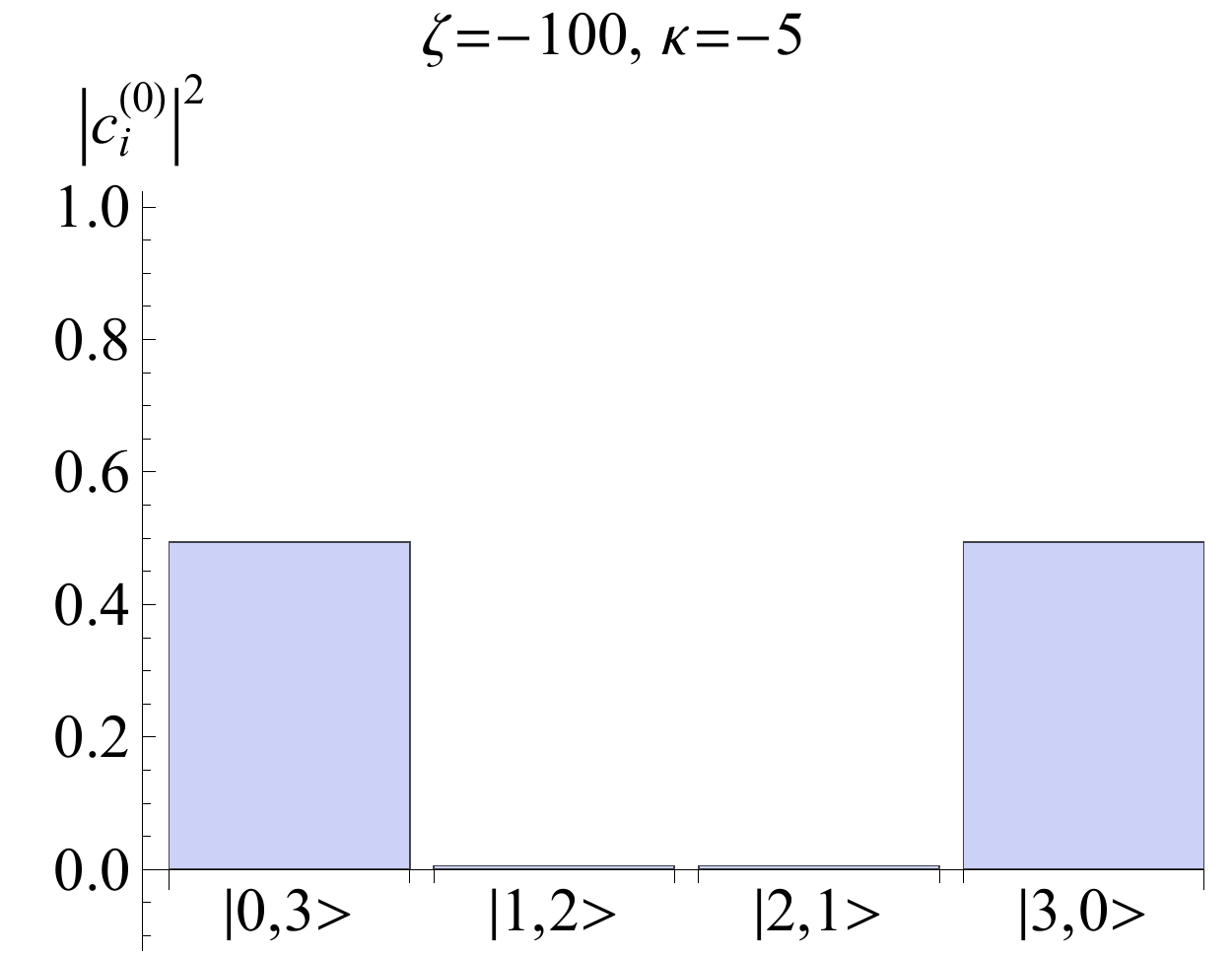}
\includegraphics[width=4.cm,clip]{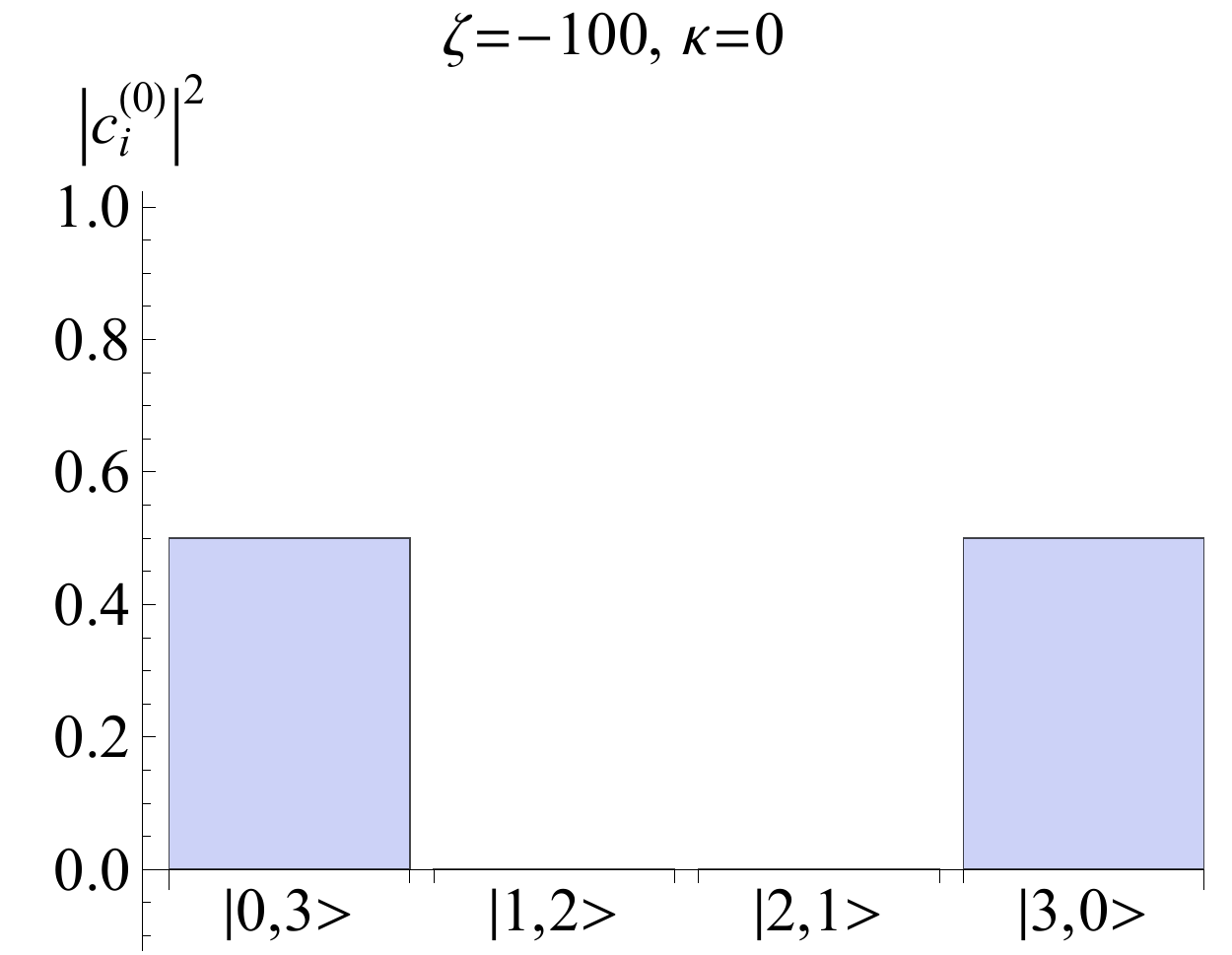}}
\centerline{
\includegraphics[width=4.cm,clip]{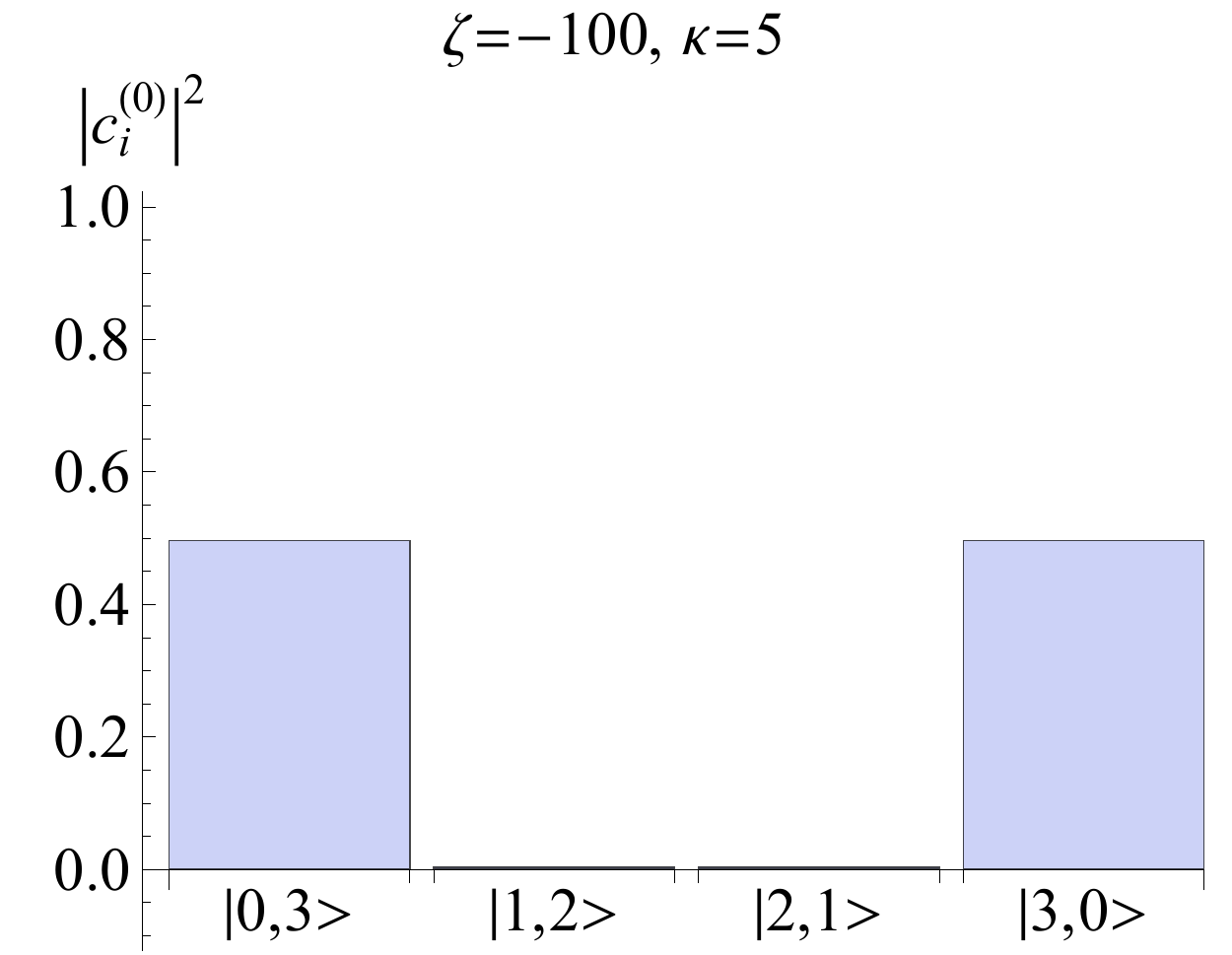}
\includegraphics[width=4.cm,clip]{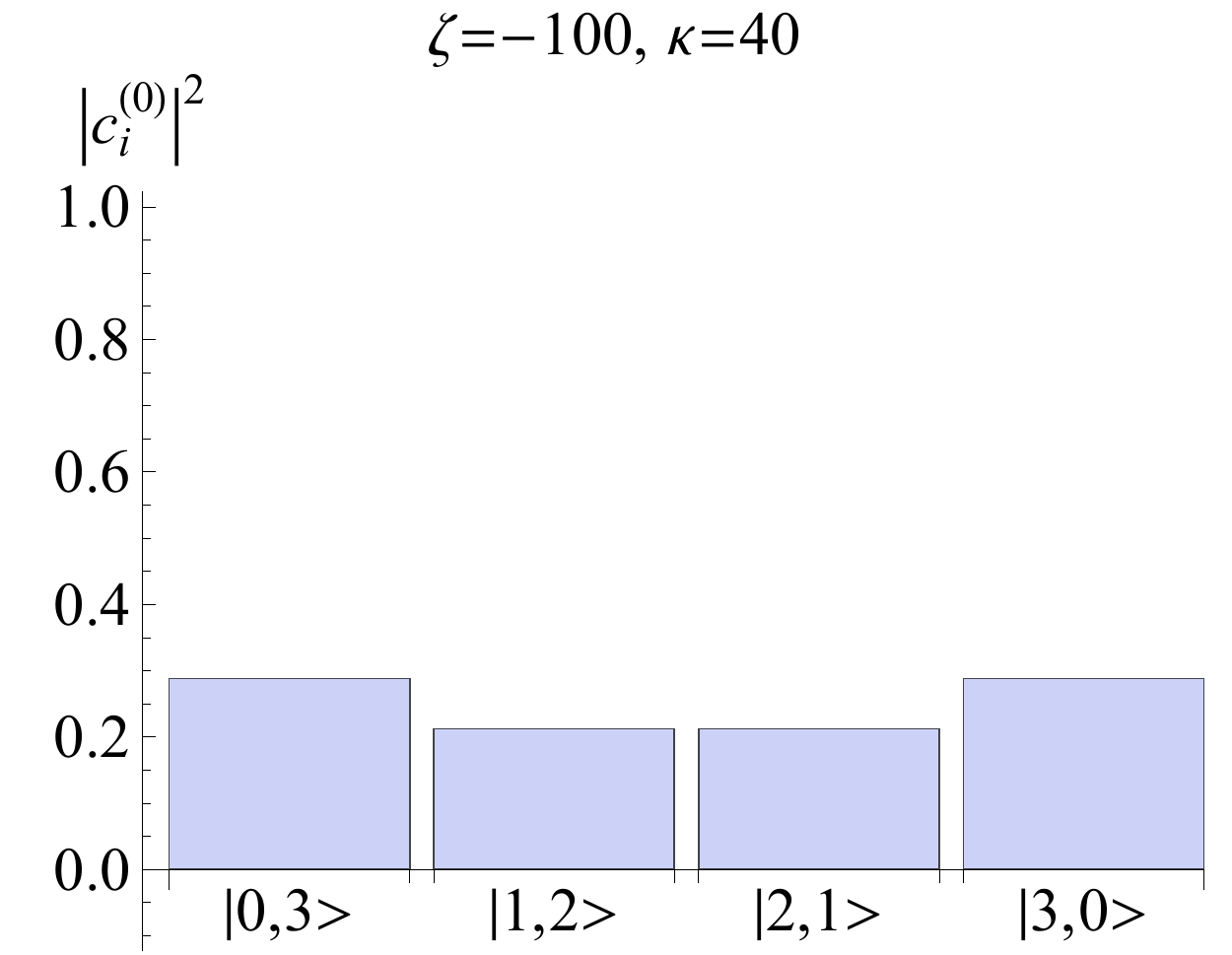}
\includegraphics[width=4.cm,clip]{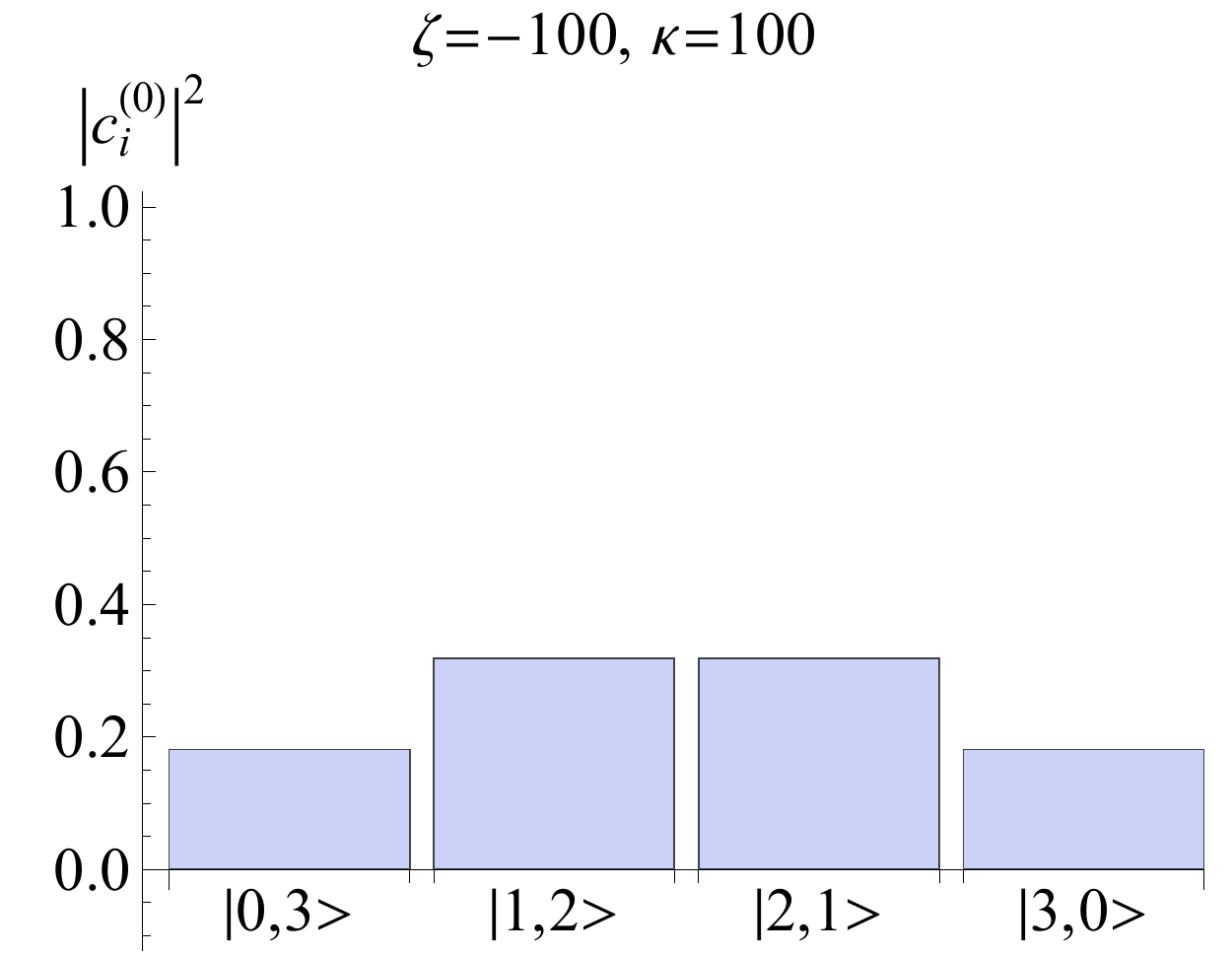}
\includegraphics[width=4.cm,clip]{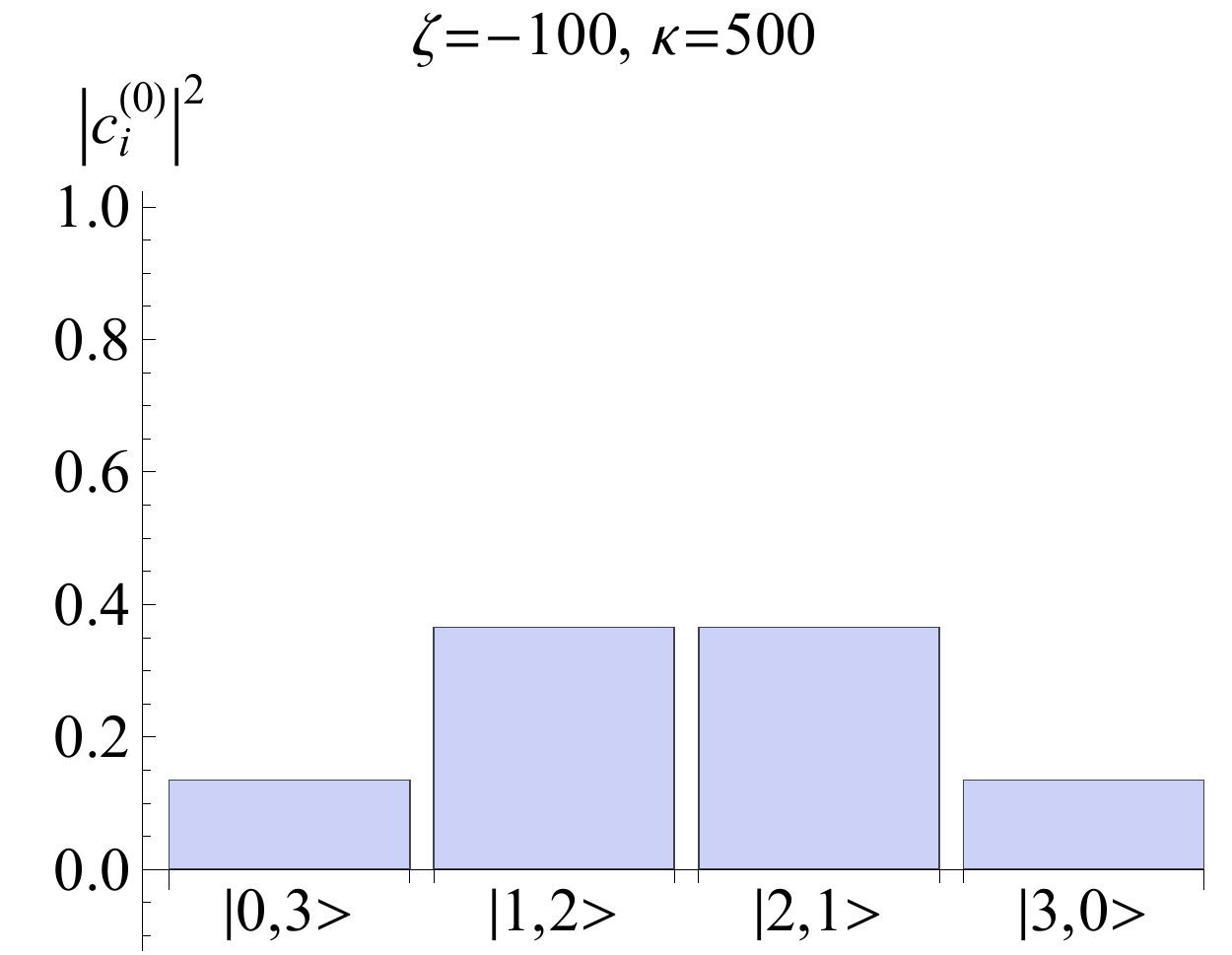}
}
\caption{$N=3$. Horizontal axis: ket $|i,N-i\rangle$. Vertical axis:$|c_{i}^{(0)}|^2$. The scaled onsite interaction $\zeta=U/J$ is fixed (strong attraction). The scaled correlated hopping $\kappa=K_c/J$ changes. Top-bottom: in the first row ($\kappa \le 0$), $|\kappa|$ decreases from left to right; in the second row ($\kappa >0$), $\kappa$ increases from left to right. All the quantities are dimensionless.}
\label{fig8}
\end{figure}

Let us continue our analysis by inspecting Figs. 5-8. In these figures we report $|c_{i}^{(0)}|^2$ when the scaled onsite interaction $\zeta$ is such that to have a strong repulsion, Figs. 5 ($N=2$) and 6 ($N=3$) and a strong attraction, Figs. 7 ($N=2$) and 8 ($N=3$). From Figs. 5-6, we can see that when $\kappa=0$, the ground state is the twin Fock state with $N=2$ (last plot [left-right] of the top row of Fig. 5) and the pseudo-Fock state with $N=3$ (last plot [left-right] of the top row of Fig. 6). In the presence of a moderate collisionally-induced tunneling ($|\kappa|=5$, the third plot [left-right] of the top row and the first plot [left-right] of the bottom row of Figs. 5 and 6), the ground-state structure does not change meaningfully. Nevertheless, we can observe that the larger $|\kappa|$ the closer the ground state to the atomic coherent state (see the plots with $|\kappa|$ ranging from $5$ to $100$ passing for $40$).
\begin{figure}[h]
\centerline{\includegraphics[width=4.5cm,clip]{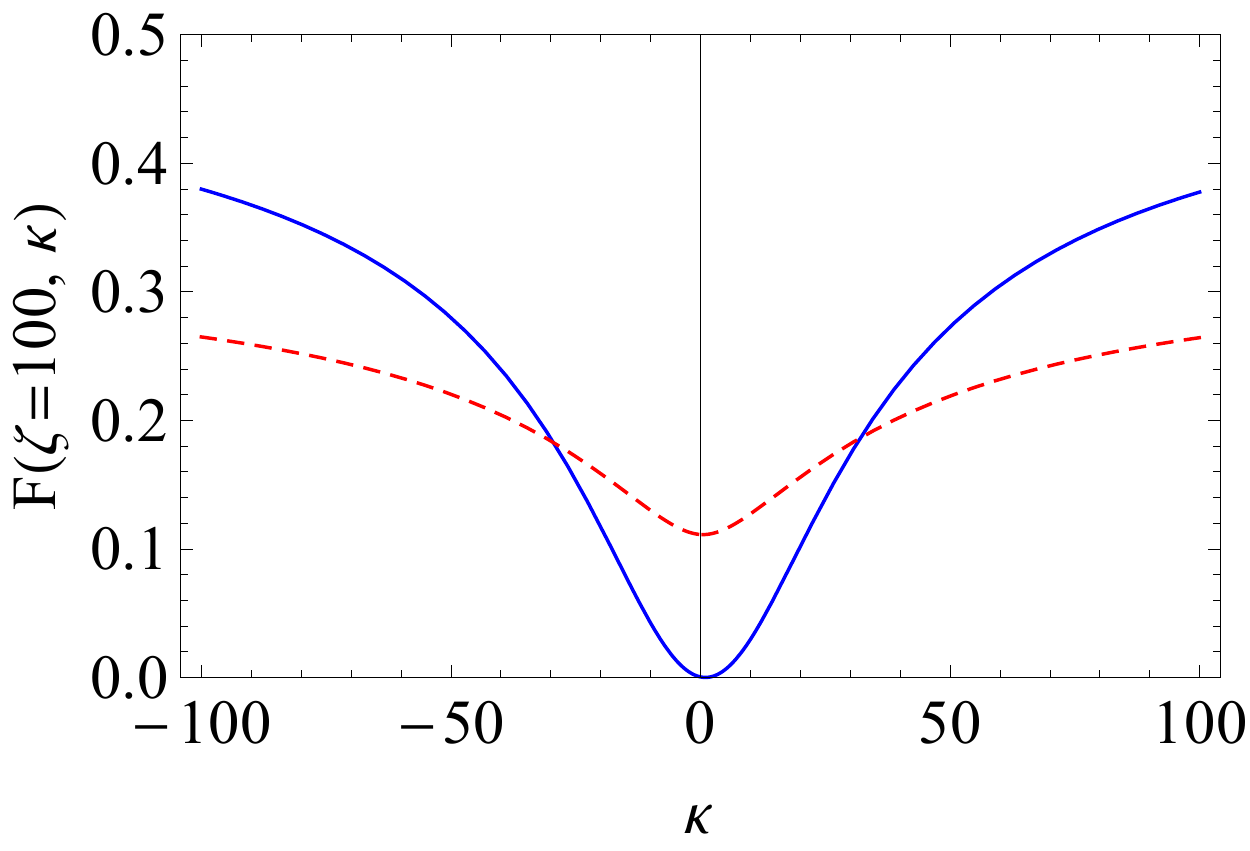}
\includegraphics[width=4.5cm,clip]{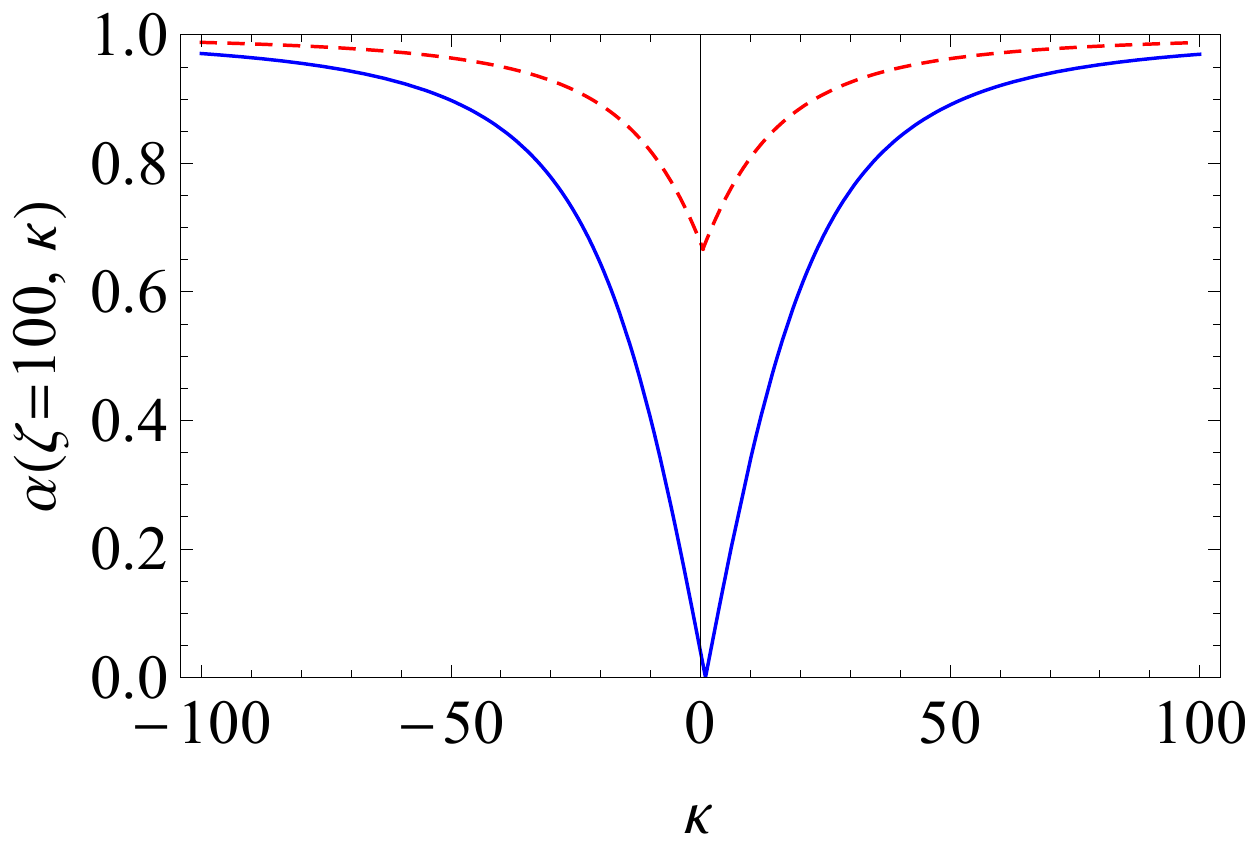}
\includegraphics[width=4.5cm,clip]{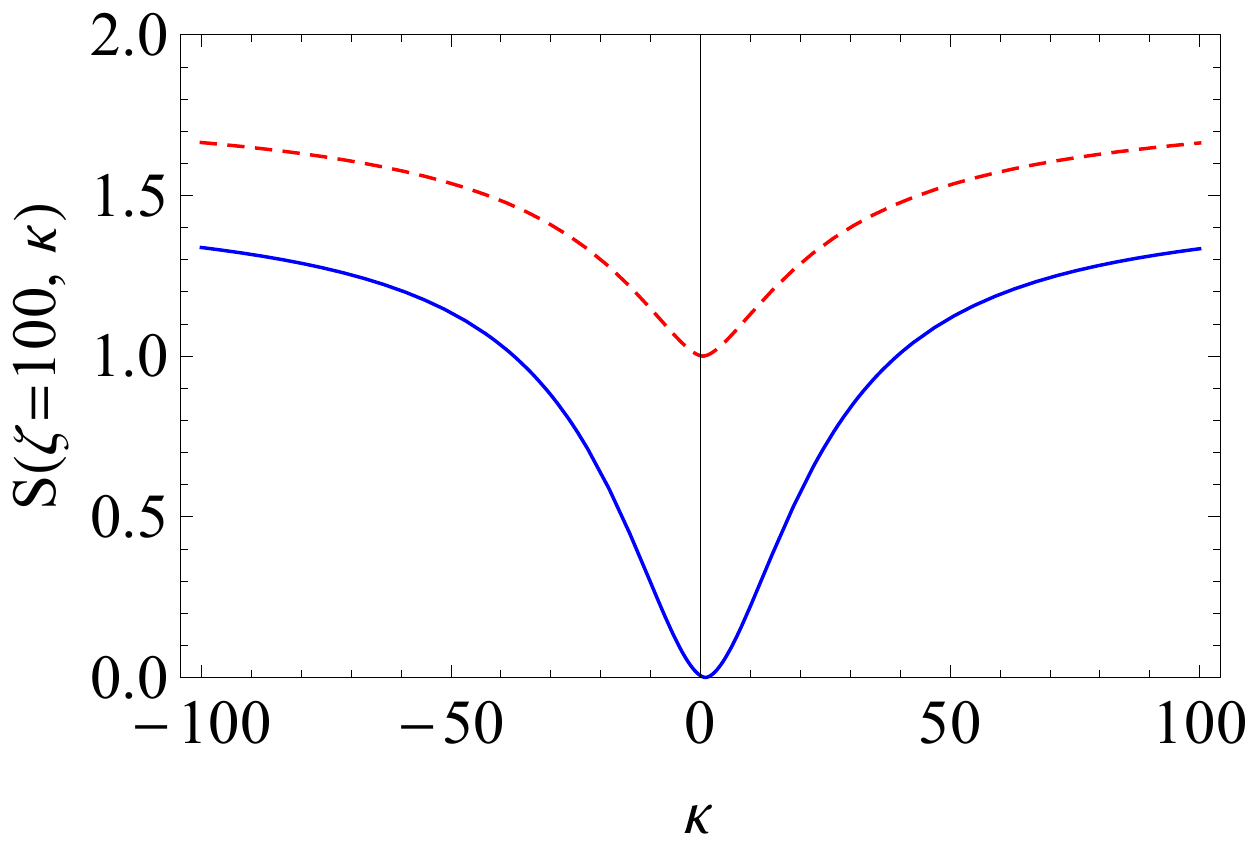}}
\caption{Fisher information $F$ (left panel), coherence visibility $\alpha$ (middle panel), entanglement entropy $S$ (right panel) vs scaled correlated hopping $\kappa=K_c/J$ with $\zeta=U/J=100$ (this corresponds to situations displayed in Figs. 5-6). In each panel the solid line corresponds to $N=2$, the dashed line to $N=3$. All the quantities are dimensionless.}
\label{fig9}
\end{figure}
\begin{figure}[h]
\centerline{\includegraphics[width=4.5cm,clip]{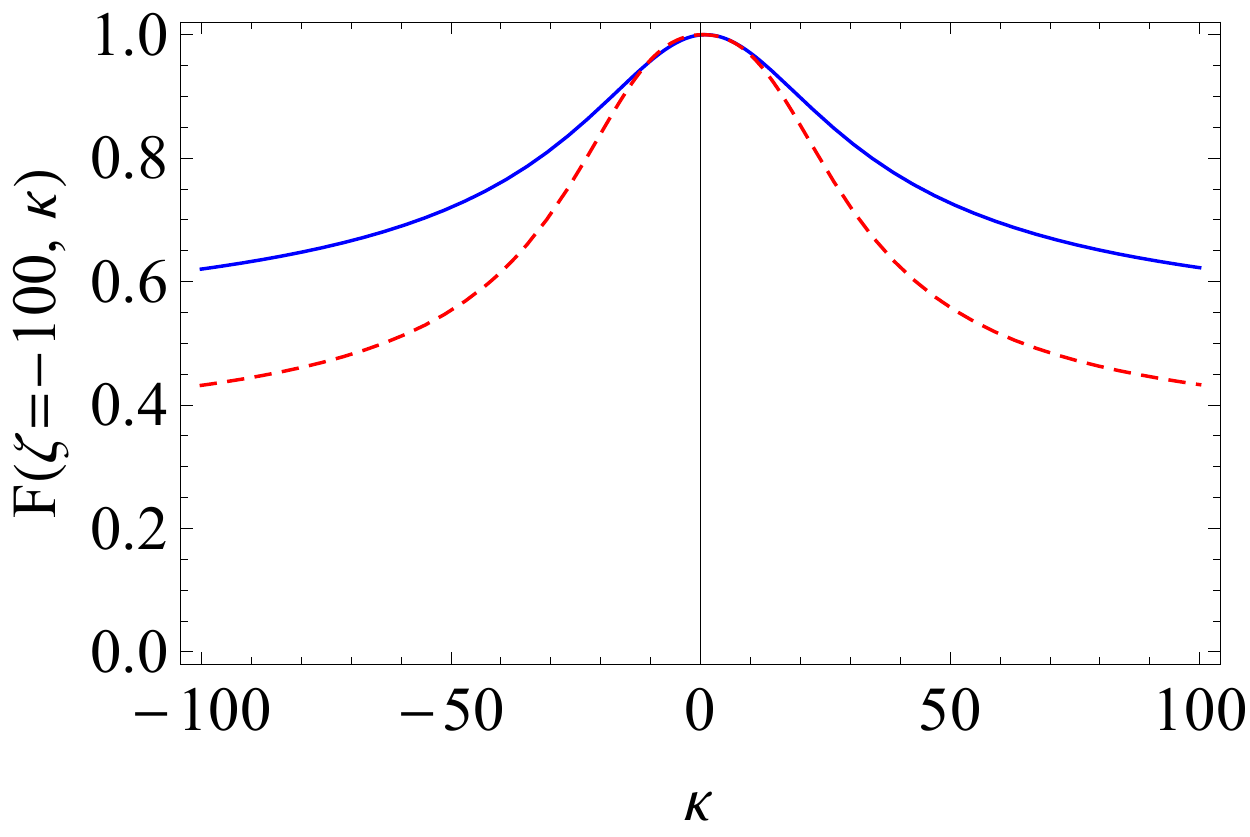}
\includegraphics[width=4.5cm,clip]{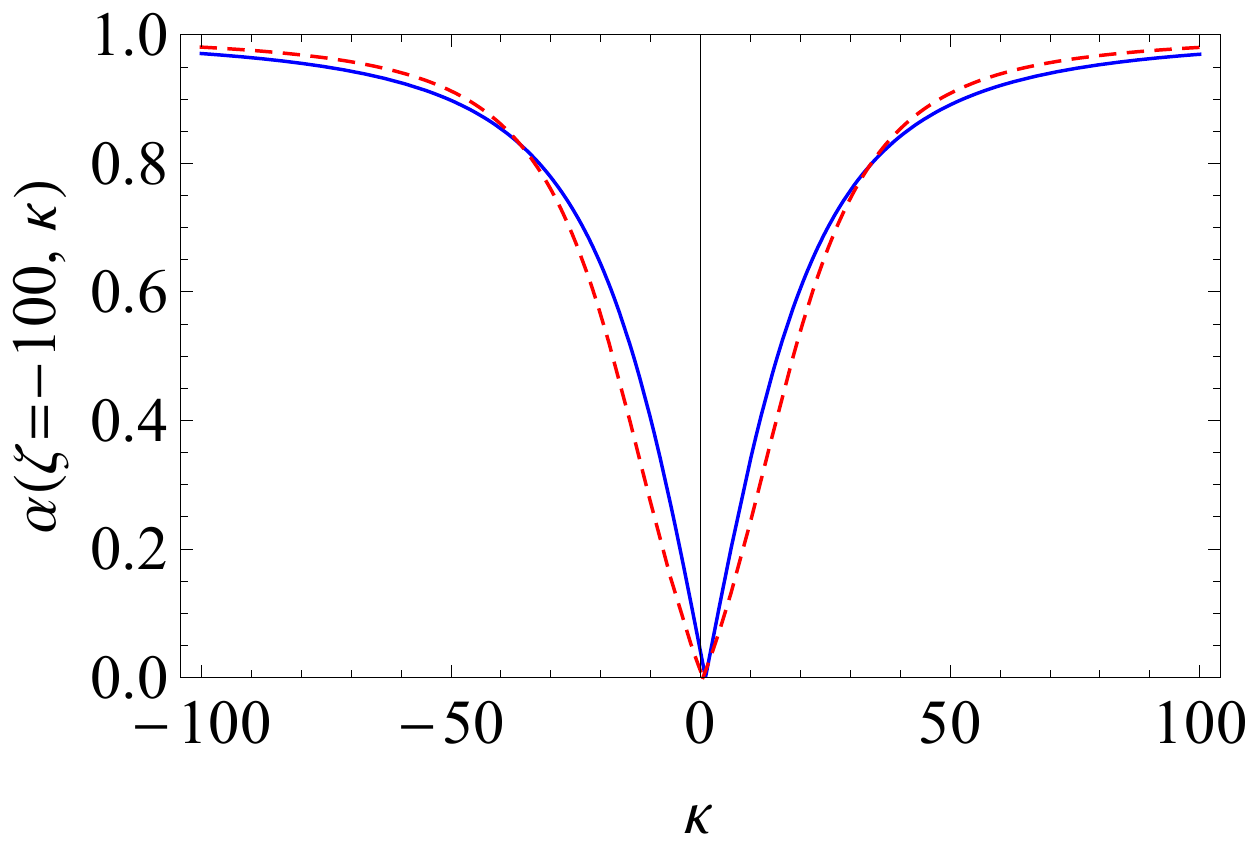}
\includegraphics[width=4.5cm,clip]{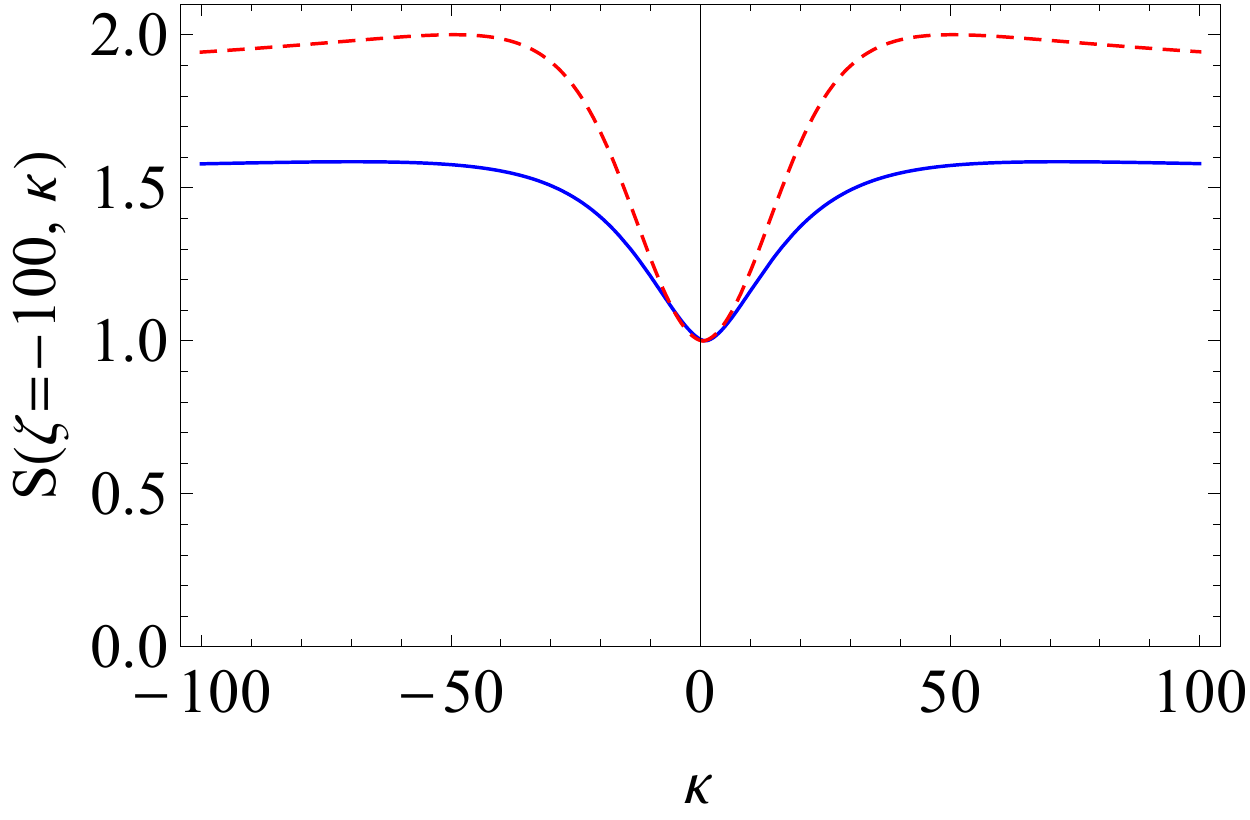}}
\caption{Fisher information $F$ (left panel), coherence visibility $\alpha$ (middle panel), entanglement entropy $S$ (right panel) vs scaled correlated hopping $\kappa=K_c/J$ with $\zeta=U/J=-100$ (this corresponds to situations displayed in Figs. 7-8). In each panel the solid line corresponds to $N=2$, the dashed line to $N=3$. All the quantities are dimensionless.}
\label{fig10}
\end{figure}
The so far discussed analysis is supported by the study of the correlation indicators plotted as functions of $\kappa$ displayed in Fig. 9. In particular, from the middle panel it can be seen that the lowest value of the coherence visibility ($\alpha=0$) is achieved when $\kappa=0$, but when one explores larger values of $\kappa$ both on the negative and positive side, $\alpha$ sensibly increases and becomes almost equal to $1$ when $|\kappa| \simeq 100$ that corresponds to the first plot (left-right) of the top row and to the third plot (left-right) of the bottom row of Figs. 5-6. Then, despite the strong onsite repulsion, the density-induced tunneling has the effect to favor a delocalized atomic coherent state at disadvantage of the (pseudo)Fock one. Similar arguments hold for Figs. 7-8 and 10 (the latter includes the plots of $F$, $\alpha$, $S$ as functions of $\kappa$), where the correlated hopping tends to destroy the cat-like state to establish an atomic coherent state.\\
So far we have analyzed the cases with $N=2$ and $N=3$ bosons. The same study could be performed also in the presence of a number of particles much larger than $3$ --as we have done in Ref. \cite{main} for nondipolar bosons-- by carrying out numerics to find the ground state of the associated Hamiltonian. Increasing $N$ might have the effect to "expedite" the crossover from the atomic coherent state to the NOON one. From the correlation properties perspective, one expects that an increase of $N$ might make smaller (than those of the cases with $N=2,3$ bosons) both the range of Hamiltonian parameters where the system exhibits a high degree of coherence (i.e., large values of $\alpha$) and the absolute value of the (effective) on-site interaction signing the maximum of the entanglement entropy.

\section{Conclusions and perspectives}

We have investigated a finite number $N$ of interacting dipolar bosonic atoms confined in a one-dimensional double-well-induced geometry. Within the two-site extended Bose-Hubbard (EBH) model framework (which takes into account, in addition to the familiar BH terms, the density-density interaction, correlated hopping and bosonic-pair hopping involving atoms in adjacent wells) we have carried out the zero-temperature analysis for $N=2$ and $N=3$ and found analytical formulas for the eigenvectors and eigenvalues of the corresponding ground states. These have been characterized, from the correlations point of view, by calculating analytically the Fisher information, the coherence visibility, and the entanglement entropy. We have analyzed the role of the correlated hopping (working in the absence of bosonic pair hopping with the nearest-neighbor interaction which is shown to be reabsorbed in an effective onsite interaction) in determining the form of the ground state. We have pointed out that even if the onsite interaction is not strong enough to guarantee the occurrence of the Schr\"odinger-cat state on the attractive side, or the formation of the twin Fock ($N=2$) and symmetric superposition of two separable Fock states ($N=3$) on the repulsive side, the presence of the collisionally-induced tunneling makes possible such circumstances. The second part of our analysis has concerned with the atomic coherent state. We have fixed the onsite interaction to a value such that to establish the cat-like (strong attraction) and Fock states (strong repulsion) and widely explored the density-induced tunneling range. We have, in such a way, observed that the correlated hopping tends to destroy the afore mentioned ground states in favor of a delocalized atomic coherent state. A detailed analysis of Fisher information, coherence visibility, and entanglement entropy has allowed us to have a complete picture of these two scenarios.

At this point, some considerations about future directions are in order. In Ref. \cite{boris}
the authors have demonstrated the crucial importance of the competition between the contact and
dipole-dipole interactions of opposite signs in determining the ground state of the system. Then,
this issue (also with the numerical diagonalization with numbers of particles much larger than $3$) and the study of the possibility to apply the present analysis to the two-site Hubbard system in the presence of two fermions --with opposite orientations-- are the ideal
candidates for our forthcoming studies.


\begin{acknowledgement}
GM and LS acknowledge Ministero dell'Istruzione, della Ricerca, e dell'Universit\`a (MIUR) for the support provided by
PRIN Grant No. 2010LLKJBX.
\end{acknowledgement}

\end{document}